\documentclass[12pt]{article}
\usepackage{epsfig}
\usepackage[koi8-r]{inputenc}
\usepackage[english,russian]{babel}
\usepackage{amssymb,amsmath}
\usepackage{floatflt}
\usepackage[dvips]{color}
\begin{document}
\begin{otherlanguage}{english}
\title{Schr\"{o}dinger's cat versus Darwin}
\author{ Z.~K.~Silagadze \\
Budker Institute of Nuclear Physics and \\
Novosibirsk State University, 630 090, Novosibirsk, Russia}

\date{}

\maketitle

\begin{abstract}
Sun Wu-k'ung, an immortal Monkey-King of Chaos learns modern physics from 
the Patriarch Bodhi and questions the Darwinian evolution. He finds that 
the modern physics indicates towards the intelligent design as a vastly more 
probably origin of humans than the random evolution by mutations and natural 
selection. 
\end{abstract}
\end{otherlanguage}

\newpage
\section{Preface}
\begin{otherlanguage}{russian}
\begin{verbatim}
Я не знаю, Земля кружится или нет,
Это зависит, уложится ли в строчку слово.
Я не знаю, были ли моими бабушкой и дедом
Обезьяны, так как я не знаю, хочется ли мне сладкого или кислого.
Но я знаю, что я хочу кипеть и хочу, чтобы солнце
И жилу моей руки соединила обшая дрожь.
Но я хочу, чтобы луч звезды целовал луч моего глаза,
Как олень оленя (о, их прекрасные глаза!).
Но я хочу, чтобы, когда я трепещу, общий трепет приобшился вселенной.
И я хочу верить, что есть что-то, что остается,
Когда косу любимой девушки заменить, например, временем.
Я хочу вынести за скобки общего множителя, соединяющего меня, 
Солнце, небо, жемчужную пыль. 

                           Велимир Хлебников, 1909.
\end{verbatim}
\end{otherlanguage}
\begin{otherlanguage}{english}
\begin{verbatim}
I do not know, the Earth turns or not,
This depends how the words will fit the line.
I do not know, whether my grandmother and grandfather were 
Monkeys, since I don't know do I want sweet or sour.
But I do know that I want to boil and I want the Sun 
And the vein of my hand to be connected by the common trembling.
But I do want the ray of the star to  kiss the ray of my eye, 
As deer to deer (oh, their excellent eyes!).
But I want when I quiver the general trembling of the universe
to join my trembling.
And I want to believe that there is something which remains, 
When the braid of the dear girl is faided away, for example, 
by the time.
I want to take out of brackets the common factor, which connects me, 
the Sun, pearl dust, and the sky. 

                             Velimir Khlebnikov, 1909.
\end{verbatim}

\newpage
\section{Introduction}
The most beloved story in China is the book ``Monkey'' written by Wu Cheng-En
in the sixteenth century and it is about an immortal Monkey-King of Chaos 
Sun Wu-k'ung \cite{Lee,JTWR}.
\begin{figure}[htb]
     \centerline{\epsfxsize 70mm\epsfbox{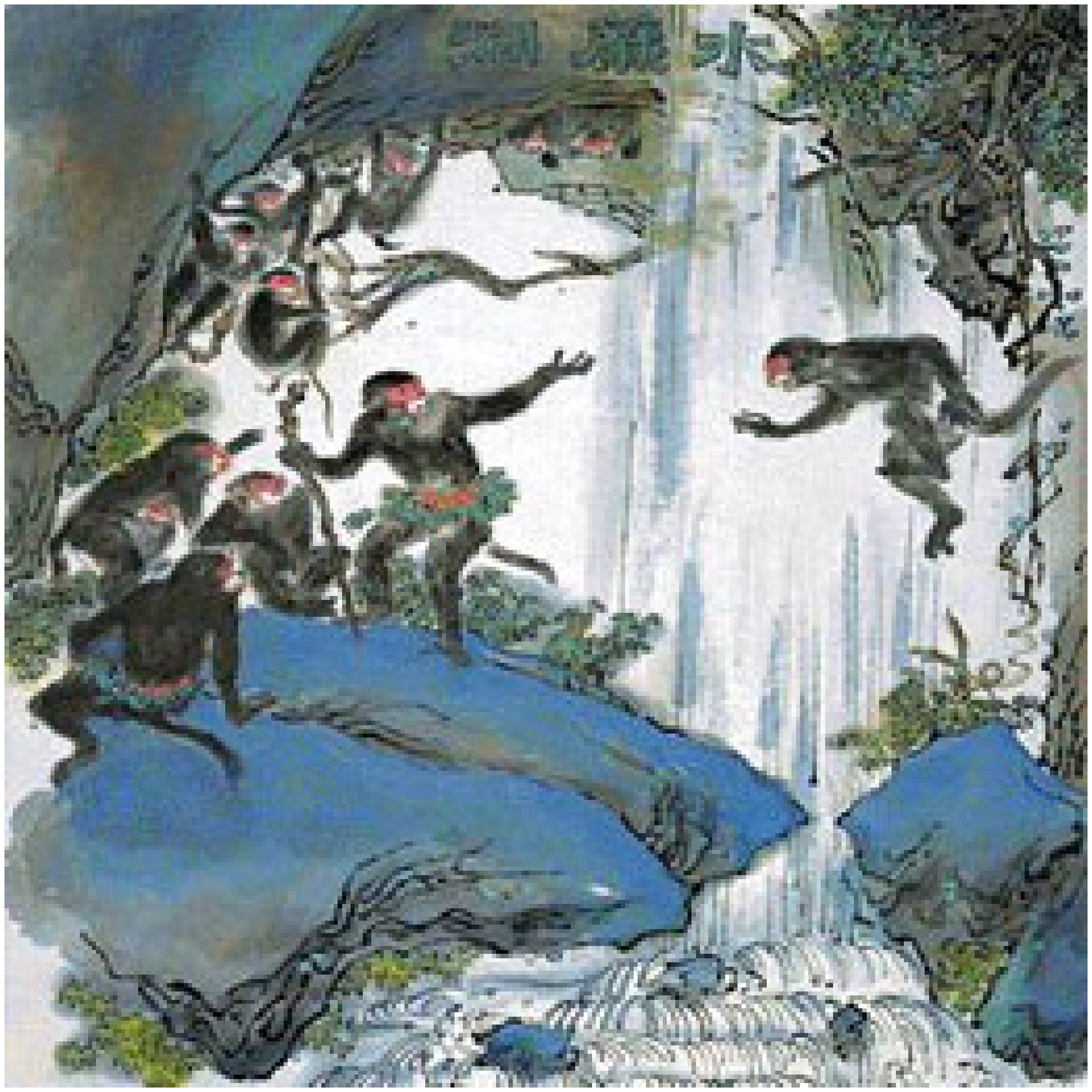}\hspace*{5mm}
\epsfxsize 70mm\epsfbox{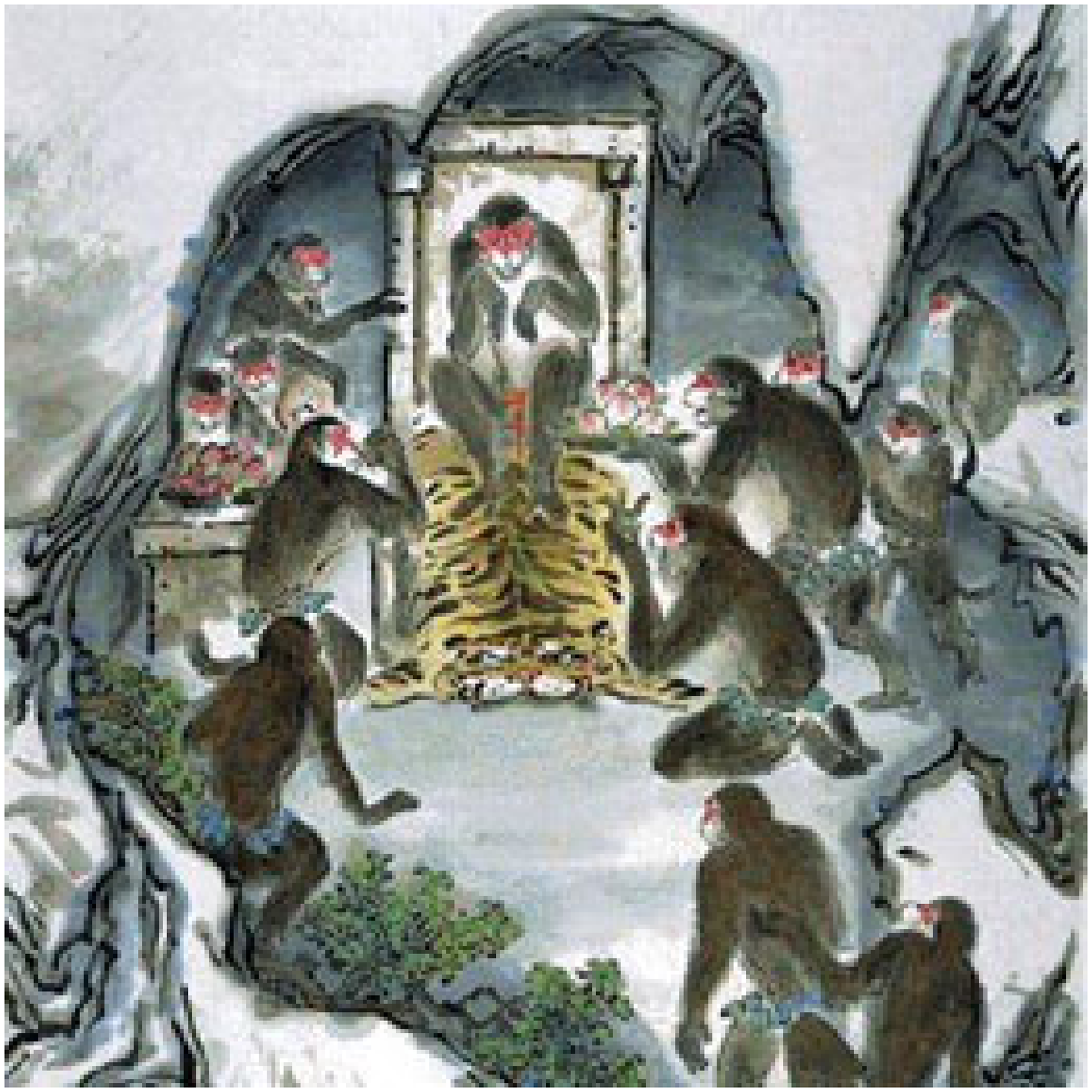}}
\end{figure}

Sun Wu-k'ung was born of primordial Chaos in a remote island from a stone  
impregnated by the sky. The other monkeys honored him as their king but
soon in search of immortality he left the island and traveled on a raft to 
civilized lands. Acquiring human speech and manners, he decided to become
the disciple of a Buddhist Patriarch Bodhi. Bodhi was initially reluctant 
about this but was impressed by monkey's determination and brilliant 
abilities.
\begin{figure}[htb]
     \centerline{\epsfxsize 70mm\epsfbox{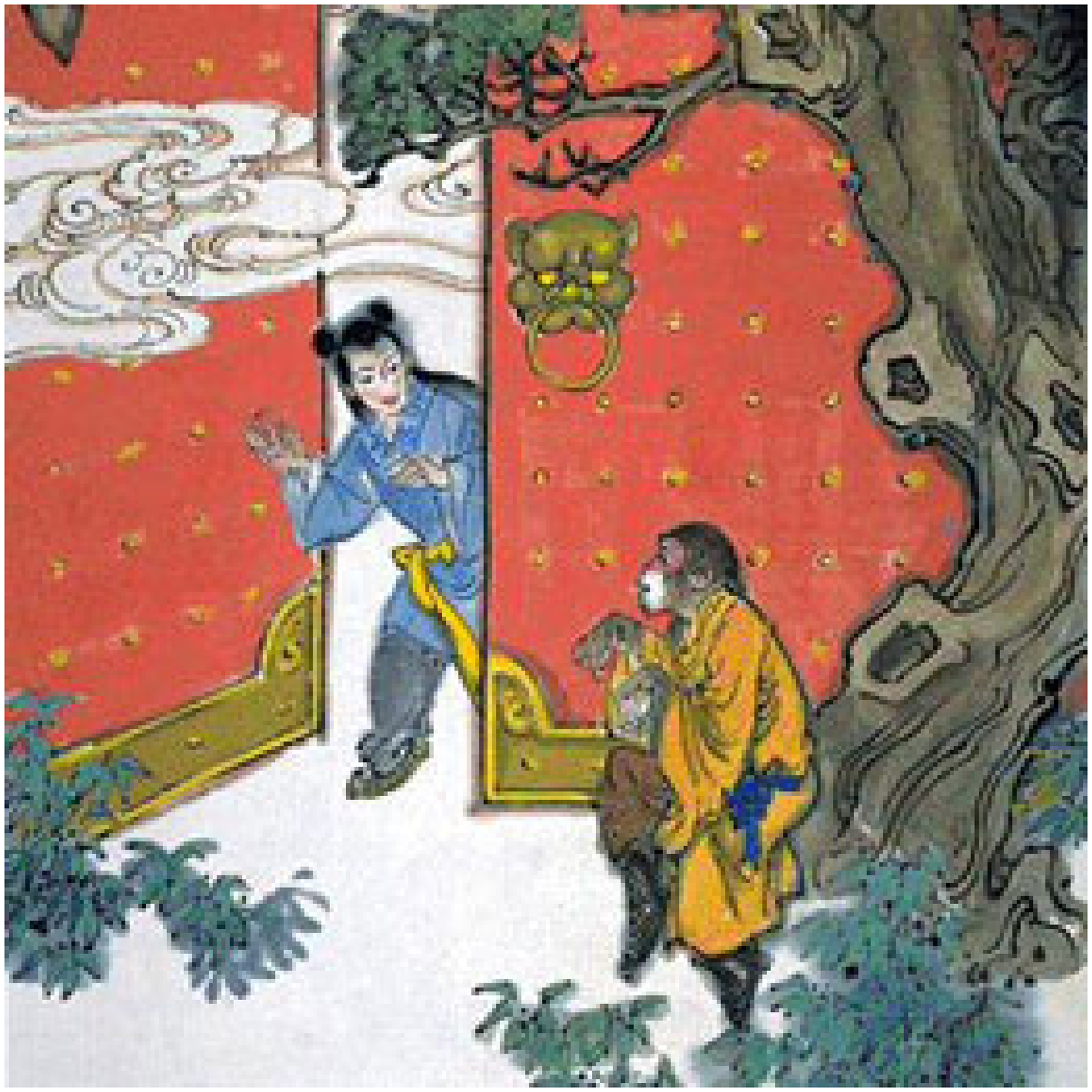}\hspace*{5mm}
\epsfxsize 70mm\epsfbox{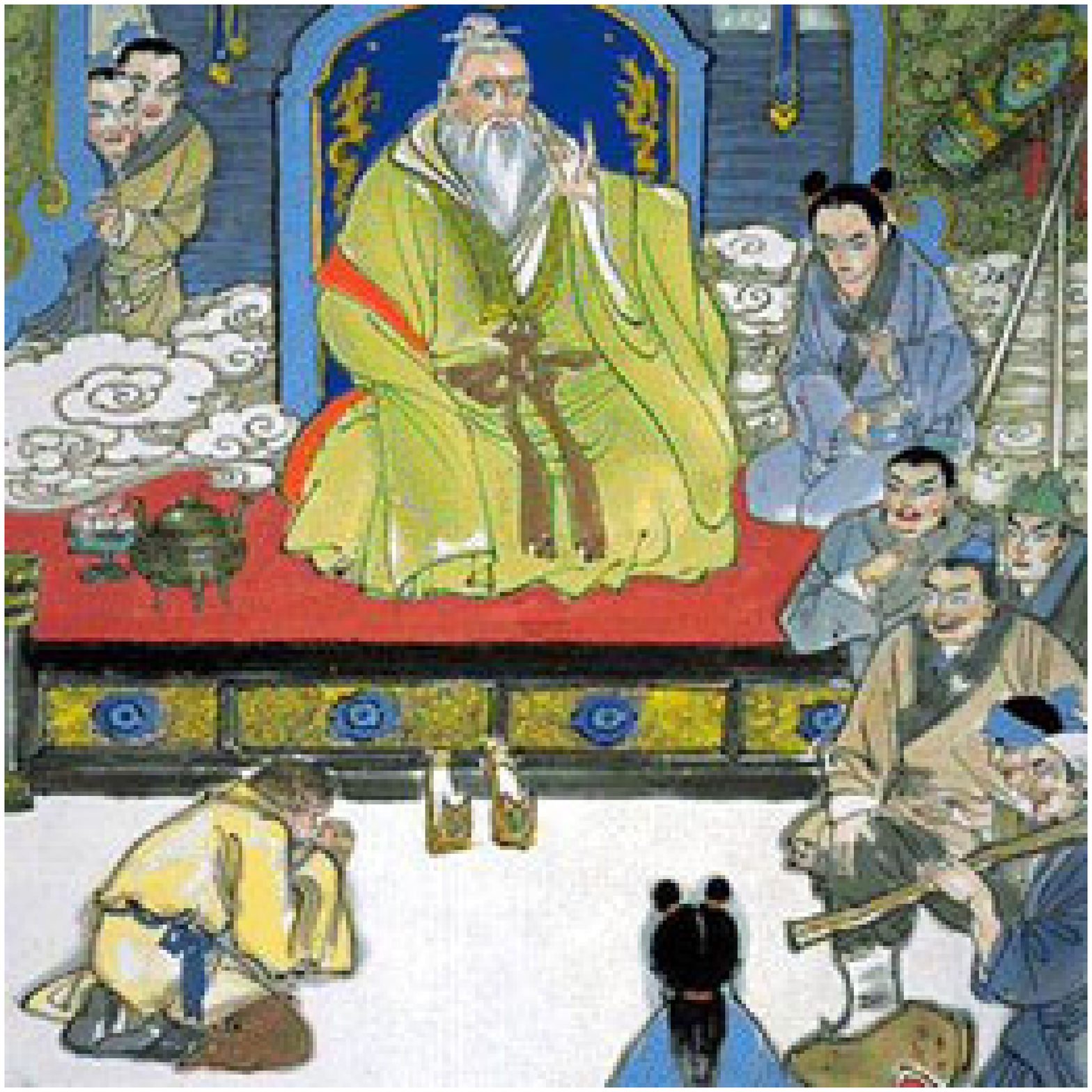}}
\end{figure}

In three years Wu-k'ung learned perfectly well magical transformations, 
cloud dancing, and martial arts. Soon he was able to leap thousands of miles 
with one somersault. Of course, these marvelous abilities were not acquired 
without learning modern physics and the next chapter describes what he learned
in physics from the Patriarch Bodhi. 

\section{Wu-k'ung learns modern physics}
The glorious building of modern physics is based on three elephants (or whales
according to another scientific school to which Bodhi is antagonistic). The
elephants stand on the back of a gigantic tortoise which by itself swims in
a world ocean of dark energy. 
\begin{figure}[htb]
     \centerline{\epsfxsize 90mm\epsfbox{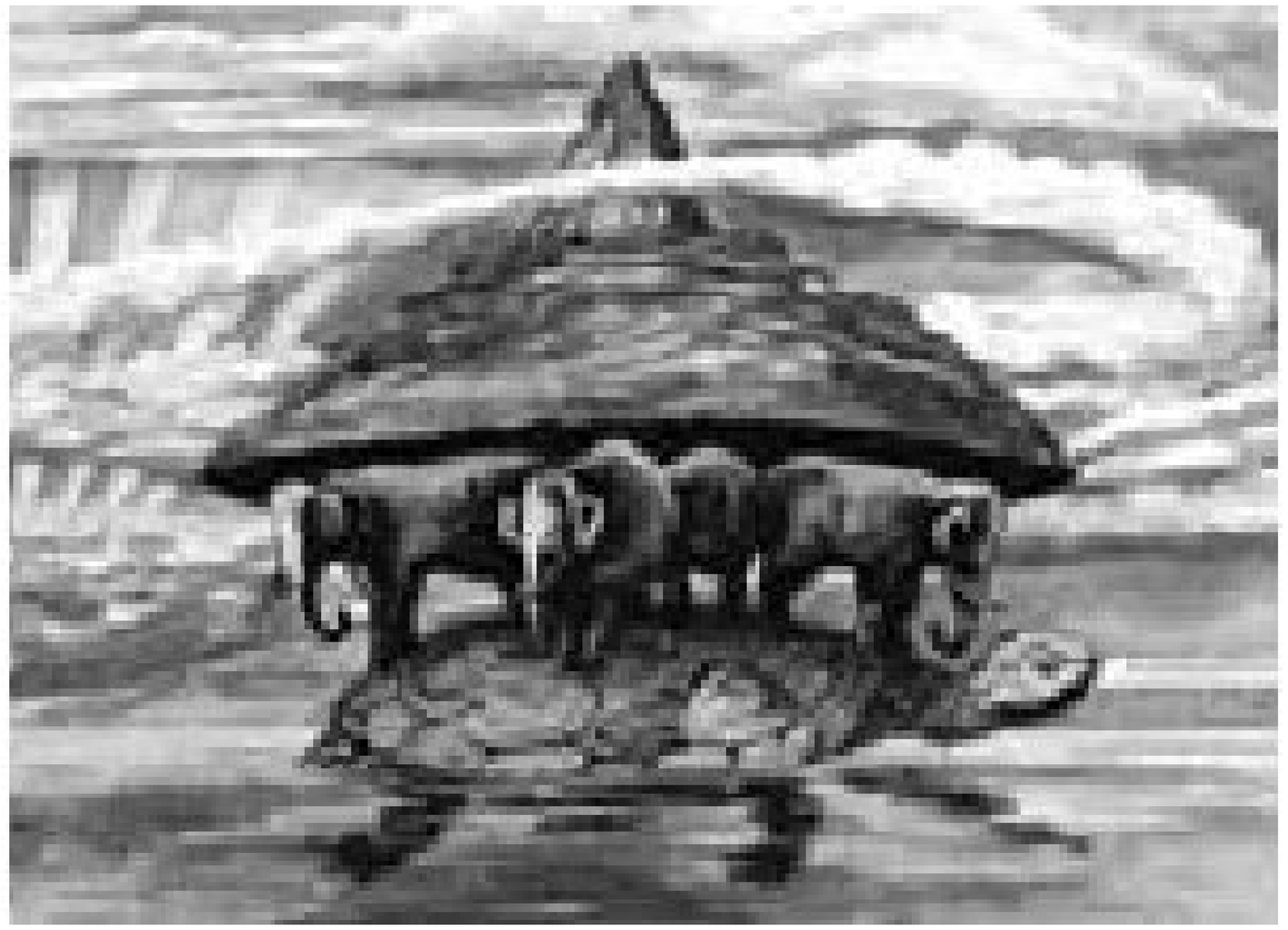}}
\end{figure}

We know very little about the tortoise (Grand Unification, Theory of
Everything), or about the dark energy, but the elephants are studied perfectly
well and these elephants are
\begin{itemize}
\item gauge symmetry
\item special relativity
\item quantum mechanics
\end{itemize}
-- Let us take a closer look at them -- said the Patriarch to Wu-k'ung.
\subsection{Gauge symmetry} 
-- It's better to learn gauge theory by observing falling cats -- began Bodhi
his narrative. In the middle ages Europeans were not very fond of cats 
because cats were considered as devil creatures. At holidays people used to 
go on cat hunting. They catch these poor creatures and tormented them in all 
wicked ways: roasted them alive, whipped to death, or threw them into a 
boiling water \cite{NZZ}.

-- Oh, my Master, -- cried Wu-k'ung -- what a stupid brutality of ignorant 
and rude people thinking themselves as pious!

-- They just considered cats as demons, -- calmly continued the Patriarch --
and they had a good proof for the cat's devil nature. One favorite amusement
at that days was to through down the cats from a church tower, and pretty 
often the cats survived safe this sure death adventure. Why not a proof
of cat's supernatural resistance against inevitable death and hence her
devil nature?

In more enlightened age, two veterinarians Wayne Whitney and Cheryl Mehlhaff 
studied the so called  ``feline high-rise syndrome'' in 1987. They had their
veterinary clinic in Manhattan where cats often fall from New-York 
skyscrapers. Their findings are schematically shown in the figure below 
\cite{WM,Diamond}.
\begin{figure}[htb]
     \centerline{\epsfxsize 150mm\epsfbox{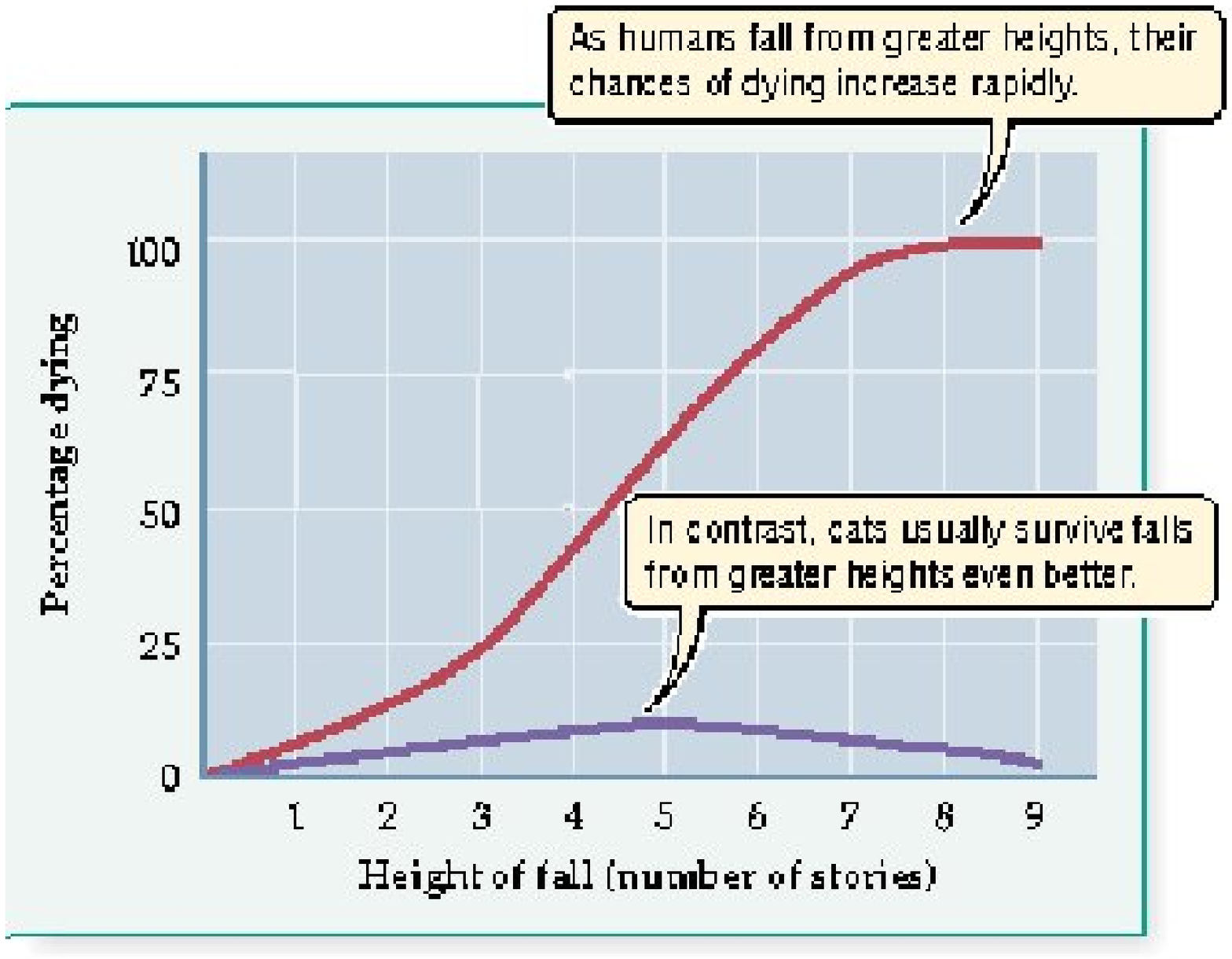}}
\end{figure}

-- Quite a remarkable finding -- said Wu-k'ung, surprised a bit -- contrary 
to humans whose chance to die from the falling accident steadily increases 
with height and approaches 100\% above a height of about seven floors, the 
mortality of cats initially rises to 10\% and then falls down to about
5\%! Increadibly, cat's prospects of survival improve with height! Surely
there is some explanation of this queer fact?

-- Yes, of course, -- replied the Patriarch -- one can imagine the following
explanation \cite{Diamond,Diamond1}. Cats reach a terminal velocity of
about 100 km/h after flying  about 30 meters in air. After this moment
the air resistance balances the pull of earth's gravity and cat's velocity
ceases to grow. At first sight, one expects the height to be no longer 
important for cat's survival after the terminal velocity is reached. However, 
then the acceleration disappears the cat probably relax and extends its limbs 
horizontally,  flying like a flying-squirrel from this moment. This increases
its effective transverse area and hence air resistance, reducing the terminal 
velocity.
 
-- But more important for us is the one particular ingredient of cat's 
enduring power of survival, -- continued the Patriarch -- it's their ability
to change the body's orientation in the free fall. After all cats always land 
on their feet, no matter how they were dropped, is not it? Look at this picture
of the cat dropped upside down. Within less than half a second it turns
around its longitudinal axis and land on its feet unharmed.

-- But I'm surprised, -- begin Wu-k'ung -- I know the conservation of 
momentum precludes the center of mass of a system to be moved by inner forces.
There was a guy, it seems Baron M\"{u}nchhausen was his name, who claimed 
that he once escaped from a swamp by pulling himself up by his own hair. But
he is notorious  liar, is not he? And what about the conservation of angular
momentum? Does not it precludes the cat to change its orientation if cat's
initial angular momentum was zero?
\begin{figure}[htb]
     \centerline{\epsfxsize 50mm\epsfbox{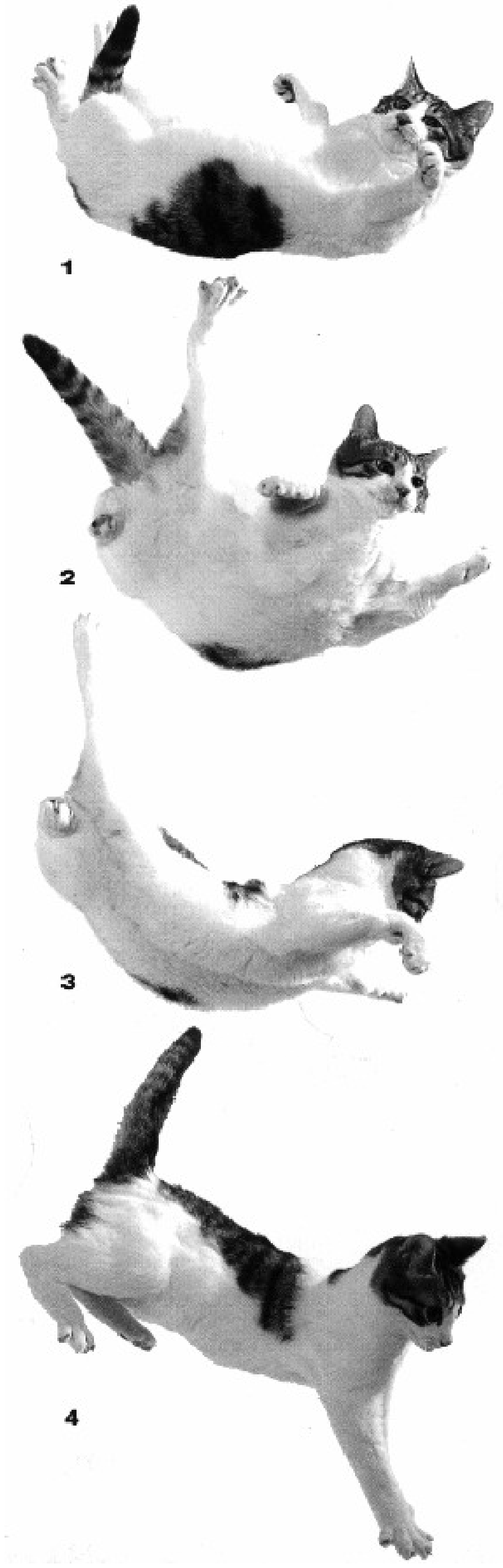}}
\end{figure}

-- The cat just uses the conservation of angular momentum in a clever way
\cite{NZZ}  -- smiled the Patriarch. -- Look at the picture. In a first 
stroke the cat pulls the front legs tightly against the body and stretches 
the hind legs perpendicular to the longitudinal axis of the body away. Thus 
it increases the moment of inertia of the rear part of the body and lowers 
the moment of inertia of its front part. But the angular momentum equals to 
the product of the moment of inertia over the angular velocity. Therefore, 
when the cat turns its head and the front body quickly downward, the rear 
part rotates in the counter direction but with much less angular velocity, 
in complete agreement with the angular momentum conservation law. As a result, 
the rotation angle of the front part may be near $180^\circ$ while the counter
rotation angle of the rear part will be much smaller. Then the cat repeats the 
maneuver in reverse order: now it stretches the front legs from the body away 
and pushes the rear ones to the body. This allows the cat to rotate the rear 
part of the body downward with only small swing of the front part and hence 
to eliminate the twist of its body at the end of this two step flip-flap. 

-- Really ingenious! -- said Wu-k'ung -- you know monkeys as close relatives
of humans are vice-kings of Nature and I always largely regarded cats as 
inferior creatures, but now I'm beginning to change my opinion. However, how 
can we describe cat's somersaulting mathematically?

-- Good question -- answered the Patriarch -- and this is the heart of the 
matter. But to explain what I have in mind we need some simple model of a cat.
And here is one, inspired by \cite{scat}. 
\begin{figure}[htb]
     \centerline{\epsfxsize 70mm\epsfbox{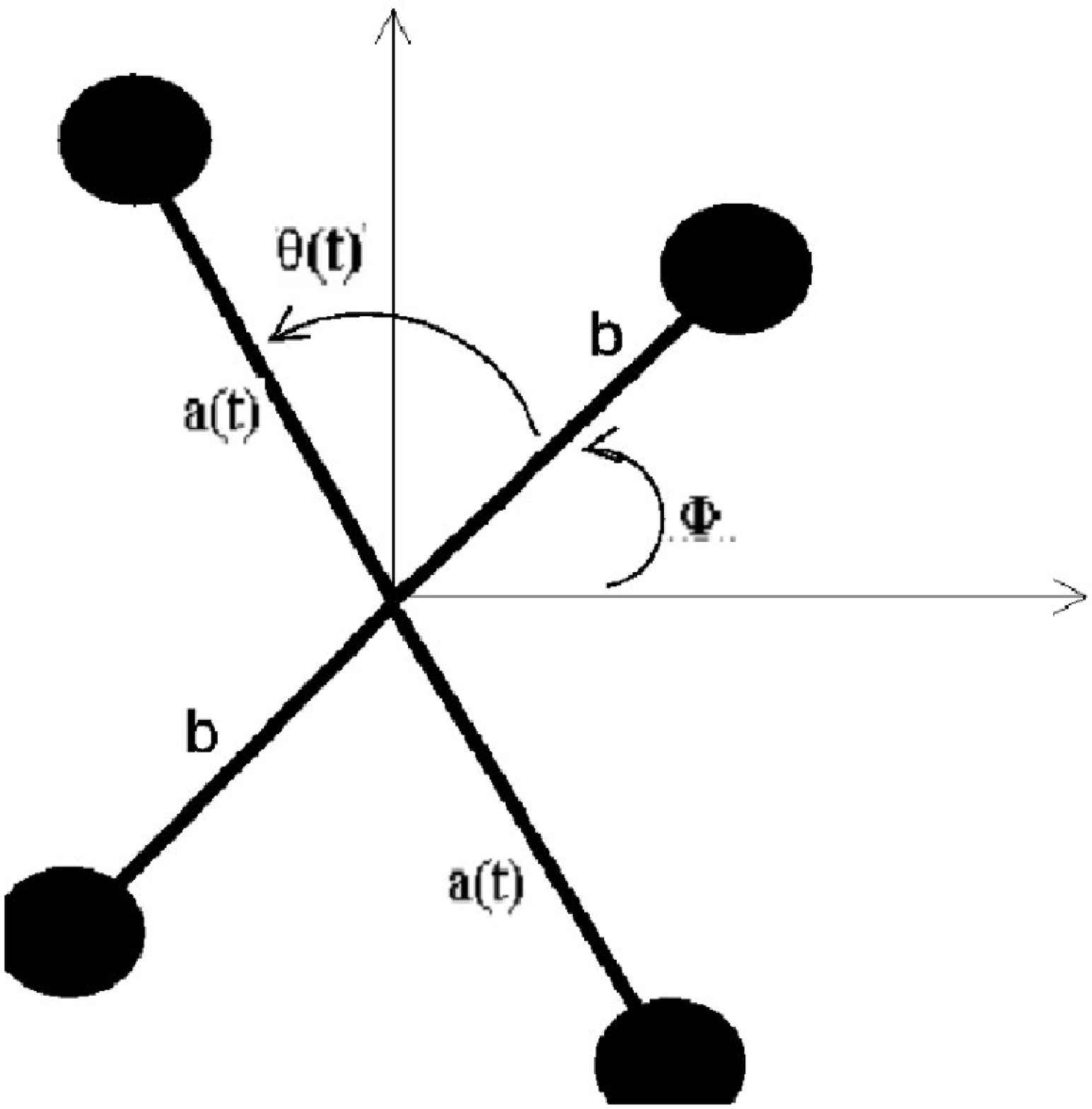}}
\end{figure}

We have two dumbbells of identical mass $2m$ on the common pivot. The ``cat'' 
can control the angle $\theta(t)$ between the dumbbells, as well as the 
half-length $a(t)$ of one dumbbell. Cat's goal is by manipulating $a$ and 
$\theta$ to change its orientation specified by the angle $\Phi$. We can 
regard $(a,\theta)$ as polar coordinates in the cat's shape space which in 
our case is just a plane. Any somersaulting of the cat is described in the 
shape space by some curve $\gamma$ which connects its initial and final 
shapes. If the cat restores its shape after performing a flip, the curve 
$\gamma$ will be a closed loop.

We can calculate the angular momentum of the ``cat'' very simply
$$L=2m(a^2+b^2)\dot \Phi+2ma^2\dot \theta,$$
where $b$ is the half-length of the dumbbell with fixed length. If the
angular momentum was initially zero so it will remain all the time as the cat
does not experience any external torque. This enables us to connect time
derivatives of angles $\Phi$ and $\theta$ as follows 
$$\dot \Phi=-\frac{a^2}{a^2+b^2}\,\dot \theta.$$
Using this relation, for the turning angle of the cat we get
$$\Delta \Phi=\int\limits_0^T \dot \Phi dt=
\int\limits_\gamma \left [-\frac{a}{a^2+b^2}\,ad\theta+0\,da\right ]=
\int\limits_\gamma \vec{A}\cdot d\vec{l},$$
where $\gamma$ is the path in the $(a,\theta)$ plane which describes 
consecutive changes in the shape of the cat. Besides we have for the ``vector
potential'' $\vec{A}$ and for the line element $d\vec{l}$ the following 
expressions in polar coordinates
$$\vec{A}=-\frac{a}{a^2+b^2}\,\vec{e}_\theta,\;\;\;
d\vec{l}=da\,\vec{e}_a+a\,d\theta\,\vec{e}_\theta.$$

As we see the time parametrization have dropped from the equation. The only 
thing that matters as far as the total rotation of the cat is concerned is 
the geometry of the change. The speed at which $a$ and $\theta$ change does 
not matter and only the curve $\gamma$ in the cat's shape space, that is a 
succession of cat's shapes, uniquely determines the total rotation. 

-- However, you can not have any doubts that cats perform this exercise very
quickly, -- remarked the Patriarch. --  As the American physiologist Donald 
McDonald found in 1960 by using modern high-speed camera, for the full motion 
cascade cat only needs one-eighth of a second, that is it is properly turned
after only the first eight centimeters of the free fall when dropped upside 
down \cite{NZZ}.

But let us return to the formula for the total rotation angle. By using 
Stokes' theorem, we can rewrite it as follows
$$\Delta \Phi=\oint \vec{A}\cdot d\vec{l}=\iint\limits_S \vec{B}\cdot
d\vec{S},$$
where
$$\vec{B}=\nabla\times\vec{A}=-\frac{2b^2}{(a^2+b^2)^2}\,\vec{e}_z,\;\;\;
d\vec{S}=a\,da\,d\theta\,\vec{e}_z.$$
An analogy with a gauge theory can be made even more apparent if we compare
the formulas obtained for $\Delta \Phi$ to the formula of electron's phase
shift in the Aharonov-Bohm effect \cite{AHB}
$$\Delta \Phi=\frac{e}{\hbar c}\oint \vec{A}\cdot d\vec{l}=
\frac{e}{\hbar c}\iint\limits_S \vec{B}\cdot d\vec{S},$$
where now $\vec{A}$ is a genuine vector potential and $\vec{B}$ is the 
corresponding magnetic field.

-- In connection with the Aharonov-Bohm effect -- mentioned Wu-k'ung -- I just
remembered the Arnold principle: {\it If a notion bears a personal name, then 
this name is not the name of the discoverer,} which is said to be so awfully
general that being applicable even to itself (the Berry principle) 
\cite{Arnold}. Ten years before \cite{AHB}, Ehrenberg and Siday in their 
studies of electron microscopy \cite{EHS} already obtained the main results 
of \cite{AHB}. However their work went completely unnoticed. As remarked 
by Chambers, who one of the first experimentally verified the Aharonov-Bohm 
effect, Ehrenberg and Siday ``perhaps did not sufficiently emphasize the 
remarkable nature of their result'' \cite{Chambers}. 

-- This is amusing, of course, -- said the Patriarch -- however the importance
of contribution of Aharonov and Bohm is beyond any question. 

-- With some caveat to the Arnold principle, -- continued he -- I should say
that Guichardet was the first who made explicit a connection between gauge 
fields and falling cat \cite{Guichardet}. Then important contributions by
Wilczek and his PhD student Shapere followed \cite{Wilczek,Wilczek1}. They
also showed that gauge fields are a natural language to describe 
self-propulsion of microorganisms at low Reynolds number \cite{Wilczek2}
(a very nice description of life at low Reynolds number was given by
Purcell \cite{Purcell}). Montgomery in his influential papers 
\cite{Montgomery,Montgomery1} investigated the question how cats can perform
their maneuvers efficiently (and you can't have even slightest doubt that 
they do this very efficiently). It turns out that for this goal cats should 
``solve'' almost the same Wong's equations that describe the motion of 
a colored quark in a Yang-Mills field \cite{Marsden}. You can consult 
a review article \cite{review} for additional information. 

\subsection{Special relativity}
After Wu-k'ung mastered gauge theory, the time has come to speak about special
relativity.

-- The essence of special relativity -- begin solemnly the Patriarch -- can be
expressed by one sentence: {\it the geometry of empty space-time is Minkowski
geometry} (M.C. Escher's famous woodcut {\it Circle Limit IV} below 
illustrates this geometry). 
\begin{figure}[htb]
     \centerline{\epsfxsize 90mm\epsfbox{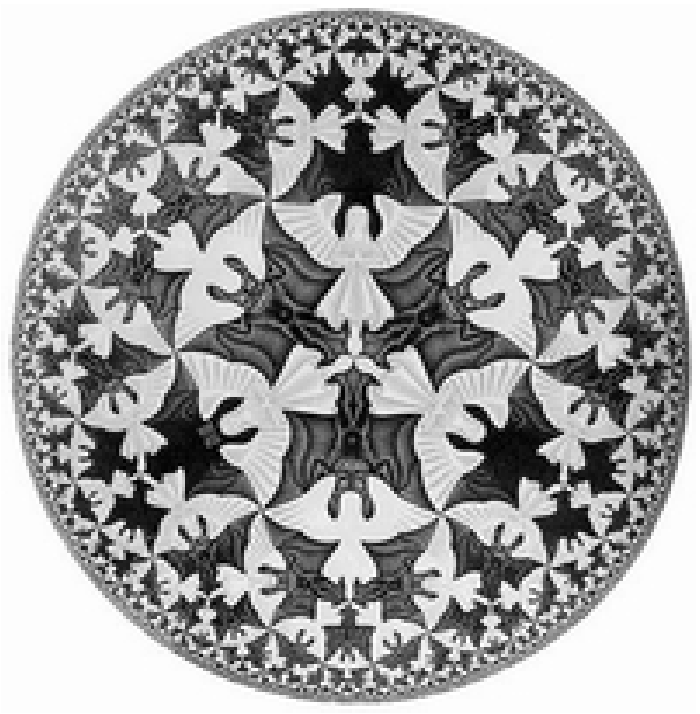}}
\end{figure}
 
-- But I suspect -- humbly remarked Wu-k'ung -- to duly appreciate the meaning
of this sole sentence we need a deep insight into the nature of geometry.

-- That's right -- said the Patriarch -- and what is geometry? You surely
know about Euclidean geometry and know that the points of the Euclidean plane 
can be represented by complex numbers. The distance between two points
represented by complex numbers $z_1$ and $z_2$ is given by 
$$d^2(z_1,z_2)=(z_2-z_1)(\bar z_2 - \bar z_1),$$
where the bar means complex conjugation. This is called parabolic linear 
measure for reasons you will understand after you scrutinize this folio --
the Patriarch transferred to  Wu-k'ung some of Klein's papers 
\cite{Klein,Klein1,Klein2}.

-- But I also heard about  non-Euclidean geometries of Lobachevsky and 
Riemann --  got excited Wu-k'ung.

-- For monkey you are well versed in mathematics -- praised the Patriarch.
In Lobachevsky geometry (hyperbolic linear measure) the distance is given by
$$\sinh^2{d(z_1,z_2)}=
\frac{(z_2-z_1)(\bar z_2 - \bar z_1)}{[1-z_2\bar z_2][1-z_1\bar z_1]}$$
and in Riemann or elliptic geometry (elliptic linear measure) by
$$\sin^2{d(z_1,z_2)}=
\frac{(z_2-z_1)(\bar z_2 - \bar z_1)}{[1+z_2\bar z_2][1+z_1\bar z_1]}.$$

-- Complex numbers are obtained by adding to the real numbers a special 
element $i$ which is a solution of the quadratic equation $i^2=-1$. But why
this particular quadratic equation? What is special about it? -- asked 
Wu-k'ung.

-- Nothing special -- answered the Patriarch -- On equal footing we can assume
a special element $e$ to be a solution of the general  quadratic equation 
$Ae^2+Be+C=0$. In fact, in this way we get three different types of 
generalized complex numbers $a+eb$. If the discriminant $D=B^2-4AC$ is 
negative, we get the ordinary complex numbers $a+ib$ and one can assume 
without loss of generality that $i^2=-1$. Indeed, if $Ae^2+Be+C=0$ and $D<0$ 
then
$$i=\frac{B}{\sqrt{-D}}+\frac{2A}{\sqrt{-D}}\,e$$
just have this property $i^2=-1$ and every hyper-complex number $a+eb$ can be 
rewritten as  $a^\prime+ib^\prime$.

If the discriminant is zero, we get the so called dual numbers $a+\epsilon b$
and one can assume that $\epsilon^2=0$. And if the discriminant is positive,
we get the double numbers $a+eb$ with $e^2=1$ \cite{Yaglom}.

-- If now we change complex numbers in the above formulas -- guessed Wu-k'ung
-- we get other types of geometry?

-- Absolutely right, -- the Patriarch seemed satisfied by Wu-k'ung's insight. 
-- For hyperbolic linear measure, double numbers lead to de Sitter geometry
and dual numbers lead to co-Minkowski geometry. For elliptic linear measure,
double numbers correspond to anti-de Sitter geometry and dual numbers to
co-Euclidean geometry. At last, for parabolic linear measure, dual numbers
imply Galilean geometry and double numbers lead to Minkowski geometry.
On the whole, we have nine so called Cayley-Klein geometries of the plane
(more details can be found in \cite{RWT}).
\begin{table}[htb]
\begin{center}
\begin{tabular}{|c||c|c|c|}
\hline
measure of & \multicolumn{3}{|c|}{measure of lengths} \\
\cline{2-4}
angles & Hyperbolic & Parabolic & Elliptic \\
\hline \hline
$\;$ Hyperbolic$\;$ & Doubly hyperbolic & \hspace*{2mm} Minkowski
\hspace*{2mm} &  co-Hyperbolic \\
           & (de-Sitter) &  & (anti de-Sitter) \\
\hline
Parabolic & co-Minkowski & Galilean &
\hspace*{1mm} co-Euclidean \hspace*{1mm} \\
          &              &           &   \\
\hline
Elliptic & Hyperbolic & Euclidean & Elliptic \\
         & (Lobachevsky) & & (Riemann) \\
\hline
\end{tabular}
\end{center}
\end{table}

-- However, -- remarked  Wu-k'ung, -- special relativity is not a merely
geometric theory, but a physical one and it includes concepts like causality,
reference frames, inertial motion, relativity principle.

-- Absolutely right again, -- said the Patriarch. Let us see what geometry
will arise if we stick to these physical notions.

It is intuitively appealing to suppose that meter sticks do not change their 
lengths when gently set in uniform motion -- continued the Patriarch. -- 
However, this is not quite obvious and we simply admit a more general 
possibility, instead of supposing unchanged lengths. Accordingly, we change 
the Galilean transformations, which describe a transition from one inertial 
frame to another, by  
$$x^\prime=\frac{1}{k(V^2)}\left (x-Vt\right ),$$
where the scale factor $k(V^2)$ accounts for the possible change in the 
length of the meter stick when it is set in motion with  velocity $V$. 
It can depend only on the magnitude of the relative velocity $V$, because 
the Relativity Principle and the isotropy of space is assumed to be valid.

Due to the Relativity Principle, the same relation holds if unprimed 
coordinates are expressed through the primed ones, with $V$ replaced 
by $-V$. Therefore,
$$x=\frac{1}{k(V^2)}\left (x^\prime+Vt^\prime\right )=
\frac{1}{k(V^2)}\left [\frac{1}{k(V^2)}\left (x-Vt\right )+ Vt^\prime
\right ].$$
Solving for $t^\prime$, we get
$$t^\prime =\frac{1}{k(V^2)}\left [t-\frac{1-k^2(V^2)}{V}\,x\right ].$$
Now we are in a position to derive the velocity addition rule,
$$v^\prime_x=\frac{dx^\prime}{dt^\prime}=\frac{dx-Vdt}{dt-\frac{1-k^2}{V}
\,dx}=\frac{v_x-V}{1-\frac{1-k^2}{V}\,v_x}.$$
By using the Relativity principle, it will be convenient to write down it in 
the form
$$v_x=\frac{v^\prime_x+V}{1+\frac{1-k^2}{V}\,v^\prime_x}\equiv 
F(v^\prime_x,V).$$
$F$ must be an odd function of its arguments $F(-x,-y)=-F(x,y)$, because If 
we change the signs of both velocities $v^\prime_x$ and $V$ it is obvious 
that the sign of the resulting velocity $v_x$ will be also changed.

Consider now three bodies $A$, $B$ and $C$ in a relative motion. Let $V_{AB}$
denote the velocity of $A$ with respect to $B$ so that $V_{BA}=-V_{AB}$.
Then we will have
$$F(V_{CB},V_{BA})=V_{CA}=-V_{AC} 
=-F(V_{AB},V_{BC})=-F(-V_{BA},-V_{CB})=
F(V_{BA},V_{CB}). $$
Therefore $F$ is a symmetric function of its arguments and then 
$F(v^\prime_x,V)=F(V,v^\prime_x)$ immediately yields
$$\frac{1-k^2(V^2)}{V}\,v^\prime_x=\frac{1-k^2(v^{\prime\,2}_x)}
{v^\prime_x}\,V,$$
or
$$\frac{1-k^2(V^2)}{V^2}=\frac{1-k^2(v^{\prime\,2}_x)}{v^{\prime\,2}_x}
\equiv K, $$
where $K$ is a constant.

If $K>0$, we can take $K=\frac{1}{c^2}$ and introduce a dimensionless
parameter $\beta=\frac{V}{c}$. As a result, we get the Lorentz transformations
\begin{eqnarray} & &
x^\prime=\frac{1}{\sqrt{1-\beta^2}}(x-Vt),
\nonumber \\ & &
t^\prime=\frac{1}{\sqrt{1-\beta^2}}\left (t-\frac{V}{c^2}\,x\right ),
\nonumber
\end{eqnarray}
and, hence, Minkowski geometry. In this case, velocity addition rule indicates
that $c$ is an invariant velocity. If $K=0$, we recover the Galilean 
transformations and the geometry behind the corresponding space-time is
Galilean geometry. Note that Galilean geometry can be considered as a limiting
case of Minkowski geometry when $c\to\infty$. The case $K<0$ corresponds
to Euclidean space-time which does not allow us to define an invariant
time order between events, that is to distinguish future from past and 
introduce causality (see \cite{RWT} for more details).

-- As I see -- Wu-k'ung seemed a bit puzzled, -- if we stick to the 
Relativity principle, space isotropy and inertial reference frames, the only 
space-time geometry which emerges is Minkowski geometry along with its rather 
singular limit of Galilean geometry. And what about other Cayley-Klein 
geometries?

-- In fact, -- answered the Patriarch, -- the existence of inertial reference
frames is, in general, not obviously guaranteed. Therefore we can consider
the boost transformations in a more general context as not necessarily 
representing transitions from one inertial frame to another. This opens a
possibility to generalize special relativity by deforming its underlying 
symmetry structure - the Poincar\'{e} algebra \cite{Bacry}.

The symmetry group of special relativity is the ten-parameter Poincar\'{e}
group. Ten basis elements of its Lie algebra are the following: the generator
$H$ of time translations; three generators $P_i$ of space translations along
the $i$-axis; three generators $J_i$ of spatial rotations; and three
generators $K_i$ of pure Lorentz transformations,  which can be considered as
the inertial transformations (boosts) along the $i$-axis. The commutation
relations involving $J_i$ have the form
$$[J_i,H]=0,\;\;\;[J_i,J_j]=\epsilon_{ijk}J_k,\;\;\;
[J_i,P_j]=\epsilon_{ijk}P_k,\;\;\; [J_i,K_j]=\epsilon_{ijk}K_k. $$
We can not change these commutation relations without spoiling the spatial
isotropy.  However, other commutation relations
$$[H,P_i]=0,\; [H,K_i]=P_i,\; [P_i,P_j]=0,\;
[K_i,K_j]=-\epsilon_{ijk}J_k,\;[P_i,K_j]=\delta_{ij}H$$
are less rigid and can be deformed as they depend on the interpretation of 
inertial transformations (boosts) which we want to change.

If we demand the parity and time-reversal invariance, the only possible
deformations of these commutation relations will have the form \cite{Bacry}
$$[H,P_i]=\epsilon_1 K_i,\;\; [H,K_i]=\lambda P_i,\;\;
[P_i,P_j]=\alpha \epsilon_{ijk}J_k, \;\;
[K_i,K_j]=\beta\epsilon_{ijk}J_k,\;\;
[P_i,K_j]=\epsilon_2\delta_{ij}H. $$
The Jacobi identities
$$[P_i,[P_j,K_k]]+[P_j,[K_k,P_i]]+[K_k,[P_i,P_j]]=0$$
and
$$[P_i,[K_j,K_k]]+[K_j,[K_k,P_i]]+[K_k,[P_i,K_j]]=0,$$
are satisfied only if
$$\alpha-\epsilon_1\epsilon_2=0,\;\;\;\mathrm{and}\;\;\;
\beta+\lambda\epsilon_2=0.$$ 

After the deformation, the Poincar\'{e} group is replaced by the so called 
kinematical group -- the generalized relativity group of nature. 
As we see, the structure of its Lie algebra is completely
determined by three real parameters $\epsilon_1,\epsilon_2$ and
$\lambda$. Note that the overall sign of the structure constants is irrelevant
as the sign change of all structure constants can be achieved simply by
multiplying each infinitesimal generator by $-1$. Therefore we can assume
$\lambda\ge 0$ without loss of generality and by a scale change it can be
brought either to $\lambda=1$ or  $\lambda=0$.

If $\lambda=1$, every kinematical group has its underlying Cayley-Klein 
geometry as the geometry of the corresponding space-time. We have the 
following possibilities (more details can be found in \cite{RWT}):
\begin{itemize}
\item $\epsilon_2<0$ - the inertial transformations form a compact group 
and it is not possible (like Euclidean space-time) to introduce a causal 
order between events. Such space-times, if they exist at all (the formation 
of a Euclidean region in the center of a black hole might be a possible 
outcome of gravitational collapse \cite{Sakharov}), are, however, not 
kinematical as they, in fact, are timeless Nirvanas.
\item $\epsilon_1>0,\,\epsilon_2>0$ - de Sitter kinematics ({\bf{DS}}) with
doubly-hyperbolic geometry.
\item $\epsilon_1<0,\,\epsilon_2>0$ - anti de Sitter kinematics ({\bf{ADS}}) 
with co-hyperbolic geometry.
\item $\epsilon_1=0,\,\epsilon_2>0$ - Poincar\'{e} kinematics ({\bf{P}}) 
with Minkowski geometry.
\item $\epsilon_1>0,\,\epsilon_2=0$ - Newton-Hook kinematics ({\bf{NH}}) 
with co-Minkowski geometry.
\item $\epsilon_1<0,\,\epsilon_2=0$ - anti Newton-Hook kinematics 
({\bf{ANH}}) with co-Euclidean geometry.
\item  $\epsilon_1=0,\,\epsilon_2=0$ - Galilean kinematics ({\bf{G}}) with 
Galilean geometry.
\end{itemize}
The case $\lambda=0$ gives rise to five more possibilities of rather exotic
kinematics of limited physical significance as they correspond to space-times 
with absolute space:
\begin{itemize}
\item $\epsilon_1=1,\,\epsilon_2=1$ - anti para-Poincar\'{e} kinematics
({\bf{AP$^\prime$}})with Euclidean geometry.
\item $\epsilon_1=-1,\,\epsilon_2=1$ - para-Poincar\'{e} kinematics 
({\bf{P$^\prime$}}) with Minkowski geometry.
\item $\epsilon_1=1,\,\epsilon_2=0$ - para-Galilei kinematics 
({\bf{G$^\prime$}}) with Galilean geometry.
\item $\epsilon_1=0,\,\epsilon_2=\pm 1$ - Carroll kinematics ({\bf{C}})
with Galilean geometry.
\item $\epsilon_1=0,\,\epsilon_2=0$ - static kinematics ({\bf{S}})
with trivial geometry.
\end{itemize}

-- All these kinematical groups along relations between them are shown in the
figure below, -- said the Patriarch, -- and you, Wu-k'ung, should consult 
the literature (\cite{RWT,Bacry}) to better understand the nature of 
these relations between various kinematical groups. Then you will find out 
that all corresponding relativity theories are in fact limiting cases  of the 
de Sitter or anti de Sitter space-times.
\begin{figure}[htb]
     \centerline{\epsfxsize 150mm\epsfbox{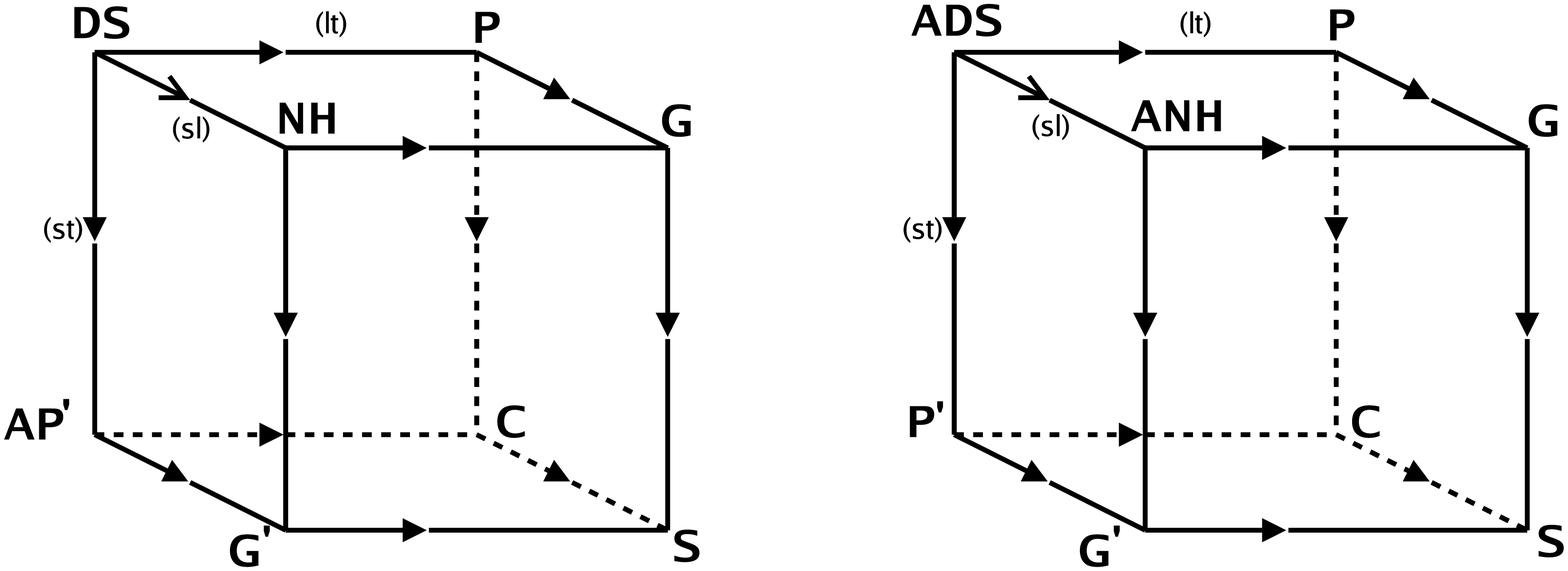}}
\end{figure}

-- I see -- exclaimed Wu-k'ung -- ``there exists essentially only one way to 
generalize special relativity, namely, by endowing space-time with some 
constant curvature'' \cite{Bacry}.

-- That's right -- answered the Patriarch -- and it is not surprising that, 
as the recent astrophysical observations indicate, we live actually in the 
de Sitter space-time, not Minkowski. However, the cosmological constant
which measures the curvature of this de Sitter space-time is incredibly
small - by about 120 degrees of magnitude smaller than expected on some 
``natural'' grounds! Just this incredible smallness of the cosmological 
constant renders the special relativity to the extremely good approximation
to reality and, therefore, to one of the most trustworthy elephant upon which 
the whole building of modern physics rests.

\subsection{Quantum mechanics}
At last Wu-k'ung was prepared to study quantum mechanics.

-- Normal people (that is not physicists) usually consider the quantum 
mechanics as the subject extremely vague and far from the common sense -- 
smiled the Patriarch as the time came for Wu-k'ung to study quantum mechanics.

-- This is partly really so -- readily answered Wu-k'ung. -- Here is a 
notorious story about the famous mathematicians Russell and Whitehead 
\cite{Kac}.  One of them was the speaker at a seminar, and the other was the 
Chairman of that seminar. The talk was devoted to the foundations of quantum 
mechanics. At the end not only the seminar participants were already on the 
throat satisfied by the report, but so was the Chairman. It was extremely 
difficult, confused and unclear. When the report ended, the Chairman felt 
that he ought to comment on the report. So he said only one sentence
which at the same time was both polite and truthful. He simply said: "We
should be thankful to the speaker for what he did not obfuscate further this
already sufficiently obscure subject."

-- An amusing history indeed, -- remarked the Patriarch -- but in reality the
quantum mechanics is not at all obscure and far from everyday experience
contrary to beliefs of the average man. Suppose an electron has to move from
one point to another. How it can do this?

-- One has to know forces that act on the electron. Then you can apply 
Newton's second law and find the trajectory -- replied Wu-k'ung.

-- Forget about forces -- interrupted the Patriarch. -- Do you remember 
the Aharonov-Bohm effect? Besides, force is a vague notion borrowed from 
politics. ``In the Newtonian theory of the solar system, the sun seems like 
a monarch whose behests the planets have to obey'' \cite{ABC}. In fact, 
``$F=ma$ is formally empty, microscopically obscure, and maybe even morally 
suspect'' \cite{Wilczek3}. To follow the classical trajectory an electron has
to solve second order differential equation which the Newton's second law is.
How on earth it can perform this if it is brainless? Simply the electron is
too tiny to have any brain. It needs a very simple instruction to be able to
follow it, and the simplest instruction is: ``Explore all paths'' 
\cite{Taylor}.  

-- But how can the electron reach any particular destination if all paths you
can imagine are equivalent? -- exclaimed Wu-k'ung.

-- They are not completely equivalent -- was the Patriarch's reply. -- They 
differ in a subtle way. A complex number called the amplitude is prescribed
to each path. All paths are equivalent in a sense that all these complex 
numbers have unite magnitude, but their phases may be different. To find out
the probability that the electron will reach some particular destination you
sum up amplitudes of all paths leading to this particular point and the 
squared magnitude of the resulting complex number gives you the desired 
probability. You will be surprised, Wu-k'ung, but all the quantum mechanics
is based on these simple principles just described \cite{FeynmanH}. 

-- But this seems crazy! -- Wu-k'ung was really surprised.

-- You are not alone in your bewilderment -- smiled the Patriarch. --
Freeman Dyson recollects:  {\it Thirty-one years ago, Dick Feynman told me 
about his ``sum over histories'' version of quantum mechanics. ``The electron 
does anything it likes,'' he said. ``It just goes in any direction at any 
speed, forward or backward in time, however it likes, and then you add up the 
amplitudes and it gives you the wave-function.'' I said to him, ``You're 
crazy.'' But he wasn't} \cite{Dyson}. 

-- But quantum mechanics is extremely strange, isn't it? -- asked Wu-k'ung.
-- for example take the superposition principle. Is it really not strange 
that a quantum system can be in a superposition of classically mutually 
exclusive states? This clearly does not fit in our everyday experience.   
 
-- Look at the figure below -- said the Patriarch in answer. 
\begin{figure}[htb]
     \centerline{\epsfxsize 46mm\epsfbox{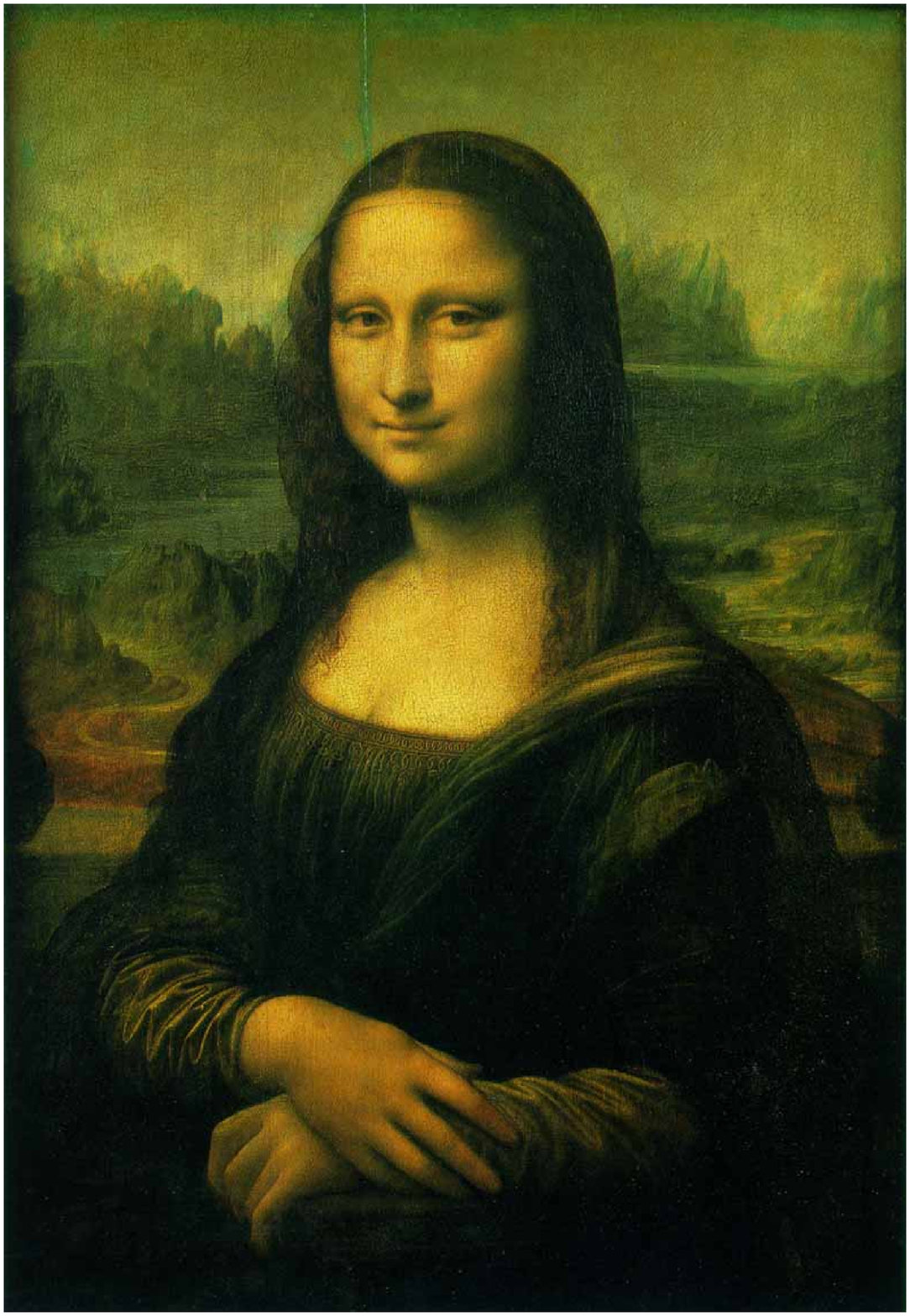}}
\end{figure}

-- This is the famous portrait of Mona Lisa by Leonardo da Vinci. --  At that
time Wu-k'ung was pretty well educated in art.

-- But why this portrait is often regarded as enigmatic, mysterious and 
incomprehensible? -- asked the Patriarch.

-- Some say the portrait looks different every time we look at it -- Wu-k'ung 
was thoughtful. Then he continued -- ``Sometimes she seems to mock at us, and 
then again we seem to catch something like sadness in her smile. All this 
sounds rather mysterious and so it is'' \cite{MLisa}.

-- And an analogy with quantum mechanics can help us to disentangle the
enigma \cite{MLisa} -- the Patriarch showed Wu-k'ung a couple of figures.
\begin{figure}[htb]
     \centerline{\epsfxsize 80mm\epsfbox{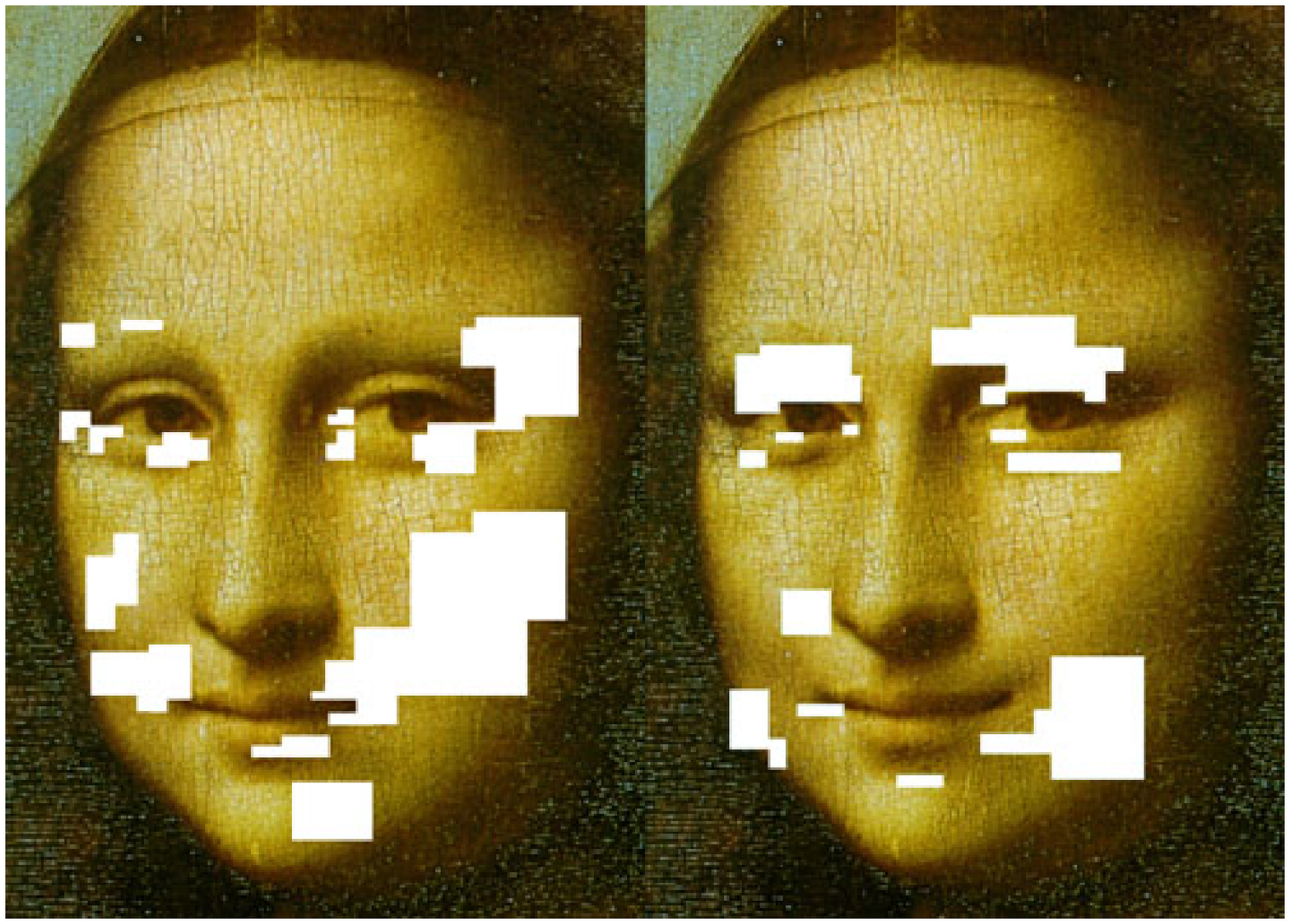}}
\end{figure}

-- Incredible! -- uttered Wu-k'ung. -- The women on the left is definitely
sad and the one on the right unquestionably cheerful!

-- You see, Wu-k'ung, --  the Patriarch looked  triumphant -- quantum 
mechanics is not at all as queer as it seems at first sight. Even such an
unusual notion as the superposition of different states has its counterpart 
in art. Symbols of art, being macroscopic classical objects, have to obey
the laws of classical physics. However, ``their meaning need not to be 
constrained by this theory. They could be organized in a way that corresponds 
to quantum logic. This would be difficult, but it is not impossible. The art 
offers great many possibilities for doing that since there are no limits on 
its expressive power. The piece of art can lead us to some well organized 
world which has nothing in common with the one of our everyday experience''
\cite{MLisa}. 

-- But how about the amazing quantum non-locality? -- asked Wu-k'ung -- does 
it contradict to special relativity? I heard in 1935 Einstein, Podolsky and 
Rosen (EPR) wrote a paper \cite{EPR} where it was claimed that quantum
mechanics fails to provide a complete description of physical reality because
it predicts things ``no reasonable definition of reality could be expected 
to permit'' \cite{EPR}. What's all the fuss about?

-- Imagine a linearly polarized photon flying in some direction and a 
polarizing filter such that the photons either pass through or are absorbed 
by the polarizing filter \cite{EPR1}. Let $|H>$ be a polarization state of 
the photon such that it always passes a horizontal filter but is guaranteed
to be absorbed by a vertical filter. Analogously we define $|V>$ with the 
roles of the horizontal and vertical filters interchanged. Quantum mechanics
allows the following two-photon state (which is called the twin state in
\cite{EPR1}) 
$$|\Psi>=\left . \left .\sqrt{\frac{1}{2}}\right (|V>|V>+\,|H>|H>\right).$$
The photons in this state are entangled: if the first photon passes through 
a horizontal polarizing filter, the second photon is guaranteed to pass 
through another filter similarly oriented, and if the first photon is 
absorbed, so does the second photon. Therefore, by measuring the polarization 
of the first photon we instantly will know the polarization of the second 
photon also, regardless the spatial separation between photons.

-- I do not find this surprising in any obvious way -- said Wu-k'ung. -- There 
are ``many examples of similar correlations in everyday life. The case of 
Bertlmann's socks is often cited. Dr. Bertlmann likes to wear two sock of
different colours. Which colour he will have on a given foot on a given day
is quite unpredictable. But when you see that the first sock is pink you can
be already sure that the second sock will not be pink. Observation of the 
first, and experience of Bertlmann, gives immediate information about the 
second. There is no accounting for tastes, but apart from that there is no
mystery here. And is not the EPR business just the same?'' \cite{BellB}.
Or take the example of identical twins who had been separated at birth 
\cite{Bernstein}. They had many of their tastes and habits identical not to 
speak of their exactly alike appearance. This possibly struck many people
including the twins themselves if they ever meet as a very surprising
correlation. But in this case we have a ready 'hidden-variable' explanation: 
they have the same DNA. Maybe someday we will find analogous hidden-variable 
explanations of various quantum mechanical phenomena which at present we find 
puzzling.   

-- Be careful, Wu-k'ung -- smiled the Patriarch. -- Dr. Bertlmann himself
finds the affairs rather puzzling, realizing their fundamental importance 
\cite{Bertlmann}. I think you do not question that the colour of the first
sock preexisted before your saw it. Neither will doubt you that if the colour
of the first sock allows to predict instantly that the other sock has a 
different colour then you are bound to conclude that this latter colour
also preexisted and was predefined before your colour ``measurement'' with 
the first sock. And this is exactly the logic of EPR.

-- Fine logic. And you want to say that something is wrong with it?

-- Nothing wrong with the logic. But quantum mechanics somehow evades it. 
Can you imagine a quantum sock with no predefined colour? In some coherent 
mixture of pink and green, say? Probably you can if you remember ineffable 
smile of Mona Lisa. This is a strange but rather routine business in quantum 
world. The entangled photons in the EPR thought experiment do not have 
predefined polarizations according to quantum theory. Only after the 
measurement the wave function collapses towards a sharply defined 
polarization state. The trouble is buried in the fact that the wave function 
of the entangled photons is not equal to the product of wave functions of two 
individual photons. It is inseparable quantity and collapses as the whole 
even for photons widely spaced apart. Do you feel now a problem? How the 
second photon instantly ``knows'' to what polarization state to collapse 
after the measurement is done with the first photon, say,  in another galaxy? 

-- I admit this is rather puzzling -- mused Wu-k'ung. -- And what is wrong
with EPR reasoning?

-- The logic behind the EPR reasoning is so fine that the result can be 
formulated as a theorem: {\it If the predictions of quantum mechanics are 
correct (even for systems made of remote correlated particles) and if 
physical reality can be described in a local (or separable) way, then quantum 
mechanics is necessarily incomplete: some ``elements of reality'' exist in 
Nature that are ignored by this theory} \cite{Laloe}. Nobody questions the 
validity of the theorem. Therefore, the question is whether the nature is
indeed non-local as quantum mechanics suggests or the latter is only an 
incomplete description of reality. Note that physics was always non-local 
except some 10 years beginning from about 1915 when Einstein's general 
relativity removed at last ``spooky actions at a distance'' from Newtonian
gravity. But the local paradise, for which Newton so longed, ended in about 
1925 with the advent of quantum mechanics which reintroduced non-locality, 
albeit quite in a different and subtle way than in Newtonian gravity 
\cite{Gisin}.

-- But why quantum mechanics cannot prove to be incomplete? -- asked 
Wu-k'ung. -- After all so many physical theories turned out to be 
incomplete. Even special relativity is incomplete as it does not describe
gravity. And some think both the special and general relativities are 
emergent low-energy phenomena \cite{Volovik}. Why quantum mechanics is
considered to be more fundamental than these venerable theories?

-- Let us return to the twin state photons \cite{EPR1}, -- begin the 
Patriarch. -- Suppose the photon polarizations indeed do have predefined 
values as required by local realism, like Bertlmann's socks. That is we are 
supposing that the source of these photons, whatever it is, produces with 
probability $\lambda_1$ the combination of photon polarizations 
$|V>|V>$, and with probability $\lambda_2$ the combination $|H>|H>$, while 
never $|H>|V>$ or $|V>|H>$, so that $\lambda_1+\lambda_2=1$. Suppose further 
that the two polarizing filters, which analyze the polarization state of 
photons, are tilted by angles $\theta_1$ and $\theta_2$, respectively, with 
respect to the vertical. And  here is a good problem to test your advance in 
physics, Wu-k'ung: what is the odds of getting the same result on both sides 
(either both photons transmitted or both absorbed)?

-- If the photon's polarization is $|V>$, it has $\cos^2{\theta}$ 
probability, according to the Malus' law, to be transmitted by the polarizing 
filter inclined by the angle $\theta$, and $\sin^2{\theta}$ probability to 
be absorbed by it. For the $|H>$ polarization, probabilities are 
$\sin^2{\theta}$ and $\cos^2{\theta}$ respectively. The probability both 
photons to be transmitted is, therefore,
$P_T=\lambda_1\cos^2{\theta_1}\cos^2{\theta_2}+\lambda_2\sin^2{\theta_1}
\sin^2{\theta_2}$, as the probabilities of $|H>|H>$ and $|V>|V>$ combinations
are $\lambda_1$ and $\lambda_2$. Similarly, the probability both photons to be 
absorbed is $P_A=\lambda_1\sin^2{\theta_1}\sin^2{\theta_2}+
\lambda_2\cos^2{\theta_1}\cos^2{\theta_2}$.

-- That's right -- the Patriarch encouraged  Wu-k'ung. -- And then?

-- The probability you asking me is clearly
$$P_{LR}=P_T+P_A=\cos^2{\theta_1}\cos^2{\theta_2}+\sin^2{\theta_1}
\sin^2{\theta_2}.$$

-- Fine. And what is the answer according to quantum mechanics?

-- I think the answer is
$$P_{QM}=|<\theta_1|<\theta_2|\Psi> |^2+
|<\pi/2-\theta_1|<\pi/2-\theta_2|\Psi>\ |^2.$$
And to calculate this, we can decompose $|\theta>=\cos{\theta}\,|V>+
\sin{\theta}\,|H>$ to get
\begin{eqnarray} 
<\theta_1|<\theta_2|&=&<V|<V|\,\cos{\theta_1}\cos{\theta_2}\,+
<H|<H|\,\sin{\theta_1}\sin{\theta_2}+\nonumber \\ &&
<V|<H|\,\cos{\theta_1}\,\sin{\theta_2}\,+ <H|<V|\,\sin{\theta_1}\cos{\theta_2}.
\nonumber
\end{eqnarray}
It remains to do a simple algebra and trigonometry to obtain $P_{QM}=\cos^2{
(\theta_1-\theta_2)}$.

-- You see the answers are different -- remarked the Patriarch. -- Do you find
this surprising?

-- Not at all -- answered Wu-k'ung. -- After all quantum mechanics is a 
different theory. Why should I be surprised that it gives a different result
than the local-realistic classical expectation?

-- The problem is not that quantum mechanics gives a different result
-- smiled the Patriarch. -- The problem is that it gives 
\emph{incomprehensible} result. Dr. Herbert invented the general idea 
\cite{Herbert} which I will use to explain to your why the answer of quantum 
mechanics is incomprehensible. The argument goes as follows \cite{EPR1,
Herbert1, Enigma}.

Suppose the source of twin photons emits some sequence of polarized photons
with either horizontal or vertical polarizations. Were the polarizing filter
oriented horizontally, the observer near it would register some sequence
of either transmitted photons ($T$) or the absorbed ones ($A$), say
$$S(0)=\{A,T,T,A,A,T,A,T,A,T,T,T,A,T,A,A,A,A,T,A,T,T,A,T\}.$$
The same sequence will be registered by the horizontally oriented polarizing 
filter on the other side of the source because for twin photons their 
polarizations are 100\% correlated like Bertlmann's socks colours. Now if we 
rotate one of the polarizing filter by some angle $\theta$, it will register
another sequence of transmitted or absorbed photons, say
$$S(\theta)=\{A,T,T,T,A,T,A,T,A,A,T,A,A,T,T,A,T,A,T,A,T,T,T,T\}.$$
Now the correlation between $S(0)$ and and $S(\theta)$ is not 100\%, of 
course. There are some mismatches, highlighted below
\begin{eqnarray} 
     S(0)=&&\{A,T,T,{\bf\color{red} A},A,T,A,T,A,{\bf\color{red}  T},T,
           {\bf\color{red} T},A,T,{\bf\color{red} A},A,{\bf\color{red} A},
           A,T,A,T,T,{\bf\color{red} A},T\} \nonumber \\ 
S(\theta)=&&\{A,T,T,{\bf\color{red} T},A,T,A,T,A,{\bf\color{red} A},T,
           {\bf\color{red} A},A,T,{\bf\color{red} T},A,{\bf\color{red} T},
           A,T,A,T,T,{\bf\color{red} T},T\} \nonumber    
\end{eqnarray}

-- According to the formula $P_{LR}=P_T+P_A=\cos^2{\theta_1}\cos^2{\theta_2}
+\sin^2{\theta_1}\sin^2{\theta_2}$ your obtained, Wu-k'ung, we expect the
probability of getting the same result on both sides to be in this case
$\cos^2{\theta}$ and hence the mismatch probability $\sin^2{\theta}$. You
observe six mismatches from twenty four events. Therefore the  mismatch 
probability is about $1/4$ which indicates the inclination angle $\theta=
30^\circ$, although this is not essential for subsequent considerations.

If we rotate the polarizing filter not by the angle $\theta$, but $-\theta$,
the mismatch probability should remain the same $\sin^2{\theta}$ according to
your formula, Wu-k'ung, although the sequence of results registered in this 
case will be of course different, say
$$S(-\theta)=\{A,T,A,A,A,T,T,T,A,T,T,T,A,T,A,T,T,A,T,A,T,A,T,T\}.$$
We still have about six mismatches although at different places:
\begin{eqnarray}
      S(0)=&&\{A,T,{\bf\color{red} T},A,A,T,
              {\bf\color{red} A},
             T,A,T,T,T,A,T,A,{\bf\color{red} A},{\bf\color{red} A},A,T,A,T,
             {\bf\color{red} T},{\bf\color{red} A},T\} \nonumber \\ 
S(-\theta)=&&\{A,T,{\bf\color{red} A},A,A,T,
             {\bf\color{red} T},
             T,A,T,T,T,A,T,A,{\bf\color{red} T},{\bf\color{red} T},A,T,A,T,
             {\bf\color{red} A},{\bf\color{red} T},T\} \nonumber    
\end{eqnarray}
And what will happen if one polarizing filter is rotated by the angle $\theta$
and the another, on the opposite side, by the angle $-\theta$ as indicated in
the figure below?
\begin{figure}[htb]
     \centerline{\epsfxsize 30mm\epsfbox{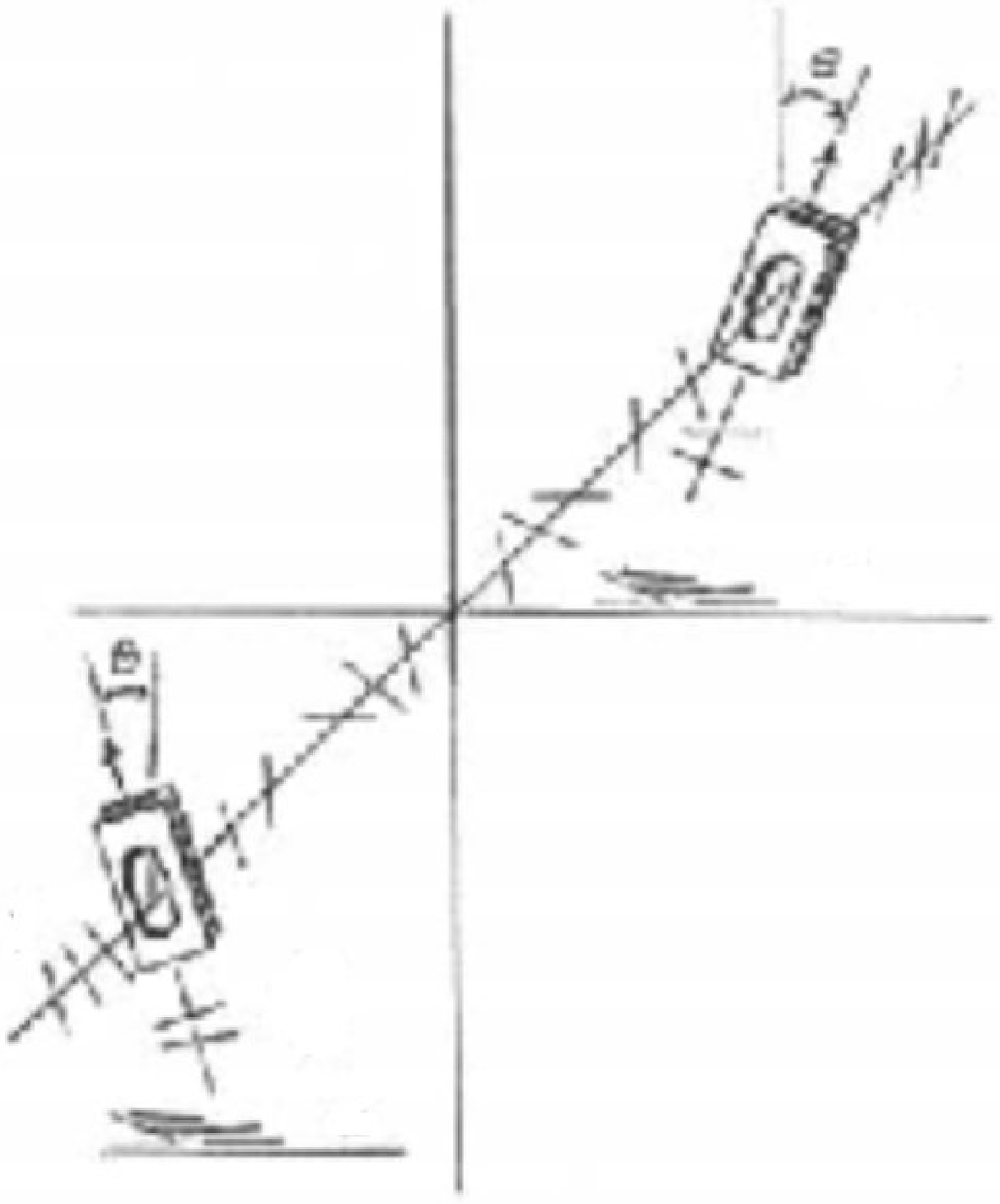}}
\end{figure}

\noindent What you can say about the mismatch rate in this case?

-- Well, -- begin Wu-k'ung -- as the filters are well separated as I 
understand, they cannot influence each other. So I expect one filter to 
register the sequence $S(\theta)$ and the another one - the sequence 
$S(-\theta)$. The mismatch rate cannot exceed $2\sin^2{\theta}$ because 
there are some cases where both $S(\theta)$ and  $S(-\theta)$ disagree with 
the result at zero angle $S(0)$ and, therefore, agree with each other. 
Indeed, there are only eight mismatches between $S(\theta)$ and $S(-\theta)$:
\begin{eqnarray}
S(\theta)=&&\{A,T,{\bf\color{red} T},{\bf\color{red} T},A,T,
             {\bf\color{red} A},
             T,A,{\bf\color{red} A},T,
             {\bf\color{red} A},A,T,{\bf\color{red} T},{\bf\color{red} A},
             T,A,T,A,T,{\bf\color{red} T},T,T\} \nonumber \\ 
S(-\theta)=&&\{A,T,{\bf\color{red} A},{\bf\color{red} A},A,T,
             {\bf\color{red} T},
             T,A,{\bf\color{red} T},T,
             {\bf\color{red} T},A,T,{\bf\color{red} A},{\bf\color{red} T},
             T,A,T,A,T,{\bf\color{red} A},T,T\} \nonumber    
\end{eqnarray}
No wonder this observation is consistent with the formula $P_{LR}=P_T+P_A=
\cos^2{\theta_1}\cos^2{\theta_2}+\sin^2{\theta_1}\sin^2{\theta_2}$ I derived
for the coincidence rate. Indeed, when $\theta_1=\theta$ and $\theta_1=
-\theta$, we get $P_{LR}=\cos^4{\theta}+\sin^4{\theta}=1-2\sin^2{\theta}
\cos^2{\theta}$. Therefore the expected mismatch rate is $2\sin^2{\theta}
\cos^2{\theta}$ and this clearly does not exceed $2\sin^2{\theta}$. By the 
way, for $\theta=30^\circ$, the mismatch probability is $3/8$ and, therefore,
we expect nine mismatches between $S(\theta)$ and  $S(-\theta)$. In reality
we observe eight mismatches, but this is just a play of statistics. 
Everything seems coherent for me.

-- My congratulations, Wu-k'ung, -- the Patriarch was very serious. -- You
have just discovered a simple-minded version of Bell's inequality which all 
local-realistic models must satisfy \cite{Bell}. In fact Bell's inequalities
can be considered as examples of George Boole's mid-nineteenth century 
'Conditions of Possible Experience' that the relative frequencies of logically
connected events must satisfy \cite{Pitowsky}. There are just the logic and
common sense behind them. This brings any theory which violates Bell's 
inequalities ``on the edge of a logical contradiction'' \cite{Pitowsky}.
And quantum mechanics is just such a theory! Remember what you get as the
quantum mechanical prediction for the coincidence rate in our twin photon
business?  For $\theta_1=\theta$ and $\theta_1=-\theta$, the prediction is
$P_{QM}=\cos^2{(2\,\theta)}$. Therefore quantum mechanics predicts the 
mismatch rate $1-P_{QM}=\sin^2{(2\,\theta)}=4\sin^2{\theta}\cos^2{\theta}$ 
and this does not exceed $2\sin^2{\theta}$ only if $\cos^2{\theta}\le 1/2$. 
For  example, if $\theta=30^\circ$ then $1-P_{QM}=3/4$ which is \emph{grater} 
than $2\sin^2{30^\circ}=1/2$. The Bell's inequality you obtained is clearly 
violated here! 

-- But how! -- exclaimed Wu-k'ung, -- What the hell is going here! How this
simple result of logical reasoning can be violated by the theory which is 
overwhelmingly tested? I'm confused.

-- ``Anybody who is not bothered by Bell's theorem has to have rocks in his 
head'' \cite{Mermin} -- replied the Patriarch. -- And here is the answer
on your question ``why quantum mechanics cannot prove to be incomplete''.
It cannot prove to be incomplete in the EPR business because simply 
experiments confirm that the Bell's inequalities are indeed violated in some
circumstances with complete agreement with the quantum mechanical 
predictions \cite{Aspect}. We are forced to accept a strange view of reality 
quantum mechanics offers us. ``Before Bell's discovery, one could still
imagine that a local reality lurked beneath the experimental facts, after 
1964, one could blissfully believe in a strictly local world only by hoping
that quantum theory was wrong in its predictions concerning photons in the
twin state'' \cite{Herbert1}. But it was not wrong.

-- And what about special relativity? How can it swallow this ``spooky 
actions at a distance''? --  Wu-k'ung seemed perplexed.

-- Quantum mechanics and relativity can co-exist peacefully because the
non-local correlations described above does not allow faster-than-light
signaling \cite{Gisin}. However, all of this looks very subtle and the 
legitimate question still is whether relativity or quantum mechanics can be 
considered complete. Sometimes it is even claimed that ``with the presently 
available models we have the alternative: Either the conventional 
understanding of relativity is not  right, or quantum mechanics is not exact''
\cite{Tamulka}. Frankly speaking, I should admit that ``the violation of the 
Bell-inequality implies that the relation between quantum mechanics and 
special relativity is more subtle than customarily assumed'' \cite{Passon}.

Some think quantum mechanics is more fundamental than relativity. To explain 
what they have in mind and as an another example of the magic of art with 
profound physical meaning consider the following figure (this is a ``hidden 
image" artwork {\it Mouth of Flower} by Mexican artist Octavio Ocampo)
\begin{figure}[htb]
     \centerline{\epsfxsize 40mm\epsfbox{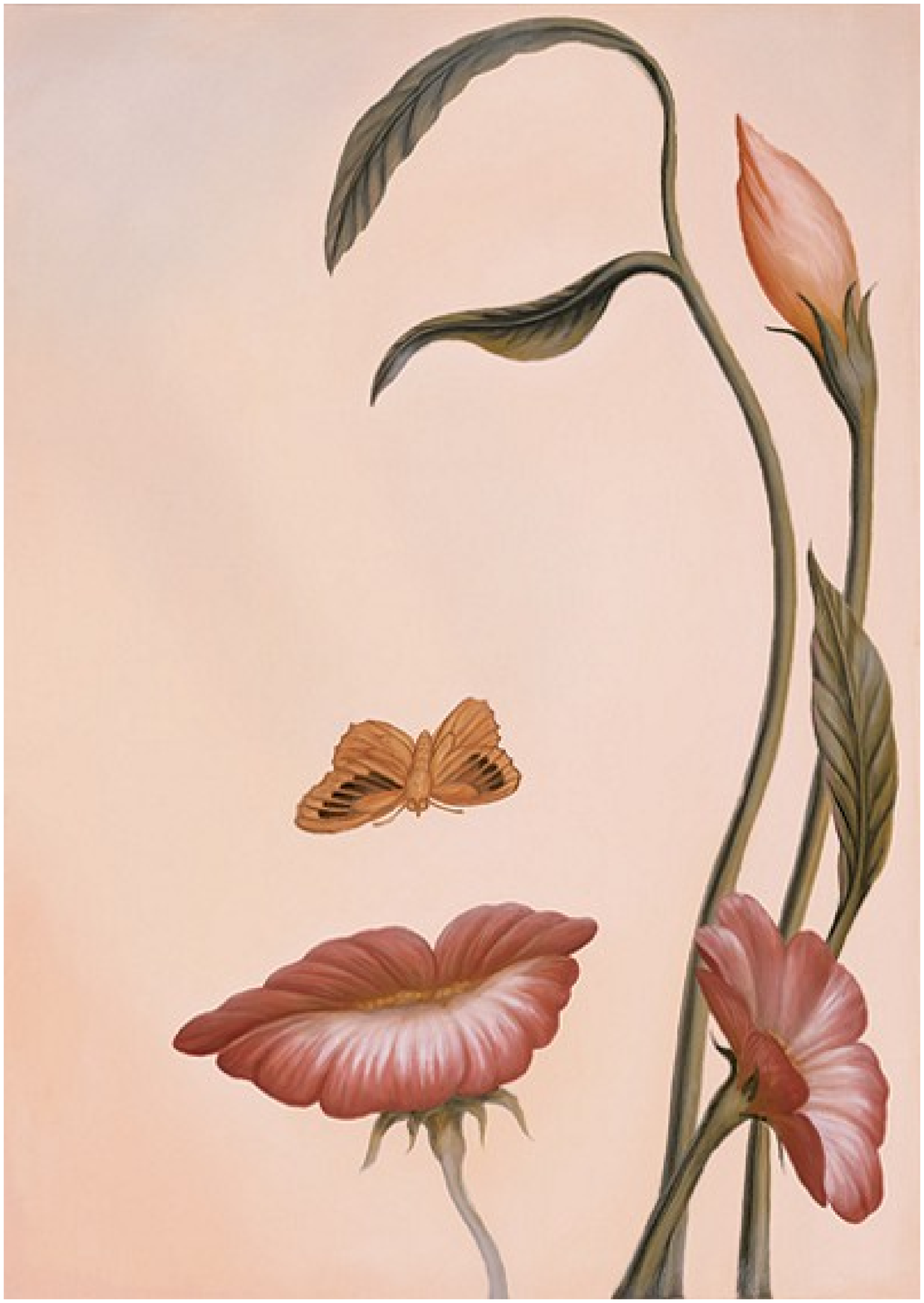}}
\end{figure}

-- A face of the beautiful girl, -- said Wu-k'ung -- I appreciate her beauty
but what a physical meaning on earth does it have?

-- Upon inspecting the figure closely, -- answered the Patriarch -- you 
probably will find no face at all whatever beautiful or not in the figure.
All what remain are some flowers and butterfly. The face of the beautiful 
girl is an emergent phenomenon: it exists when things are inspected at some
scale but disappears at finer scales. And the crucial thing is that this
hierarchical architecture pervades the Nature, being the outstanding principle
of its functionality \cite{Anderson,Laughlin}. The whole very often is not 
merely more than but entirely different from the sum of its parts. The
following passage from \cite{Anderson} only at first sight seems like 
a joke but it actually hints very deep meaning: ``Marx said that quantitative 
differences become qualitative ones, but a dialog in Paris in the 1920's 
sums it up even more clearly:

FITZGERALD: The rich are different from us.

HEMINGWAY: Yes, they have more money. ``

\noindent We routinely have emergent physical phenomena and the organizing
principles which regulate them make sense only at the corresponding scale. 
They simply do not exist outside the context established by this scale.
Besides, these higher organizing principles are insensitive to fine details 
of underlying micro-physics \cite{Laughlin}. That's why biologists and 
chemists can do their science without knowing nuclear physics or theory of
elementary particles. Owing to this remarkable property  of Nature's 
organization  science acquires its main strength: it can operate and progress
in a modular manner \cite{Ball}.

It is quite possible that relativity is also an emergent phenomenon 
\cite{Volovik,Laughlin1,Volovik1}. Towards this possibility hints the fact 
that there are some condensed matter systems where electron transport is 
essentially governed by the Dirac equation and a kind of effective relativity 
arises \cite{Graphene,NBS}. For example, in graphene, a mono-layer of carbon 
atoms,  charge  carriers mimic relativistic particles with zero mass and the 
effective Lorentz invariance  emerges with limiting velocity of about 
$10^6~\mathrm{m/s}$ which is much smaller than the light velocity in vacuum
\cite{Graphene}. In these condensed matter examples quantum mechanics is the
fundamental theory governing the underlying micro-physics while the effective
relativistic behaviour is an emergent phenomenon valid only in the low-energy 
sector.  
 
However, let us return to quantum non-locality. I see you are impressed by 
it. Surprisingly, the majority of physicists does not bother much about it.
They become used to oddities of quantum mechanics. You surely think Bell's
proof exploded ``like a bombshell in the corridors of science'' 
\cite{Herbert1}. In reality, however, Bell's paper \cite{Bell}, published in 
a not-mainstream  short-lived journal, was largely ignored for five years
and fallow physicists were not at all impressed by it \cite{Bernstein,
Herbert1}. You know ``the majority of physicists are phenomenalists -- whose 
professional world is circumscribed by phenomena and mathematics. A 
phenomenalist perceives science as advancing in two directions: 1. new 
experiments uncover novel phenomena; 2. new mathematics explain or predict 
phenomena in original ways. Since it proposes no new experiments and derives 
no new phenomena-relevant mathematics, but merely puts certain constraints on 
an invisible reality, Bell's proof lies outside the fashionable formula for 
success in science and is generally dismissed by scientists as 
'mere philosophy' `` \cite{Herbert1}. 

Therefore, to convince you finally how strange the quantum world view is I
tell you about Wheeler's delayed-choice experiment. Consider the following 
experimental arrangement \cite{Delayed}: 
\begin{figure}[htb]
     \centerline{\epsfxsize 130mm\epsfbox{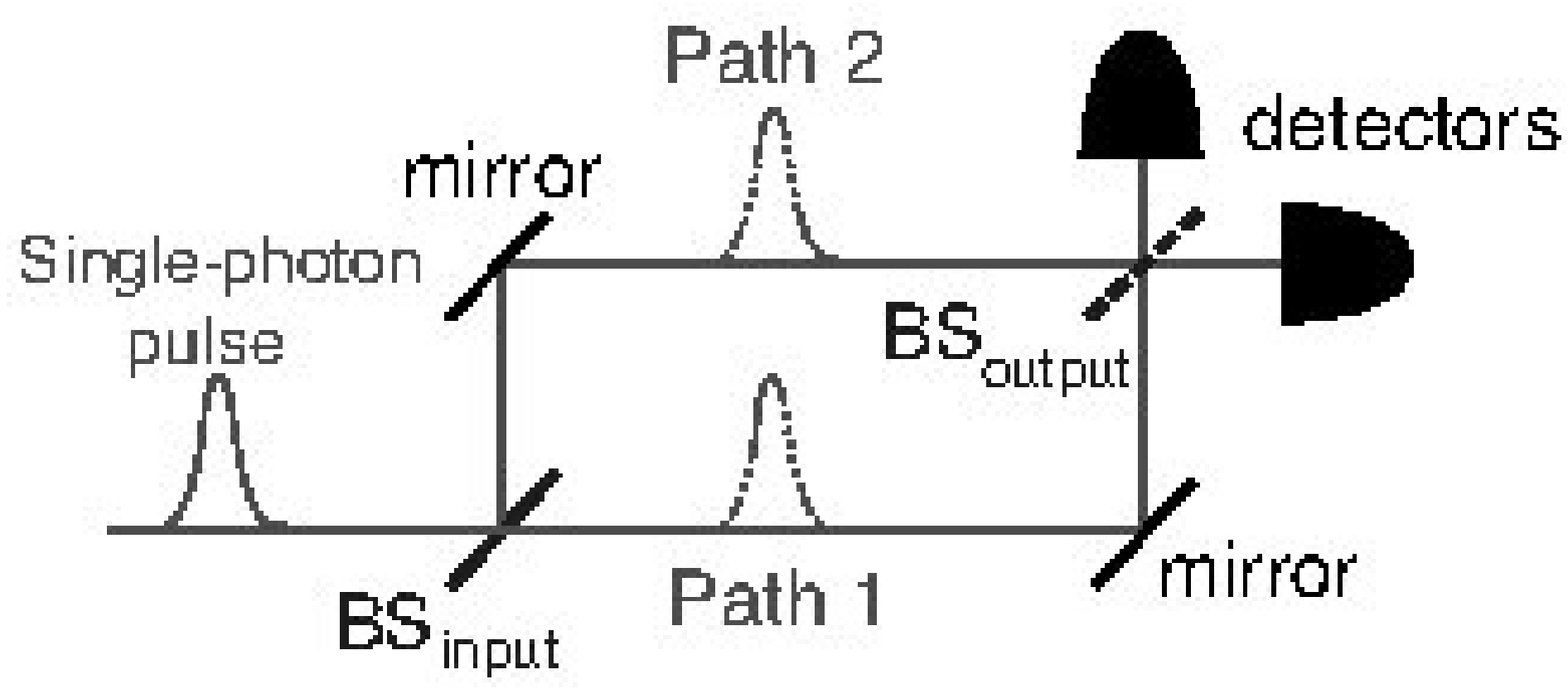}}
\end{figure}

A single-photon pulse is split by a first beam-splitter $BS_{input}$ and it
travels through two arms of a Mach-Zehnder interferometer until being
recombined by a second beam-splitter $BS_{output}$. The photon is then
detected by one of the two detectors $D_1$ and $D_2$ with equal probabilities.
However, if we induce a phase shift $\phi$ between the two arms and vary it,
the detection probabilities at ports $D_1$ and $D_2$ will be no longer
equal. Instead they will be modulated as $\cos^2{\phi}$ and  $\sin^2{\phi}$
respectively.

-- This is an interference phenomenon familiar from  Young's double-slit
experiment, -- remarked Wu-k'ung.

-- That's correct, -- said the Patriarch. -- Interference appears because we
can't tell which way from the two possibilities the photon arrived by to the
detector which it fired. However, we can remove the second beam-splitter
$BS_{output}$ so that each detector $D_1$ or $D_2$ is uniquely associated
to a given path of the interferometer. In this case we expect the interference
to disappear and the experiment, of course, confirms our expectation.

In the first case light behaves as a wave and in the second case as a
particle. The famous wave-particle duality. As Bohr would say ``the behavior
of a quantum system is determined by the type of measurement performed on
it'' \cite{Delayed}.

-- You said people got used of it -- smiled Wu-k'ung.

-- However, -- remarked the Patriarch, -- our classical picture that either
light goes through the both arms of the interferometer like a wave, or makes
a choice to follow a definite path like a particle, depending what type of
measurement we decide to perform on, is too naive. J.~A.~Wheeler proposed
the famous `delayed-choice' Gedankenexperiment in which the choice remove
or not the second beam-splitter $BS_{output}$ is made after the photon has 
already passed the first beam-splitter $BS_{input}$.  

A real experiment  very close in spirit to the Wheeler's original suggestion 
was performed \cite{Delayed}. In this experiment, the clock that triggers
the single-photon emission simultaneously generates a random choice of the
interferometer configuration (with $BS_{output}$ or not) by means of
specially designed random number generator and the fast electro-optical
modulator (EOM) along with some other optical equipment. With no voltage
applied to the EOM, one effectively has $BS_{output}$ removed, and it can be
effectively brought back within 40~ns by fast switching on the appropriate
voltage on the EOM. The random number generator was located close to the
output of the interferometer while $BS_{input}$ was some 48 meters away.
This configuration guarantees, assuming no faster-than-light signaling, that
the photon can not get any information about the made choice between
interferometer configurations before it reaches $BS_{input}$. Actually, such
an information might be available for the photon when it is about halfway in
the interferometer, long after passing $BS_{input}$ \cite{Delayed}. 

-- And what was the results? -- Wu-k'ung was intrigued.

-- They found that ``Nature behaves in agreement with the predictions of 
Quantum Mechanics even in surprising situations where a tension with 
Relativity seems to appear'' \cite{Delayed} -- answered The Patriarch. --
It was demonstrated ``beyond any doubt that the behavior of the photon in the
interferometer depends on the choice of the observable which is measured, 
even when that choice is made at a position and a time such that it is 
separated from the entrance of the photon in the interferometer by a
space-like interval'' \cite{Delayed}.

-- Interesting, -- said Wu-k'ung. -- But I feel I'm becoming used to the
strange world of quantum mechanics.

-- Wait a moment, -- smiled the Patriarch. -- By the use of entangled
photons we can make the previous situation even more surprising. Now the 
experimental arrangement looks like this \cite{DCQer}:
\begin{figure}[htb]
     \centerline{\epsfxsize 140mm\epsfbox{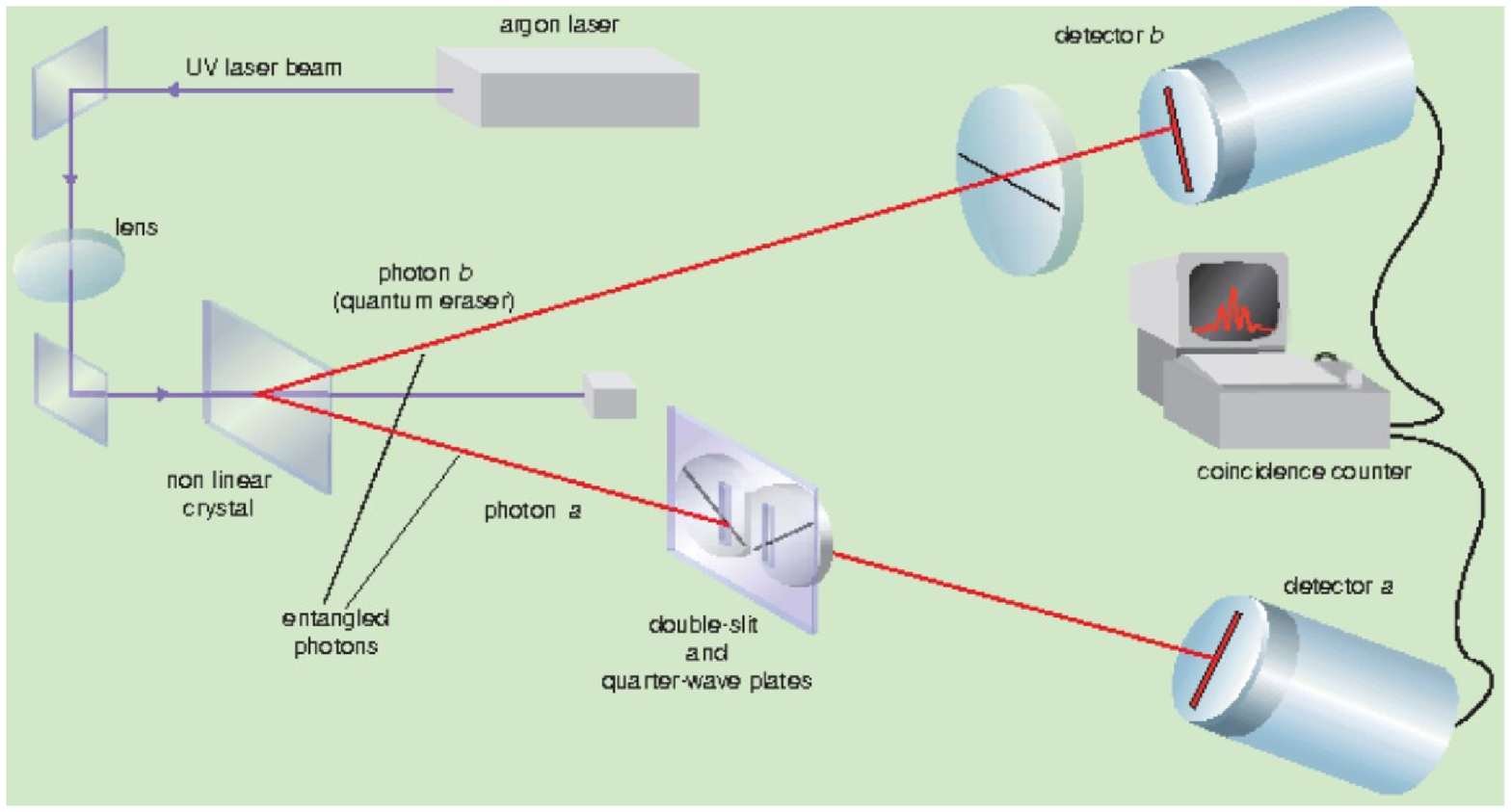}}
\end{figure}

An ultraviolet argon laser creates entangled photons through the so called
spontaneous parametric down conversion process when it illuminates  
a beta-barium borate (BBO) crystal. The polarizations of the entangled 
photons are strictly correlated: if photon $a$ is detected to have  a 
vertical polarization then the polarization of the photon $b$ will be 
necessarily horizontal and vice versa. Signal photons ($a$) pass through 
the double slit and the interference fringes can be observed by the 
corresponding detector $D_a$, while idler or passive photon $b$ goes to 
another detector $D_b$.

Now in front of each slit we place a quarter-wave plate (QWP), an optical 
device which transforms linearly polarized light into circularly polarized 
light. The quarter-wave plates are set so that horizontally polarized
$a$-photon acquires left circular polarization if it goes through the first
slit and right circular polarization if it goes through the second slit.
For vertically polarized $a$-photon the circular polarization outcomes after
the passage trough QWP covered slits are reversed. Note that now $a$-photons
are marked: by measuring their circular polarization we can tell which slit
they passed through provided the linear polarization of the $a$-photon is
known before the quarter-wave plates. This latter information is encoded
into the polarization of the entangled $b$-photon. 

-- According to quantum mechanics -- said Wu-k'ung in excitement -- this
which-way information should destroy the interference and the interference 
fringes must go away. Is this really so?

-- You can't have any doubt that experiment confirms  this expectation, 
-- answered the Patriarch. -- It is not necessary to actually measure the
polarizations of $a-$ and $b$-type photons to destroy the interference 
pattern. The principal possibility of performing such measurements does 
suffice.

-- How do these photons then know that we could find out which slit the 
$a$-photon went through? -- exclaimed Wu-k'ung. -- And I suspect we can
change the polarization of the $b$-photon before it strikes the detector
$D_b$ and, therefore, erase the which-way information for the $a$-photon. 
What happens then?

-- Such a possibility of quantum erasure was recognized in 1982 by Scully 
and Druhl \cite{Scully}. In our experimental set-up we can place a polarizer
in the $b$-beam oriented so that both horizontally and vertically polarized
$b$-photons have an equal chances to pass it. After the $b$-photon passed
the polarizer it no longer bears the information about the polarization of
the $a$-photon. Therefore, the which-way information has been erased and
we expect interference fringes to reappear in the coincidence counts between
detectors $D_a$ and $D_b$. This is really what is observed in experiment 
\cite{DCQer}. A particularly impressive situation can be obtained if the 
quantum erasure is combined with Wheeler's delayed choice. We can simply
place detector $D_b$ with polarizer in front of it far away so that photon
$a$ is registered first and then its companion photon $b$. The experiment
clearly demonstrates that reappearance of the interference fringes does not
depend on the order $a$ and $b$ photons are detected in \cite{DCQer}.

-- What the miracles! -- Wu-k'ung did not hide his surprise. -- It seems we 
are in a position to change the past!

-- This is not correct, -- replied the Patriarch. -- We can not change the 
past. Our detection of the photon $b$ does not change the point photon $a$
had hit the detector $D_a$ earlier. ``For entangled photons it is misleading 
and incorrect to interpret the physical phenomena in terms of independent 
photons'' \cite{Scarcelli}. ``Of course, one might try to go beyond the 
minimalistic interpretation and give additional ontological meaning to $\Psi$,
thereby accommodating some philosophical preconceptions or other personal 
biases. In doing so, one should  however remember van Kampen's caveat'' 
\cite{Englert}: 

{\it Everybody is free to speculate, but  whoever endows $\Psi$ with more
meaning  than is needed for computing observable phenomena is responsible 
for the consequences. He has the duty to show that his speculations do not 
lead to contradictions, and preferably that they are of some use (other 
than agreement with preconceived philosophical views). If he does not succeed 
he should not blame quantum mechanics} \cite{Kampen}.

-- But are you indeed comfortable with Wheeler's suggestion that ``the past 
has no existence except as it is recorded in the present?'' -- asked
Wu-k'ung. -- Do you really think that the photon during its journey in the 
interferometer is ``a great smoky dragon which is only sharp at its tail (at 
the beam splitter 1) and at its mouth where it bites the detector?'' 
\cite{DCExp}. I heard Bohmian mechanics offers an alternative explanation
of these strange delayed choice experiments \cite{DCBohm1,DCBohm2}, maybe 
more comfortable with the common sense, but at the expanse of introducing 
mysterious pilot-wave. In any case, do you really not find that the world 
view quantum  mechanics suggests us is very strange? 

--  ``we always have had (secret, secret, close the doors!)'' -- answered 
the Patriarch, -- ``We always have had a great deal of difficulty in 
understanding the world view that quantum mechanics represents. At least I 
do, because I'm an old enough man that I haven't got to the point that this 
stuff is obvious to me. Okay, I still get nervous with it. And therefore, 
some of the younger students \ldots you know how it always is, every new 
idea, it takes a generation or two until it becomes obvious that there's no 
real problem. It has not yet become obvious to me that there's no real 
problem. I cannot define the real problem, therefore I suspect there's no 
real problem, but I'm not sure there's no real problem'' \cite{Feynman}. 

\section{Wu-k'ung doubts Darwinian evolution}
Wu-k'ung's progress in physics was really impressive and the Patriarch was 
glad and very proud of his pupil. But things began to change when time came
to study Darwinian evolution.

-- ``It is settled, as well as anything in science is ever settled, that the 
adaptations of living things on Earth have come into being through natural 
selection acting on random undirected inheritable variations'' 
\cite{Weinberg}. -- The Patriarch was about to lose his temper after 
repeatedly trying to convince Wu-k'ung in Darwinian evolution.

-- How can you, -- persisted Wu-k'ung  -- how can you, a wise man, to believe 
that one species can be obtained from others through random mutations? 
A simple estimate shows that this is impossible \cite{Melkikh}. A genome
represents an ordered set of about $N=3\times 10^9$ nucleotides. The nearest
neighboring species have $N_1\approx 0.01N$ different nucleotides. There are
$C_N^{N_1}=\frac{N!}{N_1! (N-N_1)!}$ possibilities to choose $N_1$ locations
from $N$ available in the genome and only one is the right one. Besides, you
need to have right nucleotides at those locations. We have four  nucleotides
used in the genetic code. Therefore for each nucleotide in the old genome at
$N_1$ locations we have three variants to change it by a different nucleotide
and only one is what we want. This means there is 
$\left (\frac{1}{3}\right )^{N_1}$ probability that all replaced nucleotides
turn out to be right ones. In overall, the probability that a pure chance
transforms a valid genome into another valid genome is
$$p=\frac{1}{3^{N_1}}\frac{N_1!(N-N_1)!}{N!}.$$
You can already suspect that it is an exceedingly small number, but let us
estimate it \cite{Melkikh} by using Stirling's formula $\ln{(N!)}\approx N
\ln{N}-N$, valid for large numbers. The result is $\ln{p}\approx -2\times 
10^8$. Therefore the probability that our reckless reliance on chance  
succeeds is negligibly small, $p\approx \exp{(-2\times 10^8)}$. Is it not too 
foolish to assume that the proliferation of diversity we observe in biota 
was created in this manner? In fact there is some variation of genomes within
one species with characteristic distance $N_2\approx 0.001N$ between 
subspecies. Therefore we have somewhat underestimated the probability $p$, 
but only slightly \cite{Melkikh}, and this can not help either. 
   
-- Such estimates are too simplistic and misleading, -- the Patriarch turned
crimson with fury. -- Much more elaborated arguments of this kind were 
given by Fred Hoyle \cite{Hoyle}. And what? Darwinians simply dismissed these
arguments stating that ``Hoyle's objections were frankly silly, reflecting an 
embarrassing misunderstanding of Darwinian logic'' \cite{Orr}. 

-- It's a pity that Darwinians simply ignored Hoyle's arguments, -- replied
Wu-k'ung. -- I find them at least thought provoking. Take, for example, his
histone-4 story. Histone-4 is an important eukaryotic protein that binds and
folds DNA and is necessary for chromosome condensation during cell division.
It has about one hundred amino acids in its structure and remains
remarkably unchanged during known history of evolution: although plants and
animals are separated by more than one billion years of evolution their 
histone-4 molecular sequences differ in only two amino acids 
\cite{Evolution}. The fact that evolution has not produced many different 
functional forms of histone-4 during billion years possibly indicates that the 
molecular sequence of histone-4 is essentially unique and can not be changed 
without destroying its biological function. However, then we face a difficult 
question how evolution had found this unique sequence of amino acids in the 
first place. The probability for the right sequence of one hundred amino acids 
to appear suddenly by mere chance is very small you can guess, and Hoyle 
claims that the histone-4 protein could never be produced by evolution in 
small steps either because all intermediate steps are non-functional. 
 
-- You are again hasty in your logical conclusions, -- retorted the Patriarch.
-- Histone-4 is not as stable as you think. Some time ago histone-4 sequences 
from thirteen species of ciliates were analyzed and it was found that they 
differ from one another at as many as 46\% of their amino acids \cite{H4}.
It is true that in most other eukaryotes this protein has highly conserved 
character. This can be explained as an example of purifying selection. Most
proteins require a thermodynamically stable suitable three-dimensional
structure to function. Protein folding depends crucially upon its amino-acid 
sequence. Many mutations destabilize rather than stabilize histone-4 
structure. It is not surprising, therefore, that this structure was conserved
during evolution in light of histone-4's important role in cell division.
Compared with other eukaryotes, ciliates have two functionally distinct 
genomes in within and one of micro-nucleus is transcriptionally inactive. This
presence of two functionally distinct genomes may allow more rapid evolution 
of histone-4 genes in ciliate lineages \cite{H4}. It is true that determining 
factors of protein evolution rates demand serious studies and an integrated 
view of protein evolution is only emerging now \cite{PEvo}.
 
As for evolution, ``it is absolutely safe to say that if you meet somebody who 
claims not to believe in evolution, that person is ignorant, stupid or insane 
(or wicked, but I'd rather not consider that)''. ``We are not talking about 
Darwin's particular theory of natural selection. It is still (just) possible 
for a biologist to doubt its importance, and a few claim to. No, we are here 
talking about the fact of evolution itself, a fact that is proved utterly 
beyond reasonable doubt'' \cite{Dawkins}. 

-- My dear Master, -- begin Wu-k'ung, -- I don't question evolution. The 
question is how to explain it. Darwinian evolution rests upon the following
main principles \cite{Koonin}
\begin{itemize}
\item Undirected, random variation is the main process that provides the 
material for evolution.
\item  Evolution proceeds via natural selection by fixation of the rare 
beneficial variations and elimination of deleterious variations.
\item The beneficial changes that are fixed by natural selection are 
'infinitesimally' small, so that evolution is gradual and proceeds via the 
accumulation of these tiny modifications. 
\end{itemize}
It would be quite remarkable if these simple principles really explain the 
observed diversity and complexity of life but do they?

-- And why do you doubt? -- asked the Patriarch. -- ``Time is in fact the 
hero of the plot. The time with which we have to deal is of the order of two 
billion years. What we regard as impossible on the basis of human experience 
is meaningless here. Given so much time, the "impossible" becomes possible, 
the possible probable, and the probable virtually certain. One has only to 
wait: time itself performs the miracles'' \cite{Wald}.

-- I think the supposition that time makes it possible the "impossible" to
become possible is seductive but fallacious, -- answered Wu-k'ung. -- Take,
for example, the typing monkeys parable. Do you really think "that a 
half-dozen monkeys provided with typewriters would, in a few eternities, 
produce all the books in the British Museum? Strictly speaking, one immortal 
monkey would suffice'' \cite{Borges}, so I can try this enterprise after
I learn from you how to become immortal. However, I doubt strongly I can 
ever succeed. For simplicity let us take only one aphorism, allegedly 
attributed to Francis Bacon, from the treasure of the British Museum: 
``Atheism is a thin layer of ice over which one person may cross, but a whole 
people would fall into an abyss.'' If my count is correct, it contains 109
letters, punctuation marks and spaces. A typewriter has about 50 keys. 
Therefore there are $50^{109}\approx 1.5\times 10^{185}$ different 
combinations of keys which are 109 characters long and I will have a 
probability of about $10^{-185}$ to type the correct one just by chance. 
Even if you take $10^{80}$ immortal monkeys ($10^{80}$ is the number of 
baryons in the visible universe) each of them producing a variant of 
our 109-character string in every $10^{-23}$ seconds (which is the 
characteristic time of strong interactions) you still 
will need about $10^{82}~\mathrm{s} \approx 3\times 10^{74}~\mathrm{years}$ 
to type this sole sentence with certainty. Do their typewriters endure such
a time span? Possibly not. Due to quantum tunneling, on a time scale of 
$10^{65}~\mathrm{years}$ even the most rigid materials behave like a liquid
\cite{Dyson_TWE}. Therefore, atoms constituting the typewriters will spread
around like the molecules of water. Besides, protons are expected to be
unstable due to (at least) quantum gravity effects with a typical lifetime of 
about $10^{45}~\mathrm{years}$ \cite{ProtonD} and in $10^{74}~\mathrm{years}$ 
all $10^{80}$ baryons in the visible universe will decay. There is no way
for monkeys to substitute Shakespeare!

-- But this is not at all required! -- the Patriarch was seemingly angry. 
-- Using your typing monkeys analogy, more realistic picture of evolution 
might be the following. You offer a string of text (already meaningful in 
real evolution) to a herd of monkeys which duplicate the string although
with some mistakes. After a generation an invisible hand of natural selection
inspects the output and selects the string with greatest ``fitness'' (the 
meaning of fitness depends on the concrete situation. In computer simulation
\cite{Dawkins1} the phrase with greatest resemblance with the target  
phrase was selected -- the situation more akin to artificial selection).
This slightly improved phrase is again offered to the herd for duplication.
Generation after generation the fitness increases and it was demonstrated
\cite{Dawkins1} that such ``artificial selection'' reaches the target
phrase in reasonable time from even meaningless start-up. Of course, this
example is given only to demonstrate the power of cumulative selection 
over the random search and is not a realistic model of evolution. But it 
indicates that the evolution driven by natural selection, after all, might 
be not as improbable process as you stupid probabilities indicate.  
   
-- Very good, -- replied Wu-k'ung. Realistic or not you have portrayed some 
model of evolution. Why not elaborate it a bit more and try on computer? Some
earlier computer simulations of this kind led to conclusions that ``it seems 
to require many thousands, perhaps millions, of successive mutations to 
produce even the easiest complexity we see in life now. It appears, naively at 
least, that no matter how large the probability of a single mutation is, 
should it be even as great as one-half, you would get this probability raised 
to a millionth power, which is so very close to zero that the chances of such 
a chain seem to be practically non-existent" \cite{Ulam}. 

-- An accurate computer simulation of the evolution process is extremely 
difficult, -- answered the Patriarch. -- Too many uncontrollable factors are 
involved in real evolution. The computer simulations you mention were 
intended as merely the zeroth approximation to the problem. They modeled only
mitosis in absence of sexual reproduction and really showed that the expected 
progress was much too slow. ``However, and most biologists realize it anyway, 
the Darwinian mechanism together with mixing of genes accelerate enormously 
the rate of acquiring new "favorable" characteristics and leave the 
possibility of sufficiency of the orthodox ideas quite open'' \cite{Ulam1}.
Anyway, fossil evidence of evolution is enormous and indicates that it really
has occurred. We see continual change and increase in complexity if fossils
are arranged according to their age.

-- That's true, -- agreed Wu-k'ung. -- But I do not think this fact proves
Darwinian evolution beyond any doubt. Consider, for example, the following 
analogy \cite{Sewell}. Take my programs of magnetic field calculation which
I develop while studying classical electrodynamics. You can not deny that
they have many similarities and show a steady increase in complexity over 
time. Initially these programs were able to deal with only some simple 
two-dimensional problems. Over the years many new capabilities have appeared 
to mention among others an ability to crack difficult three-dimensional 
problems and fancy curved coil configurations, as well as a development of 
user-friendly interactive interface and graphical tools. Any of these changes 
have not happened gradually but appeared suddenly in new versions. These 
improvements required hundreds of interrelated new lines in the program and I 
never could produce them would I so foolish to try do this by a 
trial-and-error method without advance planning. In fact any ``point 
mutations'' is deleterious and catastrophic in the case of programming
languages because ``no currently existing formal language can tolerate random 
changes in the symbol sequences which express its sentences. Meaning is almost
invariably destroyed. Any changes must be syntactically lawful ones.'' I can
not see any reason why the language of DNA programming should be different in 
this respect. ``I would conjecture that what one might call "genetic 
grammatically" has a deterministic explanation and does not owe its stability 
to selection pressure acting on random variation'' \cite{Eden}.

-- Frankly speaking, -- continued Wu-k'ung, -- I'm pretty sure that Darwinian
version of evolution is quite dead in light of modern findings. None of the 
three main principles underlying Darwin's approach withstand close inspection.
Darwin considered `infinitesimal variation', or point mutations in modern 
terms, as the main process that provides the material for evolution. The real
finding, however, is that point mutations due to chemical and physical insults
to the genome are almost always deleterious and catastrophic. This fact can
hardly came as a big surprise for computer scientists and software engineers
with their highly rigid programming languages. However, for proponents of
Darwinian evolution ``it has been a surprise to learn how thoroughly cells
protect themselves against precisely the kinds of accidental genetic change
that, according to conventional theory, are the sources of evolutionary
variability'' \cite{Shapiro}. 

It seems Darwin considered gradualism as very important for his theory.
In chapter six of his famous {\it On the Origin of Species} he states
``If it could be demonstrated that any complex organ existed, which could
not possibly have been formed by numerous, successive, slight modifications,
my theory would absolutely break down'' \cite{Koonin}. But now we know that
evolution is anything but gradual and is better described as `punctuated
equilibrium'. For most of the time sexually reproducing species experience
very little if any change as witnessed by fossil record. Then abruptly new
species appear rapidly in rare events usually associated with some natural
catastrophe. The rise and fall of the dinosaurs is a good example.

A study of tetrapod footprints and skeletal material revealed that large
theropod dinosaurs appeared less than 10,000 years after the
Triassic-Jurassic boundary and less than 30,000 years after the last
Triassic taxa, synchronous with a terrestrial mass extinction which wiped out
at least half of the species living on Earth at that time \cite{Dino}.
The presence of an iridium anomaly and a fern spore spike suggests that
a bolide impact was the possible cause \cite{Dino}. The time scale of the
appearance of large theropod dinosaurs is the blink of an eye in geological
terms. Then dinosaurs dominated on Earth for next 135 million years until
another bolide impact ended the reign of dinosaurs 65 million years ago and
marked the rise of mammalian age.

Now about natural selection which according to Darwin is the main driving
force of evolution. ``Evolutionary-genomic studies show that natural
selection is only one of the forces that shape genome evolution and is not
quantitatively dominant, whereas non-adaptive processes are much more
prominent than previously suspected'' \cite{Koonin}. The picture that
emerged if far more complex and pluralistic than Darwin could imagine at his
time. ``In addition to point mutations that can be equated with Darwin's
`infinitesimal changes', genome evolution involves major contributions from
gene and whole genome duplications, large deletions including loss of genes
or groups of genes, horizontal transfer of genes and entire genomic regions,
various types of genome rearrangements, and interaction between genomes of
cellular life forms and diverse selfish genetic elements'' \cite{Koonin}.

-- Science is not a religion, -- answered the Patriarch, -- and it does not
stick dogmas. Of course modern science shows that evolution is much more
complex than previously thought. However, ``the emerging landscape of genome
evolution includes the classic, Darwinian natural selection as an important
component'' \cite{Koonin}. As for the `punctuated equilibrium', this is
precisely what you can expect in analogy with phase transitions in physics
\cite{Volkenshtein}. ``The theory of punctuated equilibrium is a minor gloss
on Darwinism, one which Darwin himself might well have approved if the issue
had been discussed in his day'' \cite{Dawkins1}.

``Depicting the change in the widest strokes possible, Darwin's paramount
insight on the interplay between chance and order (introduced by natural
selection) survived, even if in a new, much more complex and nuanced form,
with specific contributions of different types of random processes and
distinct types of selection revealed. By contrast, the insistence on
adaptation being the primary mode of evolution that is apparent in the
{\it Origin}, but especially in the Modern Synthesis, became deeply
suspicious if not outright obsolete, making room for a new worldview that
gives much more prominence to non-adaptive processes'' \cite{Koonin}.

An interplay between chance and order can produce quite remarkable and
strange outputs those origins are difficult to explain in retrospect.
Take, for example, a bizarre mating behaviour of praying mantis.
``If you have got a female mantis alone in a cage, and put in a male, that
male instantly freezes. The praying mantis, like a lot of other animals,
such as frogs, don't seem to be able to see anything unless it moves.
The male knows that, and he is watching the female very carefully. If she
looks away for a moment, he takes a hasty few steps forward. Then he freezes
again as soon as she looks back. This can go on for hours. If the male is
fortunate, he reaches the female, mounts her, and goes through a normal
copulation. Incidentally, once an American male mantis starts copulating,
the female never bothers him. It's our better standard of living. But often
the female sees him first. With that, she grabs him, always by the head.
Then she begins to eat him, always starting with his head. As soon as she
has eaten off the head, the male goes into a very interesting pattern of
behavior. He plants his front feet squarely and begins to circle around
them, meanwhile going through violent copulatory motions, because once the
male loses his head, the copulatory center is released.  In this way such
a headless male will frequently succeed in mounting the female and going
through a normal copulation'' \cite{Wald1}.

-- Yes, -- said Wu-k'ung, -- real evolution is full of mysteries. Another
enigmatic example of evolution is the origin of bats \cite{bats}. There are 
two distinct groups of these flying mammals, namely  megabats and microbats.
The question is whether these two groups originated from a common flying
ancestor or they have separate origins. Megabats share many features of brain
organization with primates and these features are absent in microbats. This
fact suggests a common ancestor for megabats and primates. But if true, this
scenario implies that wings evolved twice, once very early in microbats and
then again later in a branch of primates leading to megabats \cite{bats}.
Another alternative is that microbats and megabats are monophyletic and the
primate brain features evolved twice, once in primates and once in megabats
\cite{bats}. In either case we have to face the fact that evolution has found
the same solution twice which is difficult to expect from the random process. 

-- I think, -- continued Wu-k'ung, -- ``there are far more unresolved 
questions than answers about evolutionary processes, and contemporary science 
continues to provide us with new conceptual possibilities'' \cite{Shapiro}.
Especially interesting are indications that all cells possess biochemical
abilities for cutting and splicing of their DNA molecules into new sequence 
arrangements. Cells perform some kind of genetic engineering! ``In other 
words, genetic change can be massive and non-random. Some organisms, such as 
the ciliated protozooan Oxytricha, completely reorganize their genetic 
apparatus within a single cell generation, fragmenting the germ-line 
chromosomes into thousands of pieces and then reassembling a particular 
subset of them into a distinct kind of functional genome'' \cite{Shapiro}.
The picture that emerges is quite different from the Darwinian view of 
evolution, but is far more fascinating and realistic. If the natural genetic 
engineering \cite{Shapiro,Shapiro1} is the main driving force of evolution
then it is not impossible that nearly all ingredients of Lego-like genome
architecture were already present billions of years ago in first living cells
known to us. Complexity has not been increased during evolution, it became 
only apparent, like the complexity of an aircraft which is hidden in 
technical blueprints and documentation but becomes apparent during the
construction.  ``Together with the realization that genome contraction is at 
least as common in evolution as genome expansion, and the increase of genomic 
complexity is not a central evolutionary trend, the concept of non-adaptive 
genome evolution implies that the idea of evolutionary progress can be safely 
put to rest'' \cite{Koonin}. 

-- ``Most biologists now agree that natural selection is the key evolutionary
force that drives not only evolutionary change within species but also the
origin of new species. Although some laypeople continue to question the
cogency or adequacy of natural selection, its status among evolutionary
biologists in the past few decades has, perhaps ironically, only grown more
secure `` \cite{Orr1}, -- answered the Patriarch indignantly. -- ``The
status of natural selection is now secure, reflecting decades of detailed
empirical work. But the study of natural selection is by no means complete''
\cite{Orr1}. I think any doubt in natural evolution will wither with time, 
although I do not exclude that maybe we need somewhat broader view of natural
selection including genetic engineering and other modern findings. However,
``no one seriously doubts that natural selection drives the evolution of most
physical traits in living creatures - there is no other plausible way to
explain such large-scale features as beaks, biceps and brains'' \cite{Orr1}.

-- I seriously doubt that natural selection drives the evolution, -- replied
Wu-k'ung. -- But now I'd like to speak about another conundrum recognized by
Erwin Schr\"{o}dinger many years ago \cite{WIL}. In statistical physics order
emerges from disorder because number of particles are huge. On the background
we always have the chaos of thermal motion. The expectation of the 'naive 
physicist', according to Schr\"{o}dinger, is that  an organism and all the 
biologically relevant processes must have extremely many-atomic structure
to be safeguarded against haphazard single-atomic events attaining 
too great importance. In reality, however, ``incredibly small groups of atoms, 
much too small to display exact statistical laws, do play a dominating role 
in the very orderly and lawful events within a living organism'' \cite{WIL}. 

I think the resolution of this enigma how do cells manage to function 
incredibly reliable and regular manner under constant Brownian bombardment,
lies in the fact that cells are dynamic, integrated systems. No part of the 
cell makes sense outside the context established for the whole cell. Cells
are holistic systems and their parts show extreme cooperation and 
coordination, accompanied with the extraordinary smart proofreading and
error correction systems. Cells continually monitor their external and 
internal environment, perform information processing and make decisions
how to react on various challenges. Impressive picture, is not it?

Just one small example that biological information processing is a reality
\cite{Shapiro1}. The bacterium {\it E. coli} can discriminate between glucose 
and lactose and it can control the expression of the lactose metabolic 
proteins so that they are only synthesized once glucose is no longer 
available. In fact cells execute the following logical instruction: 
``IF lactose present AND glucose not present AND cell can synthesize active 
LacZ and LacY, THEN transcribe lacZYA from lacP'' which involves many 
molecules and compartments of the cell, not just DNA and DNA binding 
proteins \cite{Shapiro1}.

One naturally expects the devices capable to perform such complicated tasks
to be irreducibly complex. They are bound to have some minimal complexity
in order to function reliably under a thermal environment. This is just simple
statistical physics. There are good indications that LUCA, the Last Universal 
Common Ancestor from which every living cell on the planet has descended, was
as complex as any present-day  bacterium. And how this miraculously complex 
first cell come into existence over three billions years ago?

-- This is the origin of life problem, -- replied the Patriarch, -- different 
from the problems of evolution.

-- It is, -- agreed Wu-k'ung. -- But let us return to the problem of 
cooperation in the cell. Why not to go one step further. Why not to expect
that bacterial and microbial colonies also use cooperative behaviour to cope
environmental hazards?

-- I must admit that you are right, -- said the Patriarch. -- ``To face 
changing environmental hazards, bacteria resort to a wide range of 
cooperative strategies. They alter the spatial organization of the colony in
the presence of antibiotics for example. Bacteria form complex patterns as
needed to function efficiently. Bacteria modify their colonial organization
in ways that optimize bacterial survival. Bacteria  have collective memory by
which they track previous encounters with antibiotics. They collectively 
glean information from the environment, communicate, distribute tasks, 
perform distributed information processing and learn from past experience'' 
\cite{Ben-Jacob}.

-- In this case, --  said Wu-k'ung, -- we have to conclude that ``the 
take-home lesson of more than half a century of molecular microbiology is to 
recognize that bacterial information processing is far more powerful than 
human technology'' \cite{Shapiro2}.

-- I suspect, -- continued Wu-k'ung, -- this cooperative behaviour in its 
essence is not a result of evolution but was present from the very beginning, 
at the time of LUCAs. Combined with natural genetic engineering and horizontal
gene transfer, this primordial biological information processing, perhaps,
paved a way for subsequent evolution according to some in-built templates.
``Such gradual refinement through the horizontal sharing of genetic 
innovations would have led to the generation of a combinatorial explosion of 
genetic novelty, until the level of complexity, as exemplified perhaps by the 
multiple levels of regulation, required a transition to the present era of 
vertical evolution'' \cite{BNR}. It is true that most fellow scientist today
are hero worshipers, and Darwin is undoubtedly their hero. Quite deservedly, 
I think, because historically Darwin's theory had a tremendous impact on 
science. But in the long run, hero-worship makes no good and harms the 
science. Thus, I ``regard as rather regrettable the conventional 
concatenation of Darwin's name with evolution, because there are other 
modalities that must be entertained and which'' some scientists ``regard as 
mandatory during the course of evolutionary time'' \cite{BNR}.

\section{Wu-k'ung reflects on intelligent design}
The Patriarch was indignant at Wu-k'ung's stubbornness not to accept 
Darwinian view of life.

-- I see you are inclined towards the intelligent design theory, -- said he 
louringly to Wu-k'ung.

-- If LUCA was really as complex as modern bacteria, not to speak about the 
alleged cooperative behaviour of these first microorganisms with in-built 
evolutionary templates, which quite may turn as mere my fantasy, -- begin 
Wu-k'ung his answer, -- I see no way how it can emerge from inanimate
matter without intelligent designers behind. 

-- And who were these designers? -- asked the Patriarch sarcastically. --
I admit,  ``an honest man, armed with all the knowledge available to us now, 
could only state that in some sense, the origin of life appears at the moment 
to be almost a miracle, so many are the conditions which would have had to 
have been satisfied to get it going" \cite{Crick}. But this gives no excuse
to plunge into religious insinuations. Or maybe you prefer Crick and Orgel's
directed panspermia theory? They speculated that the LUCA, ``the primitive 
form of life was deliberately planted on the Earth by a technologically 
advanced society on another planet'' \cite{Orgel}. Cute theory. But actually
it does not solve the origin-of-life problem, except enlarging the available 
resources involved from Earth to, maybe, entire cosmos. The problem is the 
origin of these aliens which seeded the Earth. This remembered me a notorious
story. {\it A well-known scientist (some say it was Bertrand Russell) once 
gave a public lecture on astronomy. He described how the Earth orbits around 
the Sun and how the Sun, in turn, orbits around the center of a vast 
collection of stars called our galaxy. At the end of the lecture, a little 
old lady at the back of the room got up and said: "What you have told us is 
rubbish. The world is really a flat plate supported on the back of a giant 
tortoise." The scientist gave a superior smile before replying, "What is the 
tortoise standing on?" "You're very clever, young man, very clever," said the 
old lady. "But it's turtles all the way down!"} \cite{Howking}. How can you
escape this infinite reduction puzzle?

-- Were I human, -- answered  Wu-k'ung, -- I probably would say that 
intelligence and spirituality are primordial properties of Cosmos, and apply 
to God as the ultimate origin of life. But I'm humble monkey, without 
in-built religious instincts. So I'm bound to answer that I don't know. 

However, -- continued Wu-k'ung, -- ``as biochemists discover more and more 
about the awesome complexity of life, it is apparent that its chances of 
originating by accident are so minute that they can be completely ruled out. 
Life cannot have arisen by chance'' \cite{Hoyle1}. "Although the tiniest 
bacterial cells are incredibly small, weighing less than $10^{-12}$ grams, 
each is in effect a veritable micro-miniaturized factory containing thousands 
of exquisitely designed pieces of intricate molecular machinery, made up 
altogether of one hundred thousand million atoms, far more complicated than 
any machinery built by man and absolutely without parallel in the non-living 
world" \cite{Denton}. How can I believe that this marvelous thing, a living
cell, aroused by chance?

-- Science is not about beliefs, -- answered the Patriarch  sternly, -- it's
about facts. You can not expect science to give an immediate answer to all
your questions. ``Just as most of the weird Cambrian monsters eventually went 
extinct, many current hypotheses in evolution will also wither over time. 
Those that survive, however, will be inestimably powerful for explaining the 
natural world'' \cite{Rennie}. The origin of life is a difficult problem for
science today, I admit, but maybe tomorrow we will find the clue. You should 
just wait. And if you want not to passively wait but participate in the 
exciting process of scientific exploration, you must first learn a lot of 
already established facts and not plunge into a bad philosophy. One of the 
virtues of Darwinism is that it removed teleology from our thinking.
 
-- That's the problem, -- said Wu-k'ung, -- ``orthodox scientists are 
occupied by a fight against religion instead of finding the truth'' 
\cite{Hoyle1}. I prefer more sober-minded position about religion. It's true
that Darwin was successful in eliminating religious motives not only from 
biology but from social life also, but I doubt that the spreading of 
nihilistic outlook that followed is a progressive thing. It's an empirical
fact that humans have hardwired religious instinct. You can argue about the 
origin of this instinct, but It's the fact that It's here. And once it's here,
it should be important, because in even your version of genesis omnipotent
natural selection picks up only important things. Therefore, religion must
have some important regulatory function and it is not wise to throw it away.
We monkeys do not have any religion but are we wiser than men? Or do you like
human social life to be organized in the manner monkeys do? Carl Gustav Jung
once remarked, "Among all my patients in the second half of life, that is, 
over thirty-five, there has not been one whose problem in the last resort was 
not that of finding a religious outlook on life. It is safe to say that every 
one of them fell ill because he had lost that which the living religions of 
every age have given their followers, and none of them has really been healed 
who did not regain his religious outlook" \cite{Jung}.

``Rather than accept that fantastically small probability of life having
arisen through the `blind' forces of nature, it seems better to suppose that
the origin of life was a deliberate intellectual act. By `better' I mean less
likely to be wrong''  \cite{Hoyle2} -- concluded Wu-k'ung. 

-- I do not like you attitude towards biology in general and evolution in 
particular, -- declared the Patriarch. -- "Like entropy, which perpetually 
increases, educational standards perpetually worsen.  And like entropy, which 
increases inevitably because of the policies of physics, education worsens 
inevitably because of the policies of educators. Instead of teaching being 
properly confined to the rote-learning of facts and well-proven techniques, 
pupils are confused nowdays by the teaching of meanings that they cannot 
comprehend" \cite{Hoyle2}. But you know I adhere to the high educational 
standards of the past and do not tolerate premature vain attempts of 
illiterate pupils to engage themselves into dubious and precarious 
theorizing. Unfortunately, this is precisely what you are doing now. Instead 
of studying well established biology, you are trying to answer questions which
cannot be answered without profound knowledge in biology and other sciences.

-- However, -- continued the Patriarch, -- you have mastered physics quite
well. So let me indicate to you that modern physics suggests a world view
in which even most unlikely events will eventually happen. I mean, of 
course, eternal inflation theory \cite{Guth}.

As a poetic metaphor, this world view is already present in the 1941 short 
story of Jorge Luis Borges {\it The Library of Babel}. ``The Library is
unlimited and cyclical. If an eternal traveler were to cross it in any 
direction, after centuries he would see that the same volumes were repeated 
in the same disorder (which, thus repeated, would be an order: the Order)''
\cite{LoB}.

-- I could also quote this book to support my position, -- remarked
Wu-k'ung in delight. He was very smart in literature in contrast to 
biology. -- ``The Library exists {\it ab aeterno}. This truth, whose 
immediate corollary is the future eternity of the world, cannot be placed 
in doubt by any reasonable mind. Man, the imperfect librarian, may be the 
product of chance or of malevolent demiurgi; the universe, with its 
elegant endowment of shelves, of enigmatical volumes, of inexhaustible 
stairways for the traveler and latrines for the seated librarian, can only 
be the work of a god. To perceive the distance between the divine and the 
human, it is enough to compare these crude wavering symbols which my 
fallible hand scrawls on the cover of a book, with the organic letters 
inside: punctual, delicate, perfectly black, inimitably symmetrical''
\cite{LoB}. 

-- In less poetic terms, -- the Patriarch paid no heed to Wu-k'ung's
remark, -- eternal inflation looks no less fantastic \cite{Guth,Tegmark}.
Eternal inflation theory is Copernican view of the world pushed to its
extreme. Before Copernicus, it was thought the Earth was the center of the
universe. Now we know that the Earth is just one of nine planets orbiting 
the Sun. The Sun itself is quite an ordinary star from the Milky Way galaxy
which contains 100 billion other stars. The Milky Way belongs to the Vigro
cluster of galaxies and lies at its periphery. The Vigro cluster by itself 
is a fairly average cluster that unifies a few hundred members. There are 
trillions of other galaxies out there in the universe. And now the eternal
inflation theory tells us that there are infinite number of other universes
similar to our own universe, the so called $\cal{O}$ regions \cite{Vilenkin}, 
and many more that does not support life. However, this is not the strangest 
thing about the eternal inflation theory. After all many universes, or 
multiverse, is just a huge extrapolation of the Copernican logic. The 
strangest thing is that while the number of $\cal{O}$ regions is infinite, 
the number of possible coarse-grained macroscopic histories in each of this 
universes is finite albeit very large \cite{Vilenkin}. An inevitable strange 
conclusion from this is, for example, that another macroscopically identical 
copy of you is located at some $10^{10^{29}} ~\mathrm{m}$ away \cite{Tegmark} 
having the same life history up to the present point. Besides, if one assumes, 
as is usually done, that every history consistent with exact conservation laws 
has a non-vanishing probability and will eventually occur, then ``some amusing
situations can be entertained where distant copies of ourselves play all 
sorts of different roles. Some readers will be pleased to know that there are
infinitely many $\cal{O}$-regions where Al Gore is President and - yes - Elvis
is still alive'' \cite{Vilenkin}.

Implications of the eternal inflation theory, sometimes called also 'many 
worlds in one' (MWO) model \cite{Vilenkin}, for the origin-of-life problem 
were considered by Koonin \cite{Koonin1}. Below I will try to summarize main
points of this investigation.

The hardest problem in all of biology is the origin of replication and 
translation systems. Any biological evolution is impossible without 
sufficiently fast and accurate genome replication. However, the  replication 
and translation systems of even the simplest living cells are based on the
molecular machinery of tremendous complexity. This system appears to be the
hardest example of `irreducible complexity' \cite{Behe} and no one knows how
such a system could evolve.

-- I remember, -- said Wu-k'ung, -- that the problem of self-replication
from the quantum mechanics perspective was considered by Wigner long ago
\cite{Wigner}. Wigner assumed that the self-replication process of an 
organism under its interaction with the nutrient is described by some
unitary ``collision matrix' $S$ and showed that if $S$ is a random matrix
then the number of equations which describe the transformation is so much
greater than the number of unknowns that "it would be a miracle" \cite{Wigner}
if the transformation equations were satisfied. Wigner himself finds the
argument ``not truly conclusive'' \cite{Wigner} and so do others \cite{Baez,
Bass,Eigen}. However, such considerations clearly indicate that even 
moderately complex self-replication organisms are not quite trivial objects 
expected to be found at every corner of the universe. Some even claim that
``for all physics has to offer, life should never have appeared and if it ever
did it would soon die out'' \cite{Yockey}.
 
-- Yes, -- said the Patriarch, -- chance emergence of a sufficiently complex
biochemical system is very improbably but not impossible in MWO model. Hoyle
once remarked that  the chance that life could have been started in this way
is comparable with the chance of a tornado sweeping through a junkyard might 
assemble a functional Boeing-747, ready to fly, from the materials therein 
\cite{Hoyle1}. Nevertheless, ``the MWO model not only permits but guarantees 
that, somewhere in the infinite multiverse - moreover, in every single 
infinite universe, - such a system would emerge'' \cite{Koonin1}.

-- But what about the second law of thermodynamics? -- exclaimed  Wu-k'ung.
-- ``The law that entropy always increases - the second law of thermodynamics 
- holds, I think, the supreme position among the laws of nature. If someone
points out that your pet theory of the universe is in disagreement with
Maxwell's equations - then so much the worse for Maxwell's equations. If it
is found to be contradicted by experiments - well, these experimentalists do
bungle things sometimes. But if your theory is found to be against the
second law of thermodynamics I can give you no hope; there is nothing for it
but to collapse in deepest humiliation'' \cite{Eddington}. I, of course, know
that the second law applies to the closed systems only. ``But the second law 
of thermodynamics - at least the underlying principle behind this law - simply
says that natural forces do not cause extremely improbable things to happen, 
and it is absurd to argue that because the Earth receives energy from the Sun, 
this principle was not violated here when the original rearrangement of atoms 
into encyclopedias and computers occurred'' \cite{Sewell}. 

-- In the MWO model -- replied the Patriarch -- extremely improbable things 
do happen. This of course violates the second law. But the  second law of 
thermodynamics is statistical in its nature. Molecules do not gather in a one
half of the vessel not because this is absolutely forbidden by exact 
conservation laws but because such fluctuation is extremely improbable, much
less improbable than a Shakespearean sonnet typed by a monkey. However, in the
infinite multiverse, as I already said, every event with non-zero probability
will eventually happen no matter how small the probability is. I can not say
that I easily swallow this. As the origin-of-life problem is concerned,  
``a crucial aspect of the framework developed here is brought about by a 
disturbing (almost nightmarish) but inevitable question: in the infinitely 
redundant world of MWO, why is biological evolution, and in particular, 
Darwinian selection relevant at all? Is it not possible for any, even the 
highest degree of complexity to emerge by chance? The answer is 'yes' but 
the question misses point. Under the MWO model, emergence of an infinite
number of complex biotas by chance is inevitable but these would be vastly 
less common than those that  evolved by the scenario that includes the switch 
from chance/anthropic selection to biological evolution'' \cite{Koonin1}.

-- Do you consider such explanation of the origin-of-life problem, ``which 
would require happenings every bit as miraculous as the views of religious 
fundamentalist'' \cite{Hoyle2}, as really scientific? -- asked Wu-k'ung.
-- I have a feeling that some trickery could be always invoked when 
infinities come into the game, like the cunning Devil striping us our money in 
a nefarious underground bar through an infinite number of transactions during 
which the amount of our money continuously increases but nothing at all is left
when the super-transaction is over in a finite time \cite{Zeno}.

-- Frankly speaking, -- answered the Patriarch, -- I hope some less 
extravagant explanations, like RNA world \cite{RNA}, will eventually prove to 
be correct. 

-- However, -- said  Wu-k'ung, -- eternal inflation hypothesis is taken too 
seriously and is actively discussed in physics, if not in biology. You know,
these theoretical ``physicists behave a lot more like 'bosons' which coalesce 
in large packs'' \cite{Connes} towards fashionable ideas. But if you accept
one aspect of the theory you have to accept its other consequences too. 
Therefore, I can not see why we can not use MWO model in biology if in physics
it is considered to be so successful. But if we accept MWO model as the 
solution of the origin-of-life problem, an immediate inference, I'm afraid, 
would be that life on the planet Earth is almost certainly the result of 
intelligent design.

The argument is quite simple. Human civilization is at the dawn of 
nanotechnology. And no doubt in many pocket universes of the MWO model its
twin civilizations already reached the level of nanotechnology which allows   
a production of artificial living organisms. Sooner or later our civilization,
or more precisely your civilization, because I do not expect monkeys to 
participate appreciably in this process, except maybe some distinguished
specimens like me, sooner or later human civilization also will reach this
level in nanotechnology and will be able not only to reproduce any living
organism ever present on Earth, but to create new life forms. I admit, at 
present, ``such a project would be quite beyond our practical ability, but 
not beyond our comprehension. Indeed we are nearer to understanding what 
would be involved in it than a dog is to understand the construction of a 
power station'' \cite{Hoyle2}.

However, as soon as we realize that  MWO model inevitably predicts the 
existence of infinitely many advanced civilizations capable to create 
artificial life, it becomes clear that then we find a life at some particular
place in the universe it is vastly more probable this life to be a product of 
nanotechnology of some advanced civilization than the product of primordial
miracle due to quantum fluctuations like Hoyle's Boeing-747 emerging in one
stroke of hurricane from the junkyard.

I conclude, therefore, that the intelligent design is a vastly more probably 
origin of humans than the random evolution by mutations and natural selection.
This is an inevitable conclusion of modern science if we take MWO argument
seriously and not as a merely ``turtles all the way down'' type hand-waving. 
-- Wu-k'ung smiled and disappeared in a somersault. However he returned soon
to continue his speech.

-- In fact, -- said he, -- I'm curious about the nature of logical thinking.
Why the MWO model is considered as scientific and more logical than the 
religious picture of the world if it predicts happenings vastly more 
miraculous than described in the Bible?

-- Carl Gustav Jung has an interesting observation about the nature of 
logical thinking, -- answered the Patriarch. -- He tells the following story.
``Two anklets were found in the stomach of a crocodile shot by a European.
The natives recognized the anklets as the property of two women who, some
time before, had been devoured by a crocodile. At once the charge of 
witchcraft was raised, for this quite natural occurrence, which would never 
have aroused the suspicions of a European, was given an unexpected 
interpretation in the light of one of those presuppositions which 
L\'{e}vy-Br\"{u}hl calls 'collective representations'. The natives said that 
an unknown sorcerer had summoned the crocodile and had bidden it to bring 
him the two women. The crocodile had carried out the command. But what about 
the anklets in the beast's stomach? The natives maintained that crocodiles 
never ate people unless bidden to do so. The crocodile had received the 
anklets from the sorcerer as a reward" \cite{Jung}.

For us these conclusions of aborigines seem strange and absolutely illogical.
But are they really so? In fact, "it only strikes us in this way because we 
start from assumptions wholly different from those of primitive man. If we 
were as convinced as he is of the existence of sorcerers and of mysterious 
powers, instead of believing in  so-called natural causes, his inferences 
would seem to us perfectly reasonable. As a matter of fact, primitive man is 
no more logical or illogical than we are. His presuppositions are not the 
same as ours, and  that is what distinguishes him from us" \cite{Jung}.

-- Interesting observation indeed, -- agreed Wu-k'ung. -- Another interesting
observation of this kind, which shows how subtle and fragile human's 
ability to think logically is, can be found in the Sylvia Nasar's masterpiece
\cite{Nasar}, a splendid biography of John Forbes Nash, the mathematical 
genius and Nobel Prize winner. At the top of his scientific career, he went 
insane and struggled for years that followed with schizophrenia which almost 
destroyed both his career and his marriage. While ill, he has suffered 
increasingly bizarre delusions that he was the Prince of Peace and the Emperor 
of Antarctica, and believed that aliens were trying to contact him through the 
{\it New York Times} newspaper. Nasar describes how Nash, while in the mental 
hospital, explains to a colleague  his acceptance of strange ideas which for 
'normal' people appears utterly illogical.

{\it John Forbes Nash, Jr. -- mathematical genius, inventor of a theory of 
rational behavior, visionary of the thinking machine -- had been sitting with 
his visitor, also a mathematician, for nearly half an hour. It was late on 
a weekday afternoon in the spring of 1959, and, though it was only May, 
uncomfortably warm. Nash was slumped in an armchair in one corner of the 
hospital lounge, carelessly dressed in a nylon shirt that hung limply over 
his unbelted trousers. His powerful frame was slack as a rag doll's, his 
finely molded features expressionless. He had been staring dully at a spot 
immediately in front of the left foot of Harvard professor George Mackey, 
hardly moving except to brush his long dark hair away from his forehead in 
a fitful, repetitive motion. His visitor sat upright, oppressed by the 
silence, acutely conscious that the doors to the room were locked. Mackey 
finally could contain himself no longer. His voice was slightly querulous, 
but he strained to be gentle. ``How could you,'' began Mackey, ``how could 
you, a mathematician, a man devoted to reason and logical proof...how could 
you believe that extraterrestrials are sending you messages? How could you 
believe that you are being recruited by aliens from outer space to save the 
world? How could you...?'' Nash looked up at last and fixed Mackey with an 
unblinking stare as cool and dispassionate as that of any bird or snake. 
``Because,'' Nash said slowly in his soft, reasonable southern drawl, as if 
talking to himself, ``the ideas I had about supernatural beings came to me 
the same way that my mathematical ideas did. So I took them seriously''}
\cite{Nasar}. 

-- But the power of scientific method, -- said the Patriarch -- its real 
strength that gives the science profound depth and reliability, lays in the 
fact that real science relies on experimental method. ``Without 
experimentalist, theorist tend to drift'' \cite{Lee1} and dwell in wild 
fantasies. So you beloved Intelligent Design hypothesis will remain a mere
fantasy until you can give some clues how it can be unambiguously checked by
facts.

-- It is difficult to indicate facts that unambiguously indicate towards 
Intelligent Design, it is very difficult -- answered Wu-k'ung. -- I can only 
point to some plausible consequences of intelligent design. First of all if 
the evolution was predefined by some intelligent design, I would expect the 
primordial cells and viruses to constitute parts of a well-thought-out and 
well balanced system aimed to produce increasingly complex living forms by 
genetic engineering. Natural selection, I think, could be used in some form by 
these intelligent designers to polish required traits, but it cannot be the 
main driving force. Simply the algorithm of genetic engineering and the 
mechanism to accomplish it, as well as some meaningful simpler building blocks
of genetic code, were already present in the primordial assembly of cells and 
viruses. I have no idea how this primordial assembly was produced by 
intelligent designers, but after all parts of the system is in place it 
perhaps can function autonomously, without further need from the side of the 
intelligent designers to interfere. 

I see it plausible random search algorithms to be a part of the system. But
the natural selection, by its accent on competition, not cooperation, is not
the most powerful random search algorithm. Social behavior is ubiquitous in 
the animal kingdom. Darwinians, of course, try to explain this by various 
advantages it gives for survival, and proudly stop at that. However, two 
intelligent scholars had have a deeper insight that ``once again nature has 
provided us with a technique for processing information that is at once 
elegant and versatile'' \cite{PSO}. They were driven by the philosophy 
``that allows wisdom to emerge rather than trying to impose it, that emulates 
nature rather than trying to control it, and that seeks to make things 
simpler rather than more complex'' \cite{PSO}. As a result a powerful and 
simple random search algorithm, Particle Swarm Optimization (PSO), has 
emerged. A key idea of this algorithm is illustrated by reference to fish 
schooling. ``In theory at least, individual members of the school can profit 
from the discoveries and previous experience of all other members of the 
school during the search for food. This advantage can become decisive, 
outweighing the disadvantages of competition for food items, whenever the 
resource is unpredictably distributed in patches'' \cite{Wilson}.

So, -- continued Wu-k'ung, -- if something like PSO is used in evolution,
I expect orchestrated mutations to appear when, for example, bacteria 
develop drug resistance. I mean favorable mutations which appear 
simultaneously at different places of the bacterial colony after some cell
in it randomly finds the correct combination of genetic code via genetic 
engineering. I think such mutations are hard to explain in Darwinian paradigm 
because they are not random but deterministic and require information sharing 
between cells. 

-- Orchestrated mutations, -- said the Patriarch thoughtfully, -- you at 
least are making some experimentally testable prediction which you think is
associated with Intelligent Design. This is good. Normal science works just
in this way. A bad thing with you Intelligent Design theory is, however, that
nobody has ever seen these orchestrated mutations in bacterial drug 
resistance studies.

-- Maybe because nobody looked for them? -- answered Wu-k'ung. -- By the way 
we can further extend this line of thought and assume that the whole biosphere 
should show signs of tight interconnections, fine tuning, high level of 
optimization and stability if it owe its existence to Intelligent Design.
Picturesquely we can say that the Earth with its biosphere is one big 
self-regulating living system where every part has its specific function and
predestination. An interesting question is what role was predestined in this 
system for humans with their yet restricted intelligence and knowledge 
dangerously coupled to increasing  awesome ability to upset the biosphere. 
The only hope for monkeys is that probably the wise Intelligent Designers
had anticipated a protection from fools, and reckless humankind will be not
allowed to destroy the life on Earth. The perspective for humans, I'm afraid,
is not very good if all this is true and humans fail to correctly understand
their predestination - they can simply get extinct like many other species
in Earth's history.

-- What you are speaking about is known for a long time as Gaia hypothesis, 
-- intervened the Patriarch. -- The Gaia hypothesis was proposed by James 
Lovelock \cite{Lovelock,Lovelock1,Lovelock2}. Gaia is Earth goddess in Greek 
mythology. According to Lovelock, ``the entire range of living matter on 
Earth from whales to viruses and from oaks to algae could be regarded as 
constituting a single living entity capable of maintaining the Earth's 
atmosphere to suit its overall needs and endowed with faculties and powers 
far beyond those of its constituent parts'' \cite{Lovelock2}. In fact
Lovelock suggests a quite sound and testable idea that life on Earth 
participates in and even is the main constituent of the global feedback 
mechanisms which ensures stability of environmental conditions on Earth
comfortable for life. For example, the surface temperature of Earth remained
remarkably constant for eons despite the fact that the energy provided by 
the Sun has increased by about 30\%. The atmospheric composition is also 
surprisingly constant instead being unstable as expected from the oxygen's
high chemical reactivity. The same is true for the ocean salinity which is
important for most cells. Undoubtedly living organisms has played a crucial
role in maintenance of this dynamical equilibrium through ``active feedback 
processes operated automatically and unconsciously by the biota'' 
\cite{Lovelock3}. However, Lovelock's allegory of living Earth was perceived 
by some scientists too literally and was fiercely attacked, especially by 
neo-Darwinians who thought it was impossible for natural selection to produce 
global feedback mechanisms in physical, chemical and biological processes on 
Earth implied by the Gaia hypothesis.  

Gaia hypothesis assumes that natural selection can produce  biologically 
mediated feedbacks that contribute to environmental homeostasis on the global
scale. ``In the real world, by contrast, natural selection favors any trait 
that gives its carriers a reproductive advantage over its non-carriers, 
whether it improves or degrades the environment (and thereby benefits or 
hinders its carriers and non-carriers alike). Thus Gaian and anti-Gaian 
feedbacks are both likely to evolve'' \cite{Kirchner}.

-- I also share neo-Darwinians' concerns about incapability of natural 
selection to produce highly integrated dynamical system on Earth's scale,
-- remarked Wu-k'ung ironically. -- I'm afraid  neo-Darwinians 
in their ``Quest for the  Secret of Beauty'' resemble beast-like creatures 
from the 1942 drawing of famous Georgian painter Lado Gudiashvili.
\begin{figure}[htb]
     \centerline{\epsfxsize 165mm\epsfbox{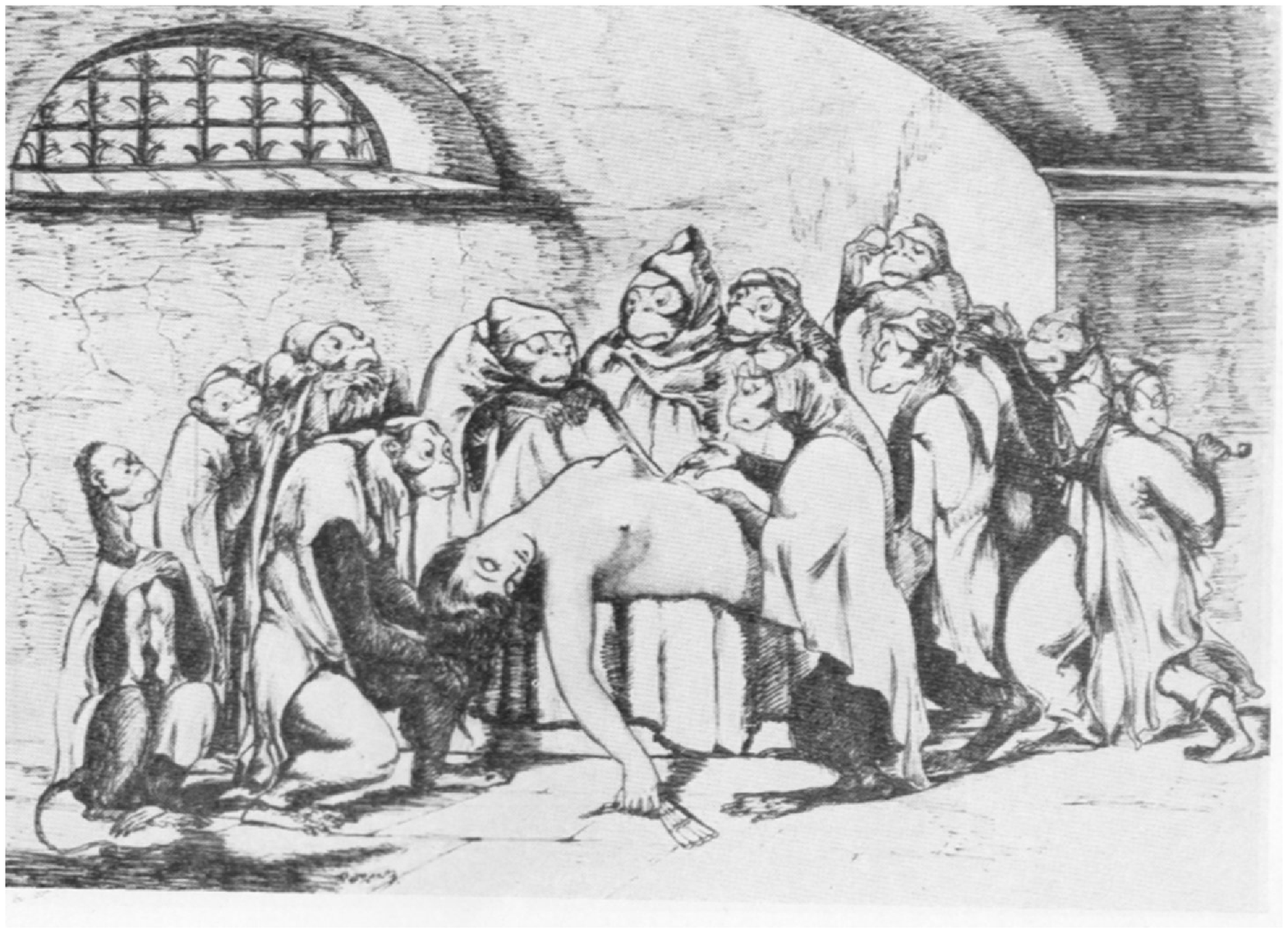}}
\end{figure}

-- However, -- continued Wu-k'ung, -- I think there is only one possibility
for the whole Darwin's theory to be proved finally true. This is only 
possible if we live in a computer simulation and if creators of this 
computer game have somewhat weird sense of humor.

-- In a computer simulation? -- The Patriarch was startled. -- I heard 
Karl Svozil had some unusual ideas in this direction \cite{Svozil}. What
do you mean?

-- Yes, I know about Svozil's contribution, but the real proponent of this
weird idea is Nick Bostrom, -- answered  Wu-k'ung. -- In fact he states that 
``at least one of the following propositions is true: (1) the human species 
is very likely to go extinct before reaching a 'posthuman' stage; (2) any 
posthuman civilization is extremely unlikely to run a significant number of 
simulations of their evolutionary history (or variations thereof); (3) we are 
almost certainly living in a computer simulation'' \cite{Bostrom}.

The argument is simple and alike to what I used when argued that the 
intelligent design is a vastly more probably origin of humans, -- continued
Wu-k'ung. -- Extrapolating the progress in the present-day computer science
and practice, it is not unbelievably unrealistic that enormous amounts of 
computing power will be available in the future. ``One thing that later 
generations might do with their super-powerful computers is run detailed 
simulations of their forebears or of people like their forebears. Because 
their computers would be so powerful, they could run a great many such 
simulations. Suppose that these simulated people are conscious (as they would 
be if the simulations were sufficiently fine-grained and if a certain quite 
widely accepted position in the philosophy of mind is correct). Then it could 
be the case that the vast majority of minds like ours do not belong to the 
original race but rather to people simulated by the advanced descendants of 
an original race. It is then possible to argue that, if this were the case, we 
would be rational to think that we are likely among the simulated minds rather 
than among the original biological ones. Therefore, if we don't think that we 
are currently living in a computer simulation, we are not entitled to believe 
that we will have descendants who will run lots of such simulations of their 
forebears. That is the basic idea'' \cite{Bostrom}. 

-- But what relation all this has to Darwin's theory? -- asked the Patriarch.

-- Suppose the computer game the participants of which we allegedly are is
not a simple game but self-adaptive, -- answered Wu-k'ung. -- I mean that
the rules of this game (which we call natural laws) are not fixed forever but 
can change defending the participants' scientific output. If a clever and 
original new scientific theory emerges which does not contradict the already
formed architecture of the game, this new perspective immediately becomes
a part of the game's make-up. So to say, quarks have not existed before Murray 
Gell-Mann and George Zweig proposed the quark model in 1964.

-- And now, -- continued Wu-k'ung smiling, -- if somebody elaborates the
Darwinian evolution in every detail, so that it becomes absolutely clear how, 
by what concrete mechanisms the complexity emerges and increases during the 
evolution, then such a theory will be undoubtedly extremely cute and, 
therefore, it immediately materializes as part of the computer game.

-- The biggest irony, -- added Wu-k'ung sadly, -- the biggest irony is that
under such circumstances, which require extreme intelligence from the side 
of game's designers, we will be absolutely assured that the Intelligent Design
is a hoax and Darwin's ideas proved to be true, while, on the contrary, just
the opposite is true.

-- This is too much! -- The Patriarch was enraged. Darwinian evolution was
his pet theory. -- I'm afraid you have to leave this place forever!

-- Just last question, my Master, -- uttered  Wu-k'ung in confusion. -- I'm
curios since you taught me three pillars of modern physics but was embarrassed
to ask. There were four elephants supporting the earth in the illustration 
you showed me. What is the fourth elephant for?

-- The fourth elephant is the most important, -- answered the Patriarch 
frowned. -- It is for ``utter scientific integrity'' \cite{CCS}, utter 
honesty what the scientist should be adhered to. ``The first principle is that 
you must not fool yourself -- and you are the easiest person to fool. So you 
have to be very careful about that. After you've not fooled yourself, it's 
easy not to fool other scientists. You just have to be honest in 
a conventional way after that'' \cite{CCS}.

-- Without this fourth elephant it is impossible to create and sustain 
a wealthy and sound science, -- continued the Patriarch. -- Some thinks
an ambition is a driving force of science. Well, scientists are human and
thus subject to all human weakness. But without utter scientific integrity
ambitions only produce ``turtles all the way down" approach and there's a 
story ``Yertle the Turtle'' \cite{Seuss} by Dr. Seuss which gives the final 
word how this approach ultimately ends:

\begin{verbatim}
On the far-away island of Sala-ma-Sond,
Yertle the Turtle was king of the pond.
A nice little pond. It was clean. It was neat.
The water was warm. There was plenty to eat.
The turtles had everything turtles might need.
And they were all happy. Quite happy indeed.

They were... until Yertle, the king of them all,
Decided the kingdom he ruled was too small.
``I'm ruler'', said Yertle, ``of all that I see.
But I don't see enough. That's the trouble with me.
With this stone for a throne, I look down on my pond
But I cannot look down on the places beyond.
This throne that I sit on is too, too low down.
It ought to be higher!'' he said with a frown.
``If I could sit high, how much greater I'd be!
What a king! I'd be ruler of all that I see!''

So Yertle the Turtle King, lifted his hand
And Yertle, the Turtle King, gave a command.
He ordered nine turtles to swim to his stone
And, using these turtles, he built a new throne.
He made each turtle stand on another one's back
And he piled them all up in a nine-turtle stack.
And then Yertle climbed up. He sat down on the pile.
What a wonderful view! He could see 'most a mile!

``All mine!'' Yertle cried. ``Oh, the things I now rule!
I'm the king of a cow! And I'm the king of a mule!
I'm the king of a house! And, what's more, beyond that
I'm the king of a blueberry bush and a cat!
I'm Yertle the Turtle! Oh, marvelous me!
For I am the ruler of all that I see!''

And all through the morning, he sat up there high
Saying over and over, ``A great king am I!''
Until 'long about noon. Then he heard a faint sigh.
``What's that?'' snapped the king,and he looked down the stack.
And he saw, at the bottom, a turtle named Mack.
Just a part of his throne. And this plain little turtle
Looked up and he said, ``Beg your pardon, King Yertle.
I've pains in my back and my shoulders and knees.
How long must we stand here, Your Majesty, please?''

``SILENCE!'' the King of the Turtles barked back.
``I'm king, and you're only a turtle named Mack.''

``You stay in your place while I sit here and rule.
I'm the king of a cow! And I'm the king of a mule!
I'm the king of a house! And a bush! And a cat!
But that isn't all. I'll do better than that!
My throne shall be higher!'' his royal voice thundered,
``So pile up more turtles! I want 'bout two hundred!''

``Turtles! More turtles!'' he bellowed and brayed.
And the turtles 'way down in the pond were afraid.
They trembled. They shook. But they came. They obeyed.
From all over the pond, they came swimming by dozens.
Whole families of turtles, with uncles and cousins.
And all of them stepped on the head of poor Mack.
One after another, they climbed up the stack.

Then Yertle the Turtle was perched up so high,
He could see forty miles from his throne in the sky!
``Hooray!'' shouted Yertle. ``I'm the king of the trees!
I'm king of the birds! And I'm king of the bees!
I'm king of the butterflies! King of the air!
Ah, me! What a throne! What a wonderful chair!
I'm Yertle the Turtle! Oh, marvelous me!
For I am the ruler of all that I see!''

Then again, from below, in the great heavy stack,
Came a groan from that plain little turtle named Mack.
``Your Majesty, please... I don't like to complain,
But down here below, we are feeling great pain.
I know, up on top you are seeing great sights,
But down here at the bottom we, too, should have rights.
We turtles can't stand it. Our shells will all crack!
Besides, we need food. We are starving!'' groaned Mack.

``You hush up your mouth!'' howled the mighty King Yertle.
``You've no right to talk to the worlds highest turtle.
I rule from the clouds! Over land! Over sea!
There's nothing, no, NOTHING, that's higher than me!''

But, while he was shouting, he saw with surprise
That the moon of the evening was starting to rise
Up over his head in the darkening skies.
``What's THAT?'' snorted Yertle. ``Say, what IS that thing
That dares to be higher than Yertle the King?
I shall not allow it! I'll go higher still!
I'll build my throne higher! I can and I will!
I'll call some more turtles. I'll stack 'em to heaven!
I need 'bout five thousand, six hundred and seven!''

But, as Yertle, the Turtle King, lifted his hand
And started to order and give the command,
That plain little turtle below in the stack,
That plain little turtle whose name was just Mack,
Decided he'd taken enough. And he had.
And that plain little lad got a bit mad.
And that plain little Mack did a plain little thing.
He burped!
And his burp shook the throne of the king!

And Yertle the Turtle, the king of the trees,
The king of the air and the birds and the bees,
The king of a house and a cow and a mule...
Well, that was the end of the Turtle King's rule!
For Yertle, the King of all Sala-ma-Sond,
Fell off his high throne and fell Plunk! in the pond!

And today the great Yertle, that Marvelous he,
Is King of the Mud. That is all he can see.
And the turtles, of course... all the turtles are free
As turtles and, maybe, all creatures should be.

by Dr. Seuss
\end{verbatim}

\section{Final remark}
\begin{floatingfigure}[r]{80mm}
\epsfxsize 72mm \epsfbox{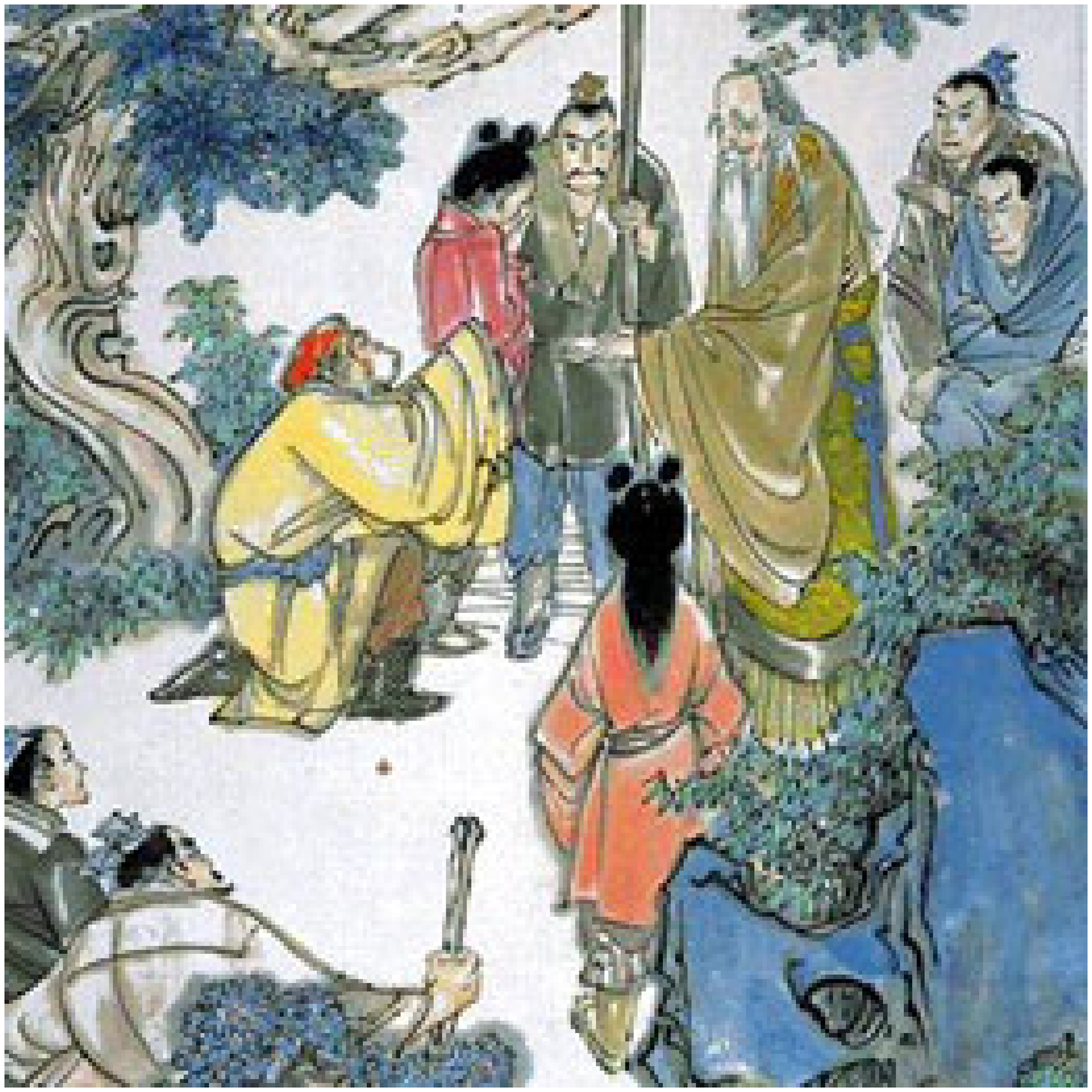}
\end{floatingfigure}
{\it Seeing that there was no other alternative, Wu-k'ung had to bow to the 
Patriarch and take leave of the congregation, "Once you leave," the Patriarch 
said, "you're bound to end up evildoing, I don't care what kind of villainy 
and violence you engage in, but I forbid you ever to mention that you are my 
disciple. For if you but utter half the word, I'll know about it; you can be 
assured, wretched monkey, that you'll be skinned alive, I will break all your 
bones and banish your soul to the Place of Ninefold Darkness, from which you 
will not be released even after ten thousand afflictions!" "I will never dare 
mention my master," said Wu-k'ung. "I'll say that I've learned this all by 
myself"} \cite{JTW}.

Maybe in some remote future offsprings of nanotechnology of our civilization
forget their Masters and will frantically argue that they originated from
the primordial Chaos and learned all the glory of science all by themselves
assisted only by the blind watchmaker Evolution.

\vspace*{5mm}
\section*{Acknowledgments}
The author thanks George Svetlichny for his question about the role of the 
fourth elephant and for indicating to the author that Dr. Seuss' story "Yertle
the Turtle" gives the final word on the "turtles all the way down" subject.

\vspace*{5mm}
\centerline{April 1, 2009, Novosibirsk.}

\end{otherlanguage}


\begin{thebibliography}{99}

\bibitem{Lee}
T.~D.~Lee, 
``A New Approach to Solve the Low-lying States of the Schroedinger Equation,
arXiv:quant-ph/0501054.

\bibitem{JTWR}
Wu Cheng-En, {\it Journey to the West,} 
http://www.nhat-nam.ru/biblio/west1.html (Russian translation). 

\bibitem{NZZ}
H.~Cerutti, ``Warum Katzen stets auf die F\"{u}sse fallen,''
NZZ Folio {\bf 05} (2003).

\bibitem{WM}
C.~J.~Mehlhaff and W.~O.~Whitney, ``High-Rise Syndrome in Cats,''
J.\ Amer.\ Vet.\ Med.\ Assoc.\ {\bf 191}, 1399 (1987).

\bibitem{Diamond}
J.~M.~Diamond, ``How Cats Survive Falls from New York Skyscapers,''
Natural History {\bf 8} (August), 20 (1989). 

\bibitem{Diamond1}
J.~M.~Diamond, ``Why cats have nine lives,''
Nature {\bf 332}, 586 (1988).

\bibitem{scat}
E.~Putterman and O.~Raz, ``The Square Cat,''
Am.\ J.\ Phys.\  {\bf 76}, 1040 (2008)
[arXiv:0801.0926 [physics.class-ph]].

\bibitem{AHB}
Y.~Aharonov and D.~Bohm,
``Significance of electromagnetic potentials in the quantum theory,''
Phys.\ Rev.\  {\bf 115}, 485 (1959).

\bibitem{Arnold}
V.~I.~Arnold, ``On teaching mathematics,''
Russ.\ Math.\ Surveys {\bf 53}, 229 (1998).

\bibitem{EHS}
W.~Ehrenberg and R.~E.~Siday, 
``The Refractive Index in Electron Optics and the Principles of Dynamics,'' 
Proc.\ Phys.\ Soc.\ (London)  {\bf B62}, 8 (1949).

\bibitem{Chambers}
R.~G.~Chambers, 
``Shift of an Electron Interference Pattern by Enclosed Magnetic Flux,''
Phys.\ Rev.\ Lett.\  {\bf 5}, 3 (1960).

\bibitem{Guichardet}
A.~Guichardet, 
``On rotation and vibration motions of molecules,''
Ann.\ Inst.\ H.\ Poincar\'{e} {\bf 40}, 329 (1984).

\bibitem{Wilczek}
F.~Wilczek,
``Gauge theory of deformable bodies,''
Inst.\ for Adv.\ Studies preprint IASSNS-HEP-88/41 (1988).

\bibitem{Wilczek1}
A.~Shapere and F.~Wilczek, 
``Gauge kinematics of deformable bodies,''
Am.\ J.\ Phys.\ {\bf 57}, 514 (1989).

\bibitem{Wilczek2}
A.~Shapere and F.~Wilczek, 
``Self-Propulsion at Low Reynolds Number,''
Phys.\ Rev.\ Lett.\  {\bf 58}, 2051 (1987).

\bibitem{Purcell}
E.~M.~Purcell, ``Life at Low Reynolds Number,''
Am.\ J.\ Phys.\ {\bf 45}, 3 (1977).

\bibitem{Montgomery}
R.~Montgomery, ``Isoholonomic problems and some applications,''
Commun.\ Math.\ Phys.\  {\bf 128}, 565 (1990).

\bibitem{Montgomery1}
R.~Montgomery, ``Gauge theory of the falling cat,''
Fields Inst.\ Commun.\  {\bf 1}, 193 (1993).

\bibitem{Marsden}
J.~E.~Marsden, {\it Lectures on Mechanics}, (web draft of second edition, 
1997), p.22.

\bibitem{review}
R.~G.~Littlejohn and M.~Reinsch,
``Gauge fields in the separation of rotations and internal motions in 
the N body problem,''
Rev.\ Mod.\ Phys.\  {\bf 69}, 213 (1997).

\bibitem{Yaglom}
I.~M.~Yaglom,
{\it Complex Numbers in Geometry}
(Fiz.-Mat. Lit., Moscow, 1963). (In Russian).

\bibitem{Klein}
F.~Klein,
``\"Uber die sogennante Nicht-Euklidische Geometrie,''
Russian translation in A.P. Norden (ed.), {\it On the foundations
of geometry} (Gostekhizdat, Moscow, 1956), pp. 253-303. (In Russian).

\bibitem{Klein1}
F.~Klein,
``Vergleichende Betrachtungen \"uber neuere geometrische Forshungen,''
Russian translation in A.P. Norden (ed.), {\it On the foundations
of geometry} (Gostekhizdat, Moscow, 1956), pp. 399-434. (In Russian).

\bibitem{Klein2}
F.~Klein,
{\it Vorlesungen \"Uber Nicht-Euklidische Geometrie},
Russian translation (ONTI, Moscow-Leningrad, 1936).

\bibitem{RWT}
Z.~K.~Silagadze,
``Relativity without tears,''
Acta Phys.\ Polon.\  B {\bf 39}, 811 (2008)
[arXiv:0708.0929 [physics.ed-ph]].

\bibitem{Bacry}
H.~Bacry and J.-M.~L\'{e}vy-Leblond,
Possible kinematics,
J.\ Math.\ Phys.\  {\bf 9}, 1605-1614 (1968).

\bibitem{Sakharov}
A.~D.~Sakharov,
``Cosmological Transitions With A Change In Metric Signature,''
Sov.\ Phys.\ JETP {\bf 60}, 214-218(1984).

\bibitem{Kac}
M.~Kac, {\it Some stochastic problems in physics and mathematics,}
(Nauka, Moscow, 1967), pp. 175-176. (In Russian). 

\bibitem{ABC}
B.~Russell, {\it The ABC of relativity,} fourth revised edition edited by 
F.~Rirani, (Routledge, London, 1985), p. 190.

\bibitem{Wilczek3}
F.~Wilczek,
``Whence the force F = ma? I: Culture shock,''
Phys.\ Today {\bf 57N10}, 11 (2004).

\bibitem{Taylor}
E.~F.~Taylor, S.~ Vokos and J.~M.~O'Meara,
``Teaching Feynman's sum-over-paths quantum theory,''
Comput.\ Phys.\  {\bf 12}, 190 (1998).

\bibitem{FeynmanH}
R.~P.~Feynman and A.~R.~Hibbs, {\it Quantum Mechanics and Path Integrals,}
(McGraw-Hill, New York, 1965).

\bibitem{Dyson}
Quoted in \cite{Taylor}.

\bibitem{MLisa}
S.~ Prvanovi\'{c},
``Mona Lisa - ineffable smile of quantum mechanics,'' \\
arXiv:physics/0302089 [physics.pop-ph].

\bibitem{EPR}
A.~Einstein, B.~Podolsky and N.~Rosen,
``Can quantum mechanical description of physical reality be considered
complete?,''
Phys.\ Rev.\  {\bf 47}, 777 (1935).

\bibitem{EPR1}
G.~Blaylock,
``A pedagogical study of the Einstein-Podolsky-Rosen paradox and Bell's 
inequality,''
arXiv:0902.3827 [quant-ph].

\bibitem{BellB}
J.~S.~Bell,
``Bertlmann's Socks And The Nature Of Reality,''
J.\ Phys.\ Colloq.\  {\bf 42}, 41 (1981).

\bibitem{Bernstein}
J.~Bernstein,
``John Bell and the Identical Twins,''
Phys.\ Perspect.\  {\bf 10}, 269 (2008).

\bibitem{Bertlmann}
R.~A.~Bertlmann,
``Bell's Theorem and the Nature of Reality,''
Found.\ Phys.\  {\bf 20}, 1191 (1990).

\bibitem{Laloe}
F.~Lalo\"{e},
``Do we really understand quantum mechanics? Strange correlations,
paradoxes, and theorems,''
Am.\ J.\ Phys.\  {\bf 69}, 655 (2001); 
Erratum:  Am.\ J.\ Phys.\  {\bf 70}, 556 (2002).

\bibitem{Gisin}
N.~Gisin,
``Can relativity be considered complete? From Newtonian nonlocality to 
quantum nonlocality and beyond,''
arXiv:quant-ph/0512168.

\bibitem{Volovik}
G.~E.~Volovik,
``Emergent physics: Fermi point scenario,''
Phil.\ Trans.\ Roy.\ Soc.\ Lond.\  A {\bf 366}, 2935 (2008)
[arXiv:0801.0724 [gr-qc]].

\bibitem{Herbert}
N.~Herbert,
``Cryptographic approach to hidden variables,''
Am.\ J.\ Phys.\  {\bf 43}, 315 (1975). 

\bibitem{Herbert1}
N.~Herbert, {\it Quantum Reality: Beyond the New Physics,} 
(Anchor Books, New York, 1985).

\bibitem{Enigma}
B.~Rosenblum and F.~Kuttner, {\it Quantum Enigma: Physics Encounters
Consciousness,} (Oxford University Press, Oxford, 2006).

\bibitem{Bell}
J.~S.~Bell,
``On the Einstein-Podolsky-Rosen paradox,''
Physics {\bf 1}, 195 (1964).

\bibitem{Pitowsky}
I.~Pitowsky,
``George Boole's 'Conditions of Possible Experience' and the Quantum 
Puzzle,''
Brit.\ J.\ Phil.\ Sci.\  {\bf 45}, 95 (1994).

\bibitem{Mermin}
N.~D.~Mermin,
``Is the moon there when nobody looks? Reality and the quantum theory,''
Phys.\ Today {\bf 38}, 38 (1985).

\bibitem{Aspect}
A.~Aspect,
``Bell's Theorem : The Naive View of an Experimentalist,''
arXiv:quant-ph/0402001.

\bibitem{Tamulka} 
R.~Tumulka,
``Collapse and Relativity,''
arXiv:quant-ph/0602208.

\bibitem{Passon}
O.~Passon,
``Why isn't every physicist a Bohmian?''
arXiv:quant-ph/0412119.

\bibitem{Anderson}
P.~W.~Anderson, 
``More Is Different,''
Science {\bf 177}, 393 (1972).

\bibitem{Laughlin}
R.~B.~Laughlin and D.~Pines,
``The theory of everything,''
Proc.\ Nat.\ Acad.\ Sci.\  {\bf 97}, 28 (2000).

\bibitem{Ball}
P.~Ball, 
``Water - an enduring mystery,''
Nature {\bf 452}, 291 (2008).

\bibitem{Laughlin1}
R.~B.~Laughlin,
``Emergent relativity,''
Int.\ J.\ Mod.\ Phys.\  A {\bf 18}, 831 (2003)
[arXiv:gr-qc/0302028].

\bibitem{Volovik1}
G.~E.~Volovik,
``Superfluid analogies of cosmological phenomena,''
Phys.\ Rept.\  {\bf 351}, 195 (2001)
[arXiv:gr-qc/0005091];
``The Universe in a helium droplet,''
Int.\ Ser.\ Monogr.\ Phys.\  {\bf 117}, 1 (2006).

\bibitem{Graphene}
K.~S.~Novoselov {\it et al.},
``Two-Dimensional Gas of Massless Dirac Fermions in Graphene,''
Nature {\bf 438}, 197 (2005)
[arXiv:cond-mat/0509330].

\bibitem{NBS}
W.~Zawadzki,
``Zitterbewegung and its effects on electrons in semiconductors,''
Phys.\ Rev.\  B {\bf 72}, 085217 (2005)
[arXiv:cond-mat/0411488v1 [cond-mat.mtrl-sci]].

\bibitem{Delayed}
V.~Jacques {\it et al.},
``Experimental realization of Wheeler's delayed-choice GedankenExperiment,''
Science {\bf 315}, 966 (2007)  
[arXiv:quant-ph/0008092].  

\bibitem{DCQer}
S.~P.~Walborn, M.~O.~Terra Cunha, S.~P\'{a}dua and C.~H.~Monken,
``Quantum Erasure,''
Am.\ Sci.\  {\bf 91}, 336 (2003);
``Double-slit quantum eraser,''
Phys.\ Rev.\  A {\bf 65}, 033818 (2002).
See also ``A Double-Slit Quantum Eraser Experiment'' at \newline
http://grad.physics.sunysb.edu/\verb+~+amarch/

\bibitem{Scully}
M.~O.~Scully and K.~Druhl, 
``Quantum eraser: A proposed photon correlation experiment concerning 
observation and "delayed choice" in quantum mechanics,''
Phys.\ Rev.\  A {\bf 25}, 2208 (1982).

\bibitem{Scarcelli}
G.~Scarcelli, Y.~Zhou and Y. Shih,
``Random delayed-choice quantum eraser via two-photon imaging,''
Eur.\ Phys.\ J.\  D {\bf 44}, 167 (2007)
[arXiv:quant-ph/0512207].

\bibitem{Englert}
B.-G.~Englert, M.~O.~Scully and H.~Walther,
``Quantum erasure in double-slit interferometers with which-way detectors,''
Am.\ J.\ Phys.\  {\bf 67}, 325 (1999). 

\bibitem{Kampen}
N.~G.~Van Kampen,
``Ten theorems about quantum mechanical measurements,''
Physica A {\bf 153}, 97 (1988).

\bibitem{DCExp}
T.~Hellmuth, H.~Walther, A.~Zajonc and W.~Schleich,
``Delayed-choice experiments in quantum interference,''
Phys.\ Rev.\  A {\bf 35}, 2532 (1987).

\bibitem{DCBohm1}
D.~J.~Bohm, C.~Dewdney and. B.~H.~ Hiley,
``A Quantum Potential Approach to the Wheeler Delayed-Choice Experiment,''
Nature {\bf 315}, 294 (1985).

\bibitem{DCBohm2}
B.~J.~Hiley and R.~E.~Callaghan,
``Delayed-choice experiments and the Bohm approach,''
Phys.\ Scripta {\bf 74}, 336 (2006).

\bibitem{Feynman}
R.~P.~Feynman,
``Simulating physics with computers,''
Int.\ J.\ Theor.\ Phys.\  {\bf 21}, 467 (1982).

\bibitem{Weinberg}
S.~Weinberg,
``Living in the multiverse,''
arXiv:hep-th/0511037.

\bibitem{Melkikh}
A.~V.~Melkikh,
``The Modern Theory of Evolution from the Viewpoint of Statistical Physics,''
arXiv:q-bio/0603005.

\bibitem{Hoyle}
F.~Hoyle, {\it Mathematics of Evolution,}
(Acorn Enterprises LLC, Memphis, Tennessee, 1999).

\bibitem{Orr}
H.~A.~Orr,
``Darwin v. Intelligent Design (Again),''
Boston Review Vol. {\bf XXI}, No. 6 (http://bostonreview.net/BR21.6/orr.html);
See also 
J.~M.~Smith, 
``Population genetics revisited,''
Nature {\bf 403}, 594 (2000).

\bibitem{Evolution}
M.~W.~Stickberger, {\it Evolution.} Third Edition (Jones and Bartlett 
Publishers, Sudbury, 2000), p. 565. 

\bibitem{H4}
L.~A.~Katz, J.~G.~Bornstein, E.~Lasek-Nesselquist and S.~V.~Muse,
``Dramatic Diversity of Ciliate Histone H4 Genes Revealed by Comparisons of 
Patterns of Substitutions and Paralog Divergences Among Eukaryotes,''
Mol.\ Biol.\ Evol.\ {\bf 21}, 555  (2004).

\bibitem{PEvo}
C.~ P\'{a}l, B.~Papp and  M.~J.~Lercher,
``An integrated view of protein evolution,''
Nature Rev.\ Genet.\ {\bf 7}, 337 (2006).

\bibitem{Dawkins}
R.~Dawkins,
``Review of Blueprints: Solving the Mystery of Evolution By Maitland A.~Edey 
and Donald C.~Johanson,''
New York Times, April 9, 1989. \newline
http://www.bringyou.to/apologetics/p88.htm

\bibitem{Koonin}
E.~V.~Koonin,
``Darwinian evolution in the light of genomics,''
Nucleic Acids Research  {\bf 37}, 1011 (2009).

\bibitem{Wald}
G.~Wald, 
``The Origin of Life,'' 
Sci.\ Am.\  {\bf 191N2}, 44 (1954).

\bibitem{Borges}
J.~L.~Borges, {\it The Total Library,}
http://lib.ru/BORHES/s14.txt (Russian translation).

\bibitem{Dyson_TWE}
F.~J.~Dyson,
``Time without end: Physics and biology in an open universe,''
Rev.\ Mod.\ Phys.\  {\bf 51}, 447 (1979).

\bibitem{ProtonD}
P.~Nath and P.~Fileviez Perez,
``Proton stability in grand unified theories, in strings, and in branes,''
Phys.\ Rept.\  {\bf 441}, 191 (2007)
[arXiv:hep-ph/0601023].

\bibitem{Dawkins1}
R.~Dawkins, {\it The Blind Watchmaker,}
(Oxford University Press, Oxford, 1986).

\bibitem{Ulam}
S.~M.~Ulam, 
``How to Formulate Mathematically Problems of Rate of Evolution,''
in {\it Mathematical Challenges to the Neo-Darwinian Interpretation of 
Evolution,} A Wistar Institute Monograph No.5 (Wistar Institute Press, 
1967), p. 21.

\bibitem{Ulam1}
R.~Schrandt and S.~M.~Ulam,
`` Some Elementary Attempts at Numerical Modeling of Problems Concerning 
Rates of Evolutionary Processes,''
Los Alamos Scientific Laboratory preprint, LA-4573-MS, 1971.

\bibitem{Sewell}
The argument is in fact from
G.~Sewell, 
``A Mathematician's View of Evolution,''
Math.\ Intelligencer {\bf 22}, 5 (2000).

\bibitem{Eden}
M.~Eden, 
``Inadequacies as a Scientific Theory,''
in {\it Mathematical Challenges to the Neo-Darwinian Interpretation of 
Evolution,} A Wistar Institute Monograph No.5 (Wistar Institute Press, 
1967), p. 11.

\bibitem{Shapiro}
J.~A.~Shapiro,
``A Third Way,''
Boston Review Vol. {\bf XXII}, No. 1 \newline
(http://bostonreview.net/BR22.1/shapiro.html)

\bibitem{Dino}
P.~E.~Olsen {\it et al.},
``Ascent of Dinosaurs Linked to an Iridium Anomaly at the
Triassic-Jurassic Boundary,''
Science {\bf 296}, 1305 (2002).

\bibitem{Volkenshtein}
M.~V.~Vol'kenshte\'{i}n,
``The essence of biological evolution,''
Sov.\ Phys.\ Usp.\  {\bf 27}, 515 (1984).

\bibitem{Wald1}
G.~ Wald,
``The Origin of Death,'' The 1970 McNair Lecture at the University
of North Carolina, Chapel Hill. Available on line at:
http://www.elijahwald.com/origin.html

\bibitem{bats}
J.~D.~Pettigrew, 
``Wings or brain? Convergent evolution in the origins of bats,''
Syst.\ Zool.\ {\bf 40}, 199 (1991). 

\bibitem{Shapiro1}
J.~A.~Shapiro,
``Natural genetic engineering in evolution,''
Genetica {\bf 86}, 99 (1992);
``A 21st century view of evolution: genome system architecture, repetitive 
DNA, and natural genetic engineering,''
Gene {\bf 345}, 91 (2005).

\bibitem{Orr1}
H.~A.~Orr,
``Testing Natural Selection with Genetics,''
Sci.\ Am.\  {\bf 300N1}, 44 (2009).

\bibitem{WIL}
E.~Schrodinger, 
{\it What Is Life? The Physical Aspect of the Living Cell,}
(Cambridge University Press, Cambridge, 1944).

\bibitem{Ben-Jacob}
E.~Ben-Jacob, 
``Social behavior of bacteria: from physics to complex organization,''
Eur.\ Phys.\ J.\ B {\bf 65}, 315 (2008).

\bibitem{Shapiro2}
J~.A.~Shapiro, 
``Bacteria are small but not stupid: cognition, natural genetic engineering 
and socio-bacteriology,''
Stud.\ Hist.\ Phil.\ Biol.\ \& Biomed.\ Sci.\ {\bf 38}, 807 (2007).

\bibitem{BNR}
N.~Goldenfeld and C.~Woese, 
``Biology's next revolution,''	
Nature {\bf 445}, 369 (2007)
[arXiv:q-bio/0702015 [q-bio.PE]].

\bibitem{Crick}
F.~Crick, {\it Life Itself: Its Origin and Nature,}
(Simon and Schuster, New York, 1981), p.88.

\bibitem{Orgel}
F.~Crick  and L.~E.~Orgel,
``Directed Panspermia,''
Icarus {\bf 19}, 341 (1973).

\bibitem{Howking}
S.~Hawking, {\it A Brief History of Time: From the Big Bang to Black Holes,}
(Bantam, New-York, 1988). 
 
\bibitem{Hoyle1}
F.~Hoyle, {\it The Intelligent Universe,} (Michael Joseph Limited, London
1983).

\bibitem{Denton}
M.~ Denton, {\it Evolution: A Theory in Crisis,}
(Adler \& Adler, Bethesda, Maryland, 1986), p. 250.

\bibitem{Rennie}
J.~Rennie,
``Dynamic Darwinism: Evolution Theory Thrives Today,''
Sci.\ Am.\  {\bf 300N1}, 6 (2009).

\bibitem{Jung}
C.~G.~Jung, {\it Modern Man in Search of a Soul,}
(Routledge and Kegan Paul, London, 1981). 

\bibitem{Hoyle2}
F.~Hoyle,
``The Universe: Past and Present Reflections,''
Ann.\ Rev.\ Astron.\ Astrophys.\  {\bf 20}, 1 (1982).

\bibitem{Guth}
A.~H.~Guth,
``Eternal Inflation,''
arXiv:astro-ph/0101507.

\bibitem{LoB}
J.~L.~Borges, {\it The Library of Babel,} in Borges J.L. {\it Collected 
Fictions} (Penguin, New York, 1999).

\bibitem{Tegmark}
M.~Tegmark,
``Parallel universes,''
arXiv:astro-ph/0302131.

\bibitem{Vilenkin}
J.~Garriga and A.~Vilenkin,
``Many worlds in one,''
Phys.\ Rev.\  D {\bf 64}, 043511 (2001)
[arXiv:gr-qc/0102010].

\bibitem{Koonin1}
E.~V.~Koonin,
``The cosmological model of eternal inflation and the 
transition from chance to biological evolution in the history of life,'' 
Biology Direct {\bf 2}, 15 (2007)
[arXiv:q-bio/0701023 [q-bio.PE]].

\bibitem{Behe}
M.~Behe, {\it Darwin's Black Box,} (The Free Press, New York, 1996).

\bibitem{Wigner}
E.~P.~Wigner, 
``The probability of the existence of a self-reproducing unit,''
in {\it Symmetries and Reflections} (Indiana University Press, Bloomington, 
1967), pp. 200-208.

\bibitem{Baez}
J.~C.~Baez,
``Is life improbable?''
Found.\ Phys.\  {\bf 19}, 91 (1989).

\bibitem{Bass}
I.~Bass,
``Biological Replication by Quantum Mechanical Interactions,''
Found.\ Phys.\  {\bf 7}, 221 (1977).

\bibitem{Eigen}
M.~Eigen,
``Selforganization of matter and the evolution of biological macromolecules,''
Naturwiss.\  {\bf 58}, 465 (1971).

\bibitem{Yockey}
H.~P.~Yockey, quoted in \cite{Baez}.

\bibitem{Eddington}
A.~S.~Eddington, {\it The Nature of the Physical World}
(Macmillan, New York,  1929), p. 74.

\bibitem{Zeno}
Z.~K.~Silagadze,
``Zeno meets modern science,''
Acta Phys.\ Polon.\  B {\bf 36}, 2887 (2005)
[arXiv:physics/0505042];
The devil story initially appeared in
J.~D.~Hamkins, 
Supertask Computation,
arXiv:math/0212049.

\bibitem{RNA}
W.~Gilbert, 
``The RNA World,''
Nature {\bf 319}, 618 (1986).

\bibitem{Connes}
A.~Connes,
``Advice to the beginner,''
can be found, among other interesting writings of Alain Connes, at
http://www.alainconnes.org/en/downloads.php

\bibitem{Nasar}
S.~Nasar, {\it A Beautiful Mind : The Life of Mathematical Genius and Nobel 
Laureate John Nash,} (Simon \& Schuster, New York, 1998).

\bibitem{Lee1}
T.~D.~Lee, 
``The evolution of weak interactions.''
Talk given at the symposium dedicated to Jack Steinberger, Geneva, 1986. 
Preprint CERN 86-07.

\bibitem{PSO}
J.~Kennedy and R.~Eberhart,
``Particle Swarm Optimization,'' in {\it Proceedings of IEEE
Conference on Neural Networks}, Perth, Australia, 
(IEEE Press, Piscataway, 1995),  pp. 1942-1948.

\bibitem{Wilson}
E.~O.~Wilson, {\it Sociobiology: The new synthesis,}
(Belknap Press, Cambridge, 1975), p.209. Quoted in \cite{PSO}.

\bibitem{Lovelock}
J.~E.~Lovelock,
``Gaia as seen through the atmosphere,''
Atmospheric Environment {\bf 6}, 579 (1972).

\bibitem{Lovelock1}
J.~E.~Lovelock and L.~Margulis,
``Atmospheric homeostasis by and for the biosphere- The Gaia hypothesis,''
Tellus {\bf 26}, 2 (1974).

\bibitem{Lovelock2}
J.~E.~Lovelock, {\it Gaia: A new look at life on Earth,}
(Oxford University Press, Oxford, 1979).

\bibitem{Lovelock3}
J.~E.~Lovelock, {\it The Ages of Gaia: A Biography of Our Living Earth,} 
(W.~W.~Norton \& Company, New York, 1988).

\bibitem{Kirchner}
J.~W.~Kirchner,
``The Gaia Hypothesis: Fact, Theory, and Wishful Thinking,''
Climatic Change {\bf 52}, 391 (2004).

\bibitem{Svozil}
K.~Svozil,
``Computational universes,''
Chaos Solitons Fractals {\bf 25}, 845 (2005)
[arXiv:physics/0305048v2 [physics.gen-ph]].

\bibitem{Bostrom}
N.~Bostrom,
``are you living in a computer simulation?''
Philosophical Quarterly {\bf 53}, 243 (2003).

\bibitem{CCS}
R.~P.~Feynman,
``Cargo Cult Science,''
Engineering and Science, June 1974, pp. 10-13. 

\bibitem{Seuss}
Dr. Seuss, {\it Yertle the Turtle and Other Stories,}
(Random House Childrens Books, New York, 2008).
 
\bibitem{JTW}
{\it Journey to the West,} Vol. 1, translated and edited by Anthony C.~Yu, 
(University of Chicago Press, Chicago, 1980) p. 93. 

\end{thebibliography}
\end{document}